\documentclass[11pt,twoside]{article}
        \usepackage{a4,hangcaption,amsmath,amssymb,ctagsplt,%
               latexsym,fancyheadings,epsfig,english }

\begin{document}
\begin{titlepage}
\null
\begin{center}
\Large\bf The CQM model
\end{center}
\vspace{0.5cm}

\begin{center}
\begin{large}
A. D. Polosa\\
\end{large}
\vspace{0.3cm} Department of Physics, P.O. Box 9, FIN-00014
University of Helsinki, Finland
\end{center}

\vspace{0.5cm}

\begin{center}
\begin{large}
{\bf Abstract}\\[0.5cm]
\end{large}
\parbox{14cm}{I review a Constituent-Quark-Meson model (CQM) for heavy meson decays, outlining
its characteristics and the calculation techniques developed for
it. The strength of this effective model, is that it enables to
evaluate heavy meson decay amplitudes through diagrams where the
heavy mesons are attached at the ends of loops containing heavy
and light quark internal lines. The phenomenological applications
are presented in detail, trying to give a self-contained operative
picture of the model.}
\end{center}

\vspace{1.0cm} \noindent PACS: 13.20.He, 12.39.Hg, 12.39.Fe\\
\vfil \noindent  HIP-2000-17/TH\\ April 2000
\end{titlepage}
\setcounter{page}{1}
\tableofcontents
\newpage

\section{\label{chap:intfis}Introduction}

During recent years, heavy meson physics has received a wide
attention both from theory and experiment. This is because it
helps the comprehension of many open problems of the standard
model and can also act as a passage in the domain of new physics.
Many experiments on $B$ physics already at work or near to be
started, BaBar, Belle, CLEO III, Hera-B, CDF-D0 and those planned
to begin after 2005, ATLAS, CMS, LHCB and BTeV confirm this
interest \cite{artuso}. $B$ physics has had an important role also
in  LEP I that has registered about  $10^6$ $Z^0\to \bar{b}b$
events \cite{perret}. $B$ decays offer the framework for
investigating in detail  the field of CP violations and for
determining CKM (Cabibbo-Kobayashi-Maskawa) matrix elements. In
particular, rare $B$ decays, those in which there is no charm in
the final state, are relevant for the research of signals of new
physics \cite{falcorari}. In fact, the Standard Model predicts
that rare $B$ decays (the Cabibbo suppressed or the penguin
induced decays) should be strongly suppressed, therefore, any
anomalous increasing of branching ratios could be due, for
example, to the existence of new particles, external to the
standard model spectrum because interacting at higher energy
scales.

The amplitudes governing heavy meson decays are theoretically
calculated mainly using lattice QCD methods and the SVZ
(Shifman-Vainshtein-Zakharov) sum rules \cite{shif}.

The lattice QCD program \cite{rothe}, is that of computing the QCD
partition functional summing over a representative ensamble of
gauge fields and fermionic field configurations; the action is
written in discrete form modelling the entire space-time as a
four-dimensional grid where the distance between nearest
neighboring sites is $a$ and the linear dimension is  $L \simeq
\Omega^{\frac{1}{4}}$, $\Omega$ being the four volume of the grid.
In principle, considering a sufficiently large number of
configurations and simulating  a very close $(a\to 0)$ and large
$(L\to \infty)$ grid on a calculator, amounts to build a
calculation framework nearly resembling that of continuous QCD. In
practice, there are many technical problems: some of them have to
do with computer power, some with the continuous limit of the
results obtained on a discrete space-time grid.

In the ordinary hadronic matter, the quarks are not very far from
each other, therefore, in ordinary circumstances, it is not
essential to consider the complex QCD dynamics giving rise to the
Abrikosov chromoelectric flux tubes thought to be responsible for
quark confinement. In this situation valence quarks are weakly
interacting with QCD vacuum fluctuations. The SVZ method aims at
determining the parameters and the regularity of ordinary mesons
and baryons through an expansion of the correlation functions,
written in terms of dispersion intergrals, in a power series
controlled by the  $\alpha_s$ parameter (the strong coupling
constant), plus power corrections expressed through the vacuum
condensates ($G_{\mu\nu}^2$, $\bar{q}q$, $\bar{q}\sigma Gq,..$).
It is believed that the vacuum condensates contain the most
relevant non perturbative effects of the QCD vacuum. Invoking the
concept of {\it parton-hadron duality}, this expansion must be
compared to the phenomenological expressions for the correlation
functions. It is this comparison that allows to extract
quantitative information on $2,3,..$-points correlators, {\it
i.e.}, on all possible observables.  One of the main drawbacks of
SVZ sum rules is the difficulty one meets in computing the
theoretical error due to the ambiguous choice concerning the
truncation point of the series expansion.

This work is devoted to introduce an effective
Constituent-Quark-Meson model based on a Lagrangian incorporating
the symmetries of heavy quark effective theory, the chiral
symmetry in the light quark sector, see section {\bf
\ref{chap:primo}}, and, as is discussed in section {\bf
\ref{chap:secondo}}, dynamical information derived from an
underlying Nambu-Jona-Lasinio interaction. In section {\bf
\ref{chap:terzo}}, together with the discussion of calculation
techniques used for computing some relevant loop-integrals, it is
shown how the determination of  strong coupling constants,
parameterizing the low energy effective hadron Lagrangian,
proceeds through a comparison of the low energy matrix elements
with the CQM computed amplitudes: CQM plays the role of a
fundamental model (since it contains, besides meson fields,  also
the elementary heavy and light quark fields) with which the hadron
theory must match at higher energy, see discussion in section {\bf
\ref{sec:fotonefotone}}. With respect to lattice QCD and SVZ sum
rules, CQM is a rough approach that, anyway, has shown to be a
quite reliable and easy-to-use method.

One of the very common problems of quark models \cite{falcorass},
is that of associating theoretical errors to predictions. This
topic is discussed for CQM in section {\bf \ref{chap:secondo}},
together with the problem of defining the light constituent quark
mass. The constituent quark mass is typically heavier than the
current mass, appearing in the QCD Lagrangian (and related to the
Higgs field VEV): one can think of a constituent quark as of a
current (bare) quark dressed by a cloud of virtual particles
generated by strong interactions \cite{okun}. The mechanism
dressing the bare quark and giving the constituent quark its mass
value, is an intrinsic feature of the model itself.

Section {\bf \ref{chap:quarto}} is devoted to the study of
exclusive semileptonic decays of $B$ mesons through the CQM model.
Here are examined  processes involving $b\to c\ell\nu$ and $b\to
u\ell\nu$ transitions, the former being related to $V_{cb}$, the
latter to $V_{ub}$. In particular, CQM has allowed to obtain a
prediction for the branching ratio of the semileptonic process
$B\to a_1$.

All existing evaluations of exclusive semileptonic $B$ decays are
strongly model-dependent or are affected by problems related to
the estimation of the theoretical error. Anyway an agreement among
diverse models, {\it e.g.}, on the determination of a particular
form factor, gives rise to a theoretical platform useful for a
comparison with experimental data. This could also be  an
alternative approach to the study of rare $B$ decays, considering
that the most commonly used method to extract $V_{ub}$ through a
comparison with data, is the so called end-point-method, see, {\it
e.g.}, \cite{martinelli}. The idea of the end-point-method is that
of eliminating the background due to $b\to c\ell\bar{\nu}$ decays
while examining the inclusive leptonic spectrum
$\frac{d}{dE_\ell}\Gamma (b\to u\ell\nu)$ in the $E_\ell$ region
where the invariant mass $M_X$ of the hadron system emerging from
the decay is such to avoid decays in a charmed final state:
$M_X\leq M_D$. But, in this region of the energy spectrum, one
meets technical difficulties related to the Wilson expansion of
$\frac{d\Gamma}{dE_\ell}$: one can only compute the first terms of
this expansion. Higher order terms depend on matrix elements of
local operators having higher dimensionality, and can at most be
estimated by phenomenological models. It is possible to show that,
in the proximity of the end-point, {\it i.e.}, in the proximity of
a certain critical value $M_{X,c}$, all terms in the Wilson
expansion are equally important and, for even higher values of
$E_\ell$, the decay cannot anymore be analyzed by
Operator-Product-Expansion. In the experiments devoted to the
determination of $V_{ub}$, a kinematic cut on $M_X$, very near to
the critical value $M_{X,c}$, is used. This means that the
determination of $V_{ub}$ is model-dependent since it is necessary
to be able to estimate the terms having higher dimensionality in
the Wilson expansion. To avoid this problem, one could think of
enlarging the $E_\ell$ region experimentally examined. This could
give the possibility of being far from $M_{X,c}$, but the price to
pay is that of a strong growth of the background of events
containing charm in the final state.

CQM gives the possibility of further investigating the exclusive
channels $B\to\rho$, $B\to a_1$ and $B\to \pi$ in such a way to
enlarge and give more solidity to the platform of model-dependent
results I mentioned before.
\section{\label{chap:primo} Introduction to the formalism}

        \subsection{\label{sec:efficaci}Effective theories}

In this section I will discuss briefly the general topic of
effective theories in particle physics with the aim of introducing
the basic ideas and tools of the CQM model in the subsequent
sections.

When one calculates the energy levels for an hydrogen atom, the
problem to face is that of solving the Schr\"odinger equation for
an electron moving in the Coulomb potential generated by the
positive proton charge: it is  not relevant to take account of the
inner quark structure of the proton. The low energy dynamics of
the hydrogen atom does not depend in any relevant way on the high
energy finer details of the proton inner structure. The proton can
be simply considered  as the static source of Coulombic potential
and, in a first approximation, we can ignore also its spin and
magnetic moment. Doing in such a way, the problem of determining
hydrogen energy levels presents essentially only one energy scale
$m_e$ (the electron mass) and the dimensionless fine structure
constant $\alpha$: we have separated out higher energy scales.
This can be done essentially because of the large separation of
the energy scales that usually enter into a physical problem. A
physical system in which there are different but close to each
other energy scales, cannot be treated in the same way because
even small perturbations can allow the system to explore all these
scales with similar probabilities.

A finer calculation of the hydrogen energy levels requires to
include in the calculation the effect of the spin and of the
magnetic moment of the proton. These details are responsible of
the well known hyperfine structure of the energy levels. We can
state that the energy levels of the hydrogen atom can be computed
ignoring the dynamics acting at scales larger than $\Lambda$, with
$\Lambda>>m_e \alpha$, with an error of order $m_e \alpha
/\Lambda$. The more the desired precision, the higher is $\Lambda
$, the smaller is the error one makes ignoring the high energy
($>\Lambda$) dynamics. For example, parity violation effects at
the atomic level are very small since the weak interaction energy
scale is $M_W$, extremely larger than the atomic energy scale.

{\it Effective theories}  \cite{rass1}-\cite{rass6} are those {\it
models} conceived to describe the physics of a certain system at
the energy scale of the experiment through which one studies it,
{\it i.e.}, at the level of accuracy chosen to experimentally
examine the system. In this sense, the atomic physics of the
hydrogen atom is an effective theory of the hydrogen.

Effective models succeed in giving reliable phenomenological
predictions where fundamental theories have many more technical
and sometimes principle problems. Quantum-Chromo-Dynamics (QCD) is
the most important example of a fundamental theory, {\it i.e.}, a
theory derived from first principles, describing the intimate
nature of strong interactions and the building fields of matter,
that has deep troubles in dialing with the low energy hadron
world. This is due to the still partial theoretical comprehension
of the confinement mechanism of quarks in the hadronic matter.
Therefore, to deal with hadrons, it is necessary to implement some
low energy model, effective in the energy regions where the
hadronic processes one wants to study take place.

A low energy effective theory of hadrons is anyway a relative of
QCD, since it incorporates the symmetry properties required by the
fundamental theory. The hadron effective Lagrangian must therefore
be  Lorentz invariant, the S-matrix must be unitary, the ${\cal
PCT}$ symmetry must be obeyed   and it has to show chiral symmetry
in the limit in which light current masses are sent to zero. New
symmetry properties could also emerge in the effective theory
being absent in the fundamental one: the example relevant for this
work is that of Heavy-Quark-Effective-Theory (HQET), to which is
devoted the next section.

Symmetry properties select an infinite class of Lagrangian
interaction terms, only a finite number of them being
renormalizable. The requisite of renormalizability, crucial for a
fundamental theory, is lost in the effective theory approach.

The origin of non-renormalizable interactions is due to the
absence of heavy particles from the spectrum of the effective
theory. An example comes from Fermi's $\beta$-decay theory, where
a non-renormalizable four fermion contact term, distorts the high
energy interaction mediated by the $W$ particle, absent in Fermi's
theory. Anyway Fermi's theory works extremely well at the energy
scales of nuclear processes. The masses $M$ of the heavy
particles, excluded by the effective theory spectrum, may appear
as energy cutoffs $\Lambda=M$ suppressing the non renormalizable
terms by factors of $E/M$, $E$ being the characteristic low energy
scale of the processes described by the effective theory. For
example, the typical energy scale of Quantum-Electrodynamics (QED)
processes is of order of $m_e$, that is a sufficiently small
number to explain way QED can be very well considered as a
fundamental, renormalizable theory of electrodynamic interactions.

In general terms one can associate to each mass of a known
particle a boundary between two different effective theories: the
anomalous breaking of scale invariance, manifested in the peculiar
distribution of particle mass values, gives then rise to a tower
of separate effective theories. For energies below a certain
boundary value, one can construct a low energy effective theory in
which all the particle states  above the boundary threshold are
excluded from the spectrum. Of course, the coupling constants in
the interaction terms related to the light fields should vary with
continuity at the boundaries.

As we go down in the energy ladder, we meet effective theories
containing less fields and a larger number of non-renormalizable
terms while, in the opposite direction, we find that the
non-renormalizable terms are progressively more important (less
suppressed by $E/M$) and disappear at boundaries, where they are
substituted by new renormalizable interaction terms.  The
important point is that {\it what happens at high energies doesn't
affect the low energy behaviour}. This picture is deeply explained
in \cite{rass4}, \cite{rass5}.

The renormalization group method \cite{wilson} allows to bridge
between two effective theories. The aim is that of calculating the
low energy parameters through the high energy ones. These
calculations can be explicitly performed only once the high energy
theory is weakly coupled. The QCD  case is therefore complicated
because the renormalization group method doesn't allow to bridge
continuously from the fundamental theory, the QCD, to the hadron
effective theories. This is why, many times, the hadron low energy
effective theory parameters are determined by a matching with some
other more fundamental model, {\it i.e.}, some model containing in
its spectrum also the higher energy elementary particles. These
models are not necessarily QCD derived, like lattice-QCD or SVZ
sum rules. In many cases these models contain hypotheses in
conflict with the QCD structure. Object of this work is to
introduce one of these effective models.

What is important to focus on, is that the proliferation of
non-renormalizable terms (the irrelevant terms in the modern
language), doesn't spoil the predictive power of the effective
theory. On the contrary, non-renormalizable terms can help in
determining the predictive power at disposal.

Here follows an example of how the effective theory approach could
make things very easy with respect to a fundamental theory
approach.

\subsubsection{\label{sec:fotonefotone} Photon-photon scattering}
\begin{figure}[t!]
\begin{center}
\epsfig{bbllx=5cm,bblly=10cm,bburx=15cm,bbury=30cm,height=12truecm,
        figure=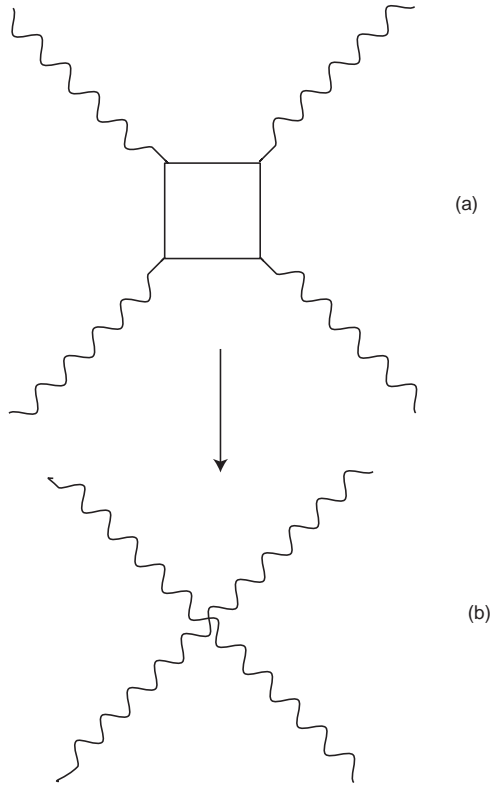}
\caption{\label{fig:fotonefotone} \footnotesize
          QED photon-photon (logarithmic divergent) scattering diagram in (a).
          The Euler-Heisenberg effective interaction in (b). }
\end{center}
\end{figure}
Let us suppose to be interested in understanding how the cross
section for the photon-photon scattering scales with the energy of
the photon in the  limit in which this is lower than the rest
energy of the electron. From an effective field theory point of
view, this means that we are interested in building an effective
theory in which the electron is excluded by the particle spectrum.
The electron mass acts as the cutoff $\Lambda=m_e$ discussed
before.

We therefore only need an interaction Lagrangian  containing four
photon fields. The symmetry principles instructing us about how to
build this low energy effective theory are: Lorentz invariance,
gauge invariance and the ${\cal P,C,T}$ symmetries. To the lowest
order we can therefore write the so called Euler-Heisenberg
Lagrangian:
\begin{equation}
{\cal L}_{\rm eff}=\frac{\alpha^2}{m_e^4} \left[ a_1
(F_{\mu\nu}F^{\mu\nu})^2+ a_2(F_{\mu\nu}{\tilde
F}^{\mu\nu})^2\right]+O(1/m_e^8),
\end{equation}
written in such a way to have the correct mass dimension.
$a_{1,2}$ are the constants multiplying respectively the scalar
squared and the pseudoscalar squared terms. The presence of
$F_{\mu\nu} {\tilde F}^{\mu\nu}$ explains why we cannot have odd
powers in higher order terms. Every gradient of the photon field,
in the lowest order Euler-Heisenberg Lagrangian, produces a factor
of $E_\gamma$. We can therefore argue that the cross section
scales as:
\begin{equation}
\sigma \propto \left(\frac{\alpha^2 E_\gamma^4}{m_e^4}\right)^2
\frac{1}{E_\gamma^2},
\end{equation}
{\it i.e.}, $\sigma\propto \frac{E_\gamma^6}{m_e^8}$. The phase
space factor $\frac{1}{E_\gamma^2}$ is needed because $\sigma$ has
dimensions of a surface and $E_\gamma$ is the only dimensional
parameter in the effective theory. The effective approach
description is the one given in fig. \ref{fig:fotonefotone}(b).
Anyway we could calculate the photon-photon scattering in QED, see
fig. \ref{fig:fotonefotone}(a), by the virtual electron box (we
should add five more diagrams renormalizing its logarithmic
divergence), {\it i.e.}, we could calculate the $ \sigma$ of the
process at high energy using the fundamental theory and then
consider the low energy limit of the result. In such a way,
through the matching of high and low energy Green's functions, we
could obtain the constants $a_1$ and $a_2$. If we didn't knew QED,
we should have fixed $a_1$ and $a_2$ by matching with the
experiment. In the case of effective hadron Lagrangians, this is
the problem: one cannot calculate the couplings (like $a_1$ and
$a_2$), essential for the phenomenological predictions, by a
matching with QCD. This is why one tries to build effective
models, intermediate in energy between QCD and the hadron world,
that could allow this matching.

        \subsection{\label{sec:hqet}Heavy quark effective theory}

The physics we are interested in, is that of mesons containing one
heavy quark ($b$ or $c$) \cite{rass7}-\cite{rass12}. These states
present a large separation of energy scales: on one hand we have
the heavy quark mass $m_Q$ and, on the other hand, we have
$\Lambda_{\rm QCD}$, the asymptotic freedom scale, which limits
the boundary between the strong coupling and the weak coupling
region. The heavy quark is surrounded by  a cloud of light quark
states interacting with it through soft gluons  having momenta of
order $\Lambda_{\rm QCD} \simeq 200-300$ MeV. Being
$m_Q>>\Lambda_{\rm QCD}$, we understand that the exchanged soft
gluons can resolve only larger distances than the heavy quark
Compton wavelenght. This means that light quark degrees of freedom
are blind to the heavy quark flavor ({\it i.e.}, to mass) and
spin. For the light degrees of freedom, the heavy quark is simply
a static chromoelectric source as the proton is simply a source of
Coulombic potential for the electron in the hydrogen atom
(chromomagnetic effects are suppressed by a factor of $1/m_Q$;
spin-spin coupling terms between light and heavy degrees of
freedom are also $1/m_Q$ terms).

We can therefore state that light degrees of freedom in an heavy
meson have a new symmetry property, not remnant of the underlying
QCD description, with respect to flavor and spin rotations of the
heavy quark with which they interact. In particular, the
excitation spectra of two heavy mesons  containing two different
heavy quarks $Q_1$ and $Q_2$ with $m_{Q_1},m_{Q_2}>>\Lambda_{\rm
QCD}$, are the same once one overlaps the ground states. Due to
flavor symmetry, it happens something like the atomic physics
independence of the electron structure  on the neutron number
contained in the nucleus. Due to spin symmetry, each excitation
level will be a doublet, degenerate in the total spin (if light
degrees of freedom are not carrying zero spin).

In a seminal paper by H. Georgi \cite{georgi}, the initial ideas
on heavy mesons and flavor-spin symmetries \cite{isgur,hill} are
translated in the effective field theory language. The aim is that
of building a low energy theory where the heavy quark mass is
considered infinite,  $m_Q \to \infty $, since it is greater than
all other energy scales appearing in the effective theory, while
the heavy quark velocity, which is practically the same of that of
the entire hadron, is a conserved contant of motion. Consider:
\begin{equation}
P^\mu=m_H v^\mu,
\end{equation}
the momentum of a meson having mass $m_H$ and containing an heavy
quark of mass $m_Q$. In the  $m_Q \to \infty$ limit, $m_H=m_Q$. Of
course, in physical situations, the infinite mass limit is not
rigorously  fulfilled and $m_H\neq m_Q$. If we suppose that the
light degrees of freedom carry a small momentum $q^\mu$, we can
define the heavy quark momentum as:
\begin{equation}
p^\mu=P^\mu-q^\mu=m_Qv^\mu+k^\mu,
\end{equation}
where the  ``residual" momentum $k^\mu$ is defined as follows:
\begin{equation}
\label{eq:smallfluctu} k^\mu=(m_H-m_Q)v^\mu-q^\mu.
\end{equation}

Let us now consider the hadron scattering from an external
potential. After the interaction we have a new hadron state,
containing the same heavy quark, and carrying momentum:
\begin{equation}
P^\mu=m_Q v^{\prime \mu} + s^\mu,
\end{equation}
where $s^\mu$ is the finite momentum exchanged with the external
potential in the $m_Q \to \infty$ limit. If  $s^\mu$ is a finite
amount of momentum, then $v^\mu=v^{\prime \mu}$ (a finite momentum
exchange cannot produce an infinite hadron momentum difference).
This means that the velocity is a {\it conserved constant of
motion}, {\it i.e.}, it isn't any more a dynamical degree of
freedom.

When the heavy quark interacts with light degrees of freedom, only
fluctuations of the residual momentum (of order $\Lambda_{\rm
QCD}$) have to be considered, while variations of velocity are
certainly excluded. QCD interactions do not vary $v$; only weak
(or electromagnetic) interactions can annihilate the heavy quark
and create a new one that can be different in flavor, spin and
velocity.

In the Georgi's paper it is therefore introduced a {\it
superselection rule} for the velocity of an heavy quark: the
fields describing the heavy quark in the effective theory should
be $h_v(x)$ fields, {\it i.e.}, depending on $x$ and  $v$. For
different $v$'s we have different heavy quark fields.

We have discussed the flavor-spin symmetry adopting the hypothesis
of an heavy quark at rest. But now we can observe that the
flavor-spin symmetry connects two heavy hadrons containing
different heavy quark flavors only if they have the same velocity.
In other words the $ SU(2 N_h)$ flavor-spin symmetry, where $N_h$
is the number of heavy flavors, transforms meson states $M_{Q_i}$
in meson states $M_{Q_j}$, having different $i\neq j $ flavors,
provided that $Q_i$ and $Q_j$ have the same velocity (not the same
momentum: therefore flavor-spin symmetry is a symmetry of certain
matrix elements, not an S-Matrix symmetry).

Importantly, flavor-spin symmetry is not an exact symmetry since
the heavy masses aren't infinite: the technology allowing to
compute the $1/m_Q$ corrections  is the  HQET
(Heavy-Quark-Effective-Theory).

In this paper we will make frequent use of the heavy quark
effective propagator. This is derived by the  QCD fermion
propagator adopting the momentum formula introduced above:
\begin{equation}
p^\mu_Q=m_Q v^\mu + k^\mu.
\end{equation}
In the  $m_Q \to \infty$ limit, the propagator:
\begin{equation}
\label{eq:otto} i\frac{\gamma \cdot p_Q + m_Q}{p_Q^2-m_Q^2},
\end{equation}
becomes:
\begin{equation}
\frac{1+\gamma \cdot v}{2} \frac{i}{v \cdot k},
\end{equation}
where we have used the relation $k\simeq \Lambda_{\rm QCD}$. The
vertex describing the heavy quark-gluon interaction in QCD is:
\begin{equation}
-ig\gamma_\mu T^a,
\end{equation}
where $T^a$ is a generator of $SU(3)_c$ and  $g$ is the coupling
constant of strong interactions. Due to the structure of the
propagator, the vertex is always intermediate between the
$\frac{1+\gamma \cdot v}{2}$ projectors and therefore the vertex
in the effective theory is:
\begin{equation}
-ig  \frac{1+\gamma \cdot v}{2}  \gamma_\mu  \frac{1+\gamma \cdot
v}{2}  T^a = -ig v_\mu T^a  \frac{1+\gamma \cdot v}{2}.
\end{equation}
The  $\frac{1+\gamma \cdot v}{2}$ projectors, that appear in
propagators and vertices, can be brought on the external lines of
Feynman graphs, where they are annihilated by the on shell heavy
quark spinors  $h_v$ having the property that $\gamma \cdot v
h_v=h_v$, see {\bf \ref{sec:1/mQ}}. We have therefore obtained the
following two Feynman rules:
\begin{eqnarray}
{\rm propagator} &=& \frac{i}{v \cdot k} \\ {\rm
vertex}&=&-igv^\mu T^a.
\end{eqnarray}
Since here the heavy quark mass is absent, the flavor symmetry is
manifested. Moreover there are no Dirac matrices, therefore also
the spin symmetry is manifested.

\subsubsection{\label{sec:1/mQ} $1/m_Q$ expansion}
The Feynman rules given above can be considered as the basic
definitions of the heavy quark effective theory. The same results
can also be obtained using a field theoretical approach
\cite{georgi} avoiding the use of QCD Feynman rules like the
propagator expression (\ref{eq:otto}). Among effective theories,
HQET has a particular role: one of the main goals of HQET is that
of describing the properties of heavy hadron decays, therefore,
even if there is the large separation of scales above mentioned,
we cannot remove completely the heavy quark state from the low
energy effective theory, see section {\bf \ref{sec:efficaci}}.
What can instead be eliminated in the effective theory, are the
components of the heavy quark spinor describing its fluctuation
around the mass shell since, in the $m_Q \to \infty$ limit, the
heavy quark is almost on shell and carries almost the entire
hadron momentum. It is therefore useful to decompose the heavy
quark QCD spinor, $Q(x)$, in its small and large components:
\begin{equation}
Q(x)=e^{-im_Q v \cdot x}(H_v(x)+h_v(x)),
\end{equation}
where:
\begin{eqnarray}
H_v(x) &=& e^{im_Q v \cdot x} \frac{1+\gamma \cdot v}{2} Q(x)\\
h_v(x) &=& e^{im_Q v \cdot x} \frac{1-\gamma \cdot v}{2} Q(x).
\end{eqnarray}
Two properties are evident: $\gamma \cdot v h_v = h_v$ and $\gamma
\cdot v H_v= -H_v$. Moreover, reminding the $\gamma_0$ structure,
one can see that in the rest frame, $v=(1,0,0,0)$, $h_v$
corresponds to the upper components, the so called large
components of the quadrispinor $Q$, while $H_v$ correspond to its
inferior components, the so called small ones. $h_v$ annihilates
an heavy quark having velocity $v$, $H_v$ creates an heavy
antiquark having velocity $v$. Let's consider the following
virtual process discussed by Neubert \cite{rass7}: an heavy quark
propagating forward in time, at the event $a$ inverts his way in
the opposite temporal direction and from the $b$ event on, it
propagates again forward in time. In $a$ we have the annihilation
of a virtual heavy quark-antiquark pair and in $b$ the creation.
The energy in the intermediate virtual state, the one propagating
between $a$ and $b$, is certainly larger, with respect to the
initial one, of about $2 m_Q$.

Therefore, this intermediate state can only propagate over
distances of order $1/2m_Q$, which are very short if compared with
the typical hadron physics distances, of order $1/\Lambda_{\rm
QCD}$. The intermediate virtual state, that is evidently connected
to the action of $H_v$ in $b$, can be simply substituted by the
propagator  $1/2m_Q$. We can therefore state that there is no
sufficient energy to create a virtual heavy quark-antiquark pair
in HQET or, in other words, this process is suppressed at order
$1/m_Q$. We must therefore proceed to the systematic elimination
of the $H_v$ field from the effective Lagrangian therefore
obtaining a non-local effective action for $h_v$. This can be
expanded in a $1/m_Q$ series of local operators.

In terms of $h$ and $H$, the QCD Lagrangian for heavy quarks takes
the form \cite{rass7}:
\begin{eqnarray}
{\cal L}_{\rm QCD} &=& \bar{Q}(i \gamma \cdot D - m_Q) Q
\nonumber\\ &=& \bar{h}_v (iv\cdot D) h_v - \bar{H}_v(iv\cdot
D+2m_Q)H_v \nonumber\\ &+& \bar{h}_v (i \gamma \cdot \tilde{D})
H_v + \bar{H}_v (i \gamma \cdot \tilde{D}) h_v,
\end{eqnarray}
where:
\begin{equation}
\tilde{D}^\mu=D^\mu-v^\mu v \cdot D.
\end{equation}
We conclude that $h$ describes massless light degrees of freedom,
while $H$ describes fluctuations having a mass of $2m_Q$. The
latter must be eliminated from the effective theory. From the last
two terms in ${\cal L}_{\rm QCD}$, describing the creation
(annihilation) of quark-antiquark pairs,  we can see that $H$ and
$h$ fields are mixed together. If we compute the functional
derivative of ${\cal L}_{\rm QCD}$ with respect to $\bar{H}$, we
obtain the following equation of motion for $H$:
\begin{equation}
(iv\cdot D + 2 m_Q) H_v = i \gamma \cdot \tilde{D} h_v,
\end{equation}
that can be formally solved for $H_v$ and the resulting expression
can be inserted in ${\cal L}_{\rm QCD}$, giving:
\begin{equation}
{\cal L}_{\rm eff}=\bar{h}_v (iv\cdot D) h_v + \bar{h}_v (i\gamma
\cdot \tilde{D} \frac{1}{2m_Q + iv\cdot D} i\gamma \cdot
\tilde{D}) h_v,
\end{equation}
where the second term describes the virtual process discussed
above. In the momentum space, derivatives acting on $h$ fields
correspond to powers of the residual momentum $k$, therefore we
can perform the following power expansion:
\begin{equation}
\label{eq:1/mQ} {\cal L}_{\rm eff}=\bar{h}_v (iv\cdot D) h_v +
\frac{1}{2m_Q} \sum_{n=0}^{\infty} \bar{h}_v i\gamma \cdot
\tilde{D} \left(-\frac{iv\cdot D}{2m_Q}\right)^n i\gamma \cdot
\tilde{D} h_v.
\end{equation}
It is not difficult to prove the following identity:
\begin{equation}
\frac{1+\gamma \cdot v}{2} (i\gamma \cdot \tilde{D})^2
\frac{1+\gamma \cdot v}{2} = \frac{1+\gamma \cdot v}{2}\left[
(i\tilde{D})^2 + \frac{g}{2} \sigma_{\mu\nu}G^{\mu\nu}\right]
\frac{1+\gamma \cdot v}{2},
\end{equation}
where $G_{\mu\nu}$ is the gluon strenght tensor field and the well
known property $[iD_\mu,iD_\nu]=igG_{\mu\nu}$ holds \cite{huang}.
Considering the $n=0$ term in the expansion (\ref{eq:1/mQ}), one
finds the interesting result:
\begin{equation}
\label{eq:hqet} {\cal L}_{\rm eff}=\bar{h}_v (iv\cdot D) h_v +
\frac{1}{2m_Q} \bar{h}_v (i\tilde{D})^2 h_v + \frac{g}{4 m_Q}
\bar{h}_v \sigma_{\mu\nu}G^{\mu\nu} h_v+O(1/m_Q^2).
\end{equation}
The $m_Q \to \infty$ limit selects only the first term in the
preceding Lagrangian. Since the heavy quark mass is large, but not
infinite, all other terms are to be considered as corrections,
that, as we can see, are included in the effective theory in a
systematic way. The first term in ${\cal L}_{\rm eff}$ allows to
write down the Feynman rules for propagator and gluon vertex
already discussed before. Let us rewrite it including a sum over
$N_h$ heavy flavors and a sum over velocities:
\begin{equation}
\label{eq:hqet1} {\cal L}_{\rm eff}^{(1)}=\sum_{i=1}^{N_h}\int
d^3{\bf v} \frac{1}{2 v_0} \bar{h}_v^{(i)} (iv\cdot D) h_v^{(i)}.
\end{equation}
Since in this Lagrangian there aren't terms containing  $m_Q$, we
can deduce that ${\cal L}_{\rm eff}^{(1)}$ is invariant under
flavor space rotations. Moreover, since no Dirac's $\gamma$ are
present, the interactions among heavy quarks and gluons conserve
heavy quark spins. This is the $SU(2N_h)$ symmetry already
discussed. To be rigorous, since Lorentz transformation can boost
the heavy quark velocity, the symmetry group should be ${\rm
Lorentz} \times SU(2 N_h)^{\infty}$.

Let's now consider the two operators of order $1/m_Q$ in
(\ref{eq:hqet1}). To understand their role it is convenient to
write them in the rest frame of the heavy quark $v=(1,0,0,0)$:
\begin{eqnarray}
\frac{1}{2m_Q}\bar{h}_v (i\tilde{D})^2 h_v &\rightarrow &
-\frac{1}{2m_Q}\bar{h}_v (i{\bf D})^2 h_v \\ \frac{g}{4 m_Q}
\bar{h}_v \sigma_{\mu\nu}G^{\mu\nu} h_v &\rightarrow & -\frac{g}{
m_Q} \bar{h}_v ({\bf S}\cdot {\bf H}_c) h_v,
\end{eqnarray}
where ${\bf H}_c^i=-\frac{1}{2} \epsilon^{ijk} G^{jk}$ and ${\bf
S}$ is a spin operator defined as a  $4\times 4$ matrix with Puli
matrices $\frac{\sigma_i}{2}$ on the diagonal. Therefore the first
operator is a kinetic energy operator connected to the residual
motion of the off-mass-shell heavy quark. The second operator is
the non Abelian extension of the Pauli interaction describing the
chromomagnetic heavy quark spin coupling with the gluon field. We
find confirmation that the heavy quark spin is decoupled by a
factor of $1/m_Q$.

\subsubsection{\label{sec:1/mQQCD} Relations with QCD }

To match HQET with QCD at high energy, one must include some
corrective effects in HQET due to high energy virtual processes.
For example, the weak current transforming the flavor from $b$ to
$c$ must be corrected at the $\alpha_s$ order both in QCD and in
HQET. We can expect that there are differences between these
corrections. These differences instruct on how one should modify
the coefficient of the  weak current in HQET and on what terms
should be added to the HQET current to guarantee the correct
matching between the low energy and the fundamental theory. Let us
consider for example the case of the current $\bar{b}\gamma^\mu
c$. The $\gamma^\mu$ of QCD has to be substituted by the
$\Gamma^\mu$ of HQET where \cite{neub}:
\begin{equation}
\Gamma^\mu = \left(1+C_0 \frac{\alpha_s}{\pi}\right)\gamma^\mu +
\frac{\alpha_s}{\pi}\sum_{i}C_i \Gamma_i^\mu,
\end{equation}
and $\Gamma_i^\mu$ are new structures containing  $v$ and
$v^\prime$, {\it i.e.}, the velocities of the heavy quark before
and after the weak interaction vertex. In practice, this type of
calculation is made by comparing the vertex diagrams where the
fermionic heavy quark current is coupled to the weak current, up
to order $\alpha_s$. We have therefore to compare four QCD
diagrams with four HQET diagrams. The difference between the two
set of diagrams lays in the Feynman rules describing the strong
vertices and the heavy quark propagators. The first of the
mentioned four diagrams is the three level diagram (the simple
tree weak vertex). In the remaining three diagrams, one should
close the gluon line on the heavy quark line according to the
three diagrammatically possible ways.
        \subsection{\label{sec:chirali}Chiral lagrangians}
The QCD Lagrangian with three massless quark flavors incorporates
the $U(3)_L\times U(3)_R$ global symmetry \cite{mano}. The left-
and right- handed components  $q_L=(u_L, d_L, s_L)$ and $q_R=(u_R,
d_R, s_R)$ transform respectively as the fundamental
representations of $U(3)_L$ and $U(3)_R$ respectively. Anyway, the
symmetry group of the quantum theory is a subgroup of
$U(3)_L\times U(3)_R$, this is what is usually called anomalous
breaking at the quantum level of a symmetry of the classical
Lagrangian. The reason for this phenomenon is that $U(1)_A$ is not
a good symmetry of the theory since its generator, $Q_5$, is not a
time independent quantity due to the presence of instantonic
configurations of gauge fields in Yang-Mills theories. The quantum
theory has therefore the following symmetry group: $SU(3)_L\times
SU(3)_R\times U(1)_V$, {\it the chiral symmetry}. Anyway the
physical states are invariant only under $SU(3)_V \times U(1)_V$;
for example the baryon spectrum is not doubled in two spectra
having opposite parity, but it is well described as an {\it octet
representation} of $SU(3)_V$ having baryon number $1$: this is the
phenomenon of spontaneous symmetry breaking. The mechanism
underlying the spontaneous symmetry breaking of chiral symmetry is
most likely of non perturbative nature. The energy scale
associated to this phenomenon is $\Lambda_\chi \simeq 1$ GeV (this
point will be discussed with greater detail later).

Each broken global symmetry implies the exitence of a
Nambu-Goldstone (NG) boson emerging as a scalar massless particle
induced by the non symmetric structure of theory's vacua. The
chiral symmetry is not an exact symmetry because light quark
masses are not exactly zero: they are simply small if compared
with $\Lambda_{\rm QCD}$. Therfore we should expect to have
pseudo-NG-bosons having small masses.

Light quark masses are slightly dissimilar to each other causing
that NG bosons will also have slightly dissimilar masses. On the
other hand, since light quark masses are much smaller than
$\Lambda_{\rm QCD}$, the light baryons are almost all degenerate
in mass (because, differently from NG bosons, sending to zero the
light quark masses, baryon masses should tend to a value different
from zero).

Due to the lightness of pseudo NG bosons, it emerges a hierarchy
of energy scales allowing to decide that NG bosons interactions at
energies much lower than $\Lambda_\chi$ can be described within a
{\it chiral effective theory}. Even if in this case the separation
among energy scales is not as large as in the case of heavy quark
effective theory, the chiral effective theory, developed in
seminal papers by Weinberg \cite{rass6}, Manohar and Georgi
\cite{geomano} and by Gasser and Leutwyler \cite{galeut}, is a
great success.

Since chiral symmetry is spontaneously broken, we have  a {\it
chiral condensate} different from zero that we can write as
\cite{shuryak}:
\begin{equation}
\label{eq:chiralradius} \langle \bar{\psi}_{jR} \bar{\psi}_{iL}
\rangle = c \Sigma{ij},
\end{equation}
where $c$ is the value of the condensate, while $\Sigma$ defines a
direction of the condensate in the flavor space. All  $\Sigma$'s
orientations are degenerate vacua that are mapped one in the other
through the $SU(3)_L \times SU(3)_R$ transformations:
\begin{equation}
\label{eq:lr} \Sigma \rightarrow L\Sigma R^+.
\end{equation}
$\Sigma$ is normalized in such a way that $\Sigma^+ \Sigma=1$.

When $\Sigma$ has a spatial dependency, then we are dealing with a
Goldstone excitation: low energy excitation are in fact
characterized by a vacuum configuration that varies from point to
point in the space, being the orientation of the vacuum state a
function having a smooth dependence on the position (think of spin
waves in a ferromagnet). The excitation energy is as small as one
likes when one considers very small  $\Sigma(x)$ variations over
large lenght scales: this is the case of NG bosons.

In order to construct a chiral effective theory, one needs to
follow the instructions that we have already described for a
general effective theory. One must write the most general
Lagrangian, containing the $\Sigma$ field, consistent with
relativistic invariance, ${\cal PCT}$, QCD chiral symmetry.

In the chiral effective Lagrangian there aren't terms not
containing derivatives \cite{rass5}. If there was such a term, it
could only be constant (think of ${\rm Tr}[\Sigma^+\Sigma]$ where
$\Sigma^+ \Sigma=1$). Every invariant function of $\Sigma$ without
derivatives, is just a constant. If we don't exclude terms not
containing derivatives, we could have a Lagrangian containing only
zero momentum $\Sigma$ fields. But a NG boson with zero momentum
simply does not exist, it is equivalent to the vacuum.

The first non trivial term that we are going to consider in the
chiral Lagrangian construction is the one having two derivatives:
\begin{equation}
\label{eq:primo} {\cal L}_2=\frac{f_\pi^2}{4}{\rm Tr}
[\partial_\mu \Sigma
\partial^\mu\Sigma^+],
\end{equation}
this is the only allowed one with two derivatives. $f_\pi$ is the
pion decay constant, {\it i.e.}, one of the energy scales
characterizing the pion (the other one is its mass). The $\Sigma$
field is described as the exponential of NG boson fields:
\begin{equation}
\Sigma(x)=e^{\frac{2 \pi(x)i}{f_\pi}}, \label{eq:usha}
\end{equation}
where:
\begin{equation}
\pi(x)=\pi^a(x)T^a,
\end{equation}
$T^a$ being the $8$ generators of $SU(3)$-colour. Such an
expression for the $\Sigma$ field amounts to consider the NG
fields as the angular variables describing the vacuum rotations.
Moreover let's observe that under the transformation
(\ref{eq:lr}), the $\pi$ fields transform {\it non linearly} as
complicated functions of $L$ and $R$. The exponential
representation (\ref{eq:usha}), is only a particular one, the
simplest, among all the possible non linear representations of
$\pi$ fields. All of those give the same S-matrix \cite{coleman}.

Let us now take the first term introduced in (\ref{eq:primo}).
This can be expanded in powers of the pion momentum in the
following way:
\begin{equation}
{\cal L}_2={\rm Tr}[\partial_\mu \pi \partial^\mu \pi] +
\frac{1}{2f_\pi^2} {\rm Tr}[[\pi,\partial_\mu \pi]^2]+.....,
\end{equation}
where higher order terms give non-linear interactions with NG
bosons. Besides ${\cal L}_2$ we can introduce higher dimensional
operators ${\cal L}_n$, {\it i.e.}, containing a larger number of
derivatives. Manohar and Georgi have shown that the energy scale
$\Lambda$ controlling this expansion in an increasing number of
derivatives $\frac{\partial}{\Lambda}$ is $\Lambda=\Lambda_\chi$
and they estimate $\Lambda_\chi \simeq 4\pi f_\pi\simeq 1$ GeV.
This limits the range of validity of the Lagrangian to the low
energy domain since the pion momenta have to be small compared to
$\Lambda_\chi$ (otherwise the derivative expansion gets divergent
very soon). We will make these points clear in the next
sub-session.

\subsubsection{\label{sec:lambdachi}$\Lambda_\chi$ }

As stated above,  besides (\ref{eq:primo}), the chiral effective
Lagrangian may also contain terms with an higher number of
derivatives. Let's consider ${\cal L}_4$ containing four
derivatives, {\it i.e.},  proportional to:
\begin{equation}
\label{eq:secondo} {\cal L}_4 \simeq{\rm Tr}[  \partial_\mu \Sigma
\partial_\nu
\Sigma^+ \partial^\mu \Sigma \partial^\nu \Sigma^+].
\end{equation}
An higher number of derivatives means an higher number of pion
momenta. As already stressed, in a low energy effective theory a
term such as ${\cal L}_4$ needs to be multiplied by some inverse
power of $\Lambda$, an energy scale characterizing the upper
energy bound of the effective theory and representing the
parameter that controls the convergence of the pion momentum
expansion. This $\Lambda$ can be associated to the energy scale of
spontaneous chiral symmetry breaking,  $\Lambda_\chi=\Lambda$, and
this association will allow for an estimation of $\Lambda_\chi$
that is going to be used throughout this work. We will follow an
argument due to  Manohar and Georgi \cite{geomano}. We can
normalize ${\cal L}_4$ relatively to the lowest order term
(\ref{eq:primo}), adding two $\Lambda_\chi^{-1}$ powers for each
additional derivative. The term having the correct mass dimension
is:
\begin{equation}
\label{eq:ssecondo} {\cal L}_4 =
\frac{f_\pi^2}{\Lambda_\chi^2}{\rm Tr}[
\partial_\mu\Sigma\partial_\nu
\Sigma^+ \partial^\mu \Sigma \partial^\nu \Sigma^+].
\end{equation}
If $\Lambda_\chi$ was much higher than pseudoscalar masses, then
all orders having an higher number of derivatives with respect to
(\ref{eq:primo}) would be negligible. But we cannot formulate this
hypothesis since radiative corrections to (\ref{eq:primo}) produce
a term as (\ref{eq:ssecondo}) and also higher order terms having
an infinite coefficient. Thus, even if these terms are absent or
negligible for a certain choice of the renormalization scale, they
can be present and important for a different one. On the other
hand, one could reasonably think that the $f_\pi^2/\Lambda_\chi^2$
coefficient should be numerically larger or equal to the variation
induced in it by an $O(1)$ shift in the renormalization scale of
radiative corrections to (\ref{eq:primo}).

These considerations get particularly clear if one examines the
$\pi-\pi$ scattering process. If we refer back to
(\ref{eq:primo}), we see that each four pion vertex has the
following structure:
\begin{equation}
\frac{p^2 \pi^4}{f_\pi^2}.
\end{equation}
\begin{figure}[t!]
\begin{center}
\epsfig{bbllx=128pt,bblly=303pt,bburx=466pt,bbury=706pt,height=12truecm,
        figure=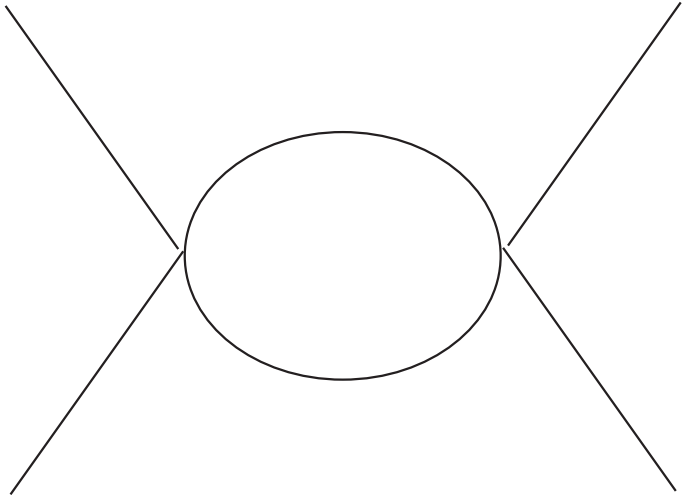}
\caption{\label{fig:caramella} \footnotesize
          One loop diagram for $\pi-\pi$ scattering. }
\end{center}
\end{figure}
Using two of these four pion vertices to generate the one-loop
diagram shown in fig. \ref{fig:caramella}, we can obtain the
particular case of pion-pion scattering diagram with all
derivatives acting on the external legs (the same diagram in fig.
\ref{fig:caramella} may have only two derivatives acting on
external legs, while the two remaining could act on the internal
ones. In such a situation we are facing a quadratically divergent
diagram).

If instead one refers directly to (\ref{eq:ssecondo}), the
fundamental four pion vertex has the form:
\begin{equation}
\frac{p^4 \pi^4}{f_\pi^2 \Lambda_\chi^2},
\end{equation}
therefore one can also write the pion-pion scattering diagram with
all derivatives acting on the external legs simply using the tree
level diagram extracted from ${\cal L}_4$. The diagram in fig.
\ref{fig:caramella}, generated by two insertions of  ${\cal L}_2$,
is just a one-loop correction to the tree level process described
by a single insertion of ${\cal L}_4$. The diagram in fig.
\ref{fig:caramella} clearly gives:
\begin{equation}
\label{eq:loop} \frac{p^4 \pi^4}{f_\pi^4}
\frac{1}{(2\pi)^4}\int_\ell \frac{1}{(\ell^2)^2}= \frac{p^4
\pi^4}{f_\pi^4} \frac{1}{(4\pi)^2}{\rm ln}\left(\frac{\Lambda_{\rm
co}}{\mu}\right),
\end{equation}
dividing by the correct symmetry factor and introducing a cutoff
$\Lambda_{\rm co}$ and the renormalization scale $\mu$. To be
consistent with a chiral effective theory scheme, we should limit
the momenta circulating in the loop to  $\Lambda_{\rm co}\simeq
\Lambda_\chi$. Let's now rescale $\mu$ in (\ref{eq:loop}), of an
$O(1)$ quantity: let's take the Neper number. We obtain a
variation of (\ref{eq:loop}) amounting to:
\begin{equation}
\label{eq:change} \frac{1}{(4\pi)^2} \frac{p^4 \pi^4}{f_\pi^4}.
\end{equation}
The shift in the renormalization point $\mu$ can be absorbed into
a redefinition of the coefficient $f_\pi^2/\Lambda_\chi^2$ in
(\ref{eq:ssecondo}) and we have that (\ref{eq:change}) corresponds
to a change in that coefficient amounting to:
\begin{equation}
\frac{1}{(4\pi)^2}.
\end{equation}
But, as above observed, we can expect that:
\begin{equation}
\frac{f_\pi^2}{\Lambda_\chi^2} \geq \frac{1}{(4\pi)^2},
\end{equation}
therefore:
\begin{equation}
\Lambda_\chi \leq 4\pi f_\pi,
\end{equation}
which suggest to use, as an estimate of the energy scale
associated to spontaneous chiral symmetry breaking, the following
one:
\begin{equation}
\Lambda_\chi = 4\pi f_\pi \simeq 1 {\rm GeV}.
\end{equation}
In this work we will make use of a cutoff  value close to
$\Lambda_\chi$ for the same physical motivations here described.
The fact that $f_\pi$ is a measure of the strength of the symmetry
breaking is also discussed in \cite{weinbook}.

\subsubsection{\label{sec:manoge} The Manohar-Georgi Lagrangian}

The effective Lagrangian defined below the chiral symmetry
breaking scale contains, besides pion fields, also light quarks
and gluons. Let's define the $\xi$ field in the following way:
\begin{equation}
\xi=e^{i\frac{\pi}{f_\pi}},
\end{equation}
{\it i.e.}, $\xi=\sqrt{\Sigma}$. We know that under $SU(3)\times
SU(3)$ transformations of the $\Sigma$ field, the NG bosons,
represented by the $\pi$ matrix, undergo a non-linear
transformation where $\pi_a \to \pi_a^\prime$, $\pi_a^\prime$
being non-linear functions of $\pi$, $L$, and $R$. The
transformation properties of $\pi$ fields determine also the
transformation properties under $SU(3)\times SU(3)$ of the
$\xi$-fields. Since $\Sigma=\xi^2$:
\begin{equation}
\xi \to \xi^\prime = L \xi U^+ = U \xi R^+,
\end{equation}
where $U$ is a non linear function of  $L$, $R$ and $\pi$ that can
be written as an ordinary $SU(3)$ matrix in the following way:
\begin{equation}
U=e^{i v},
\end{equation}
and the Hermitian matrix $v$ contains ``non-linearly" the
$SU(3)\times SU(3)$ symmetry. The $U$ matrix is invariant under
parity transformations exchanging $L$ with $R$ and $\pi$ with
$-\pi$. If $L=R$, the chiral transformation reduces to a simple
$SU(3)$ transformation and  $U=L=R$. Through $\xi$ matrices we can
construct two auxiliary fields:
\begin{eqnarray}
{\cal V}^\mu &=& \frac{1}{2} (\xi^+\partial^\mu
\xi+\xi\partial^\mu \xi^+)=\frac{1}{f_\pi^2}[\pi,\partial_\mu
\pi]+.....\\ \label{eq:aa} {\cal A}^\mu &=&-\frac{i}{2}
(\xi^+\partial^\mu \xi-\xi\partial^\mu
\xi^+)=\frac{1}{f_\pi}\partial_\mu\pi+.....,
\end{eqnarray}
which, under chiral transformation, behave like this:
\begin{eqnarray}
{\cal V}^\mu & \to & U{\cal V}^\mu U^+ + U\partial^\mu U^+
\\
{\cal A}^\mu  &\to & U {\cal A}^\mu U^+.
\end{eqnarray}
In particular, due to the transformation property of ${\cal
V}^\mu$, we can treat it as a gauge field in a covariant
derivative:
\begin{equation}
\label{eq:dcov} D^\mu=\partial^\mu+ {\cal V}^\mu.
\end{equation}

Let us now form the Lagrangian terms related to  $u$, $d$ and $s$
quarks considered together in a unique triplet  $\psi$ of
flavor-$SU(3)$. $\psi$ is supposed to transform under $SU(3)\times
SU(3)$ in the following way:
\begin{equation}
\psi \to U \psi.
\end{equation}
The only term without derivatives is:
\begin{equation}
-m \bar{\psi}\psi,
\end{equation}
where $m$ is not the current mass of the QCD Lagrangian. $m$ is a
constituent mass whose origin can be related to the spontaneous
breaking of chiral symmetry (we will come back on this point when
we will introduce the constituent light quark mass in the CQM
model).

We also have two derivative terms. The first one is the kinetic
term:
\begin{equation}
i\bar{\psi} \gamma \cdot D \psi,
\end{equation}
the other one is  the interaction term between the light quarks
and the pion. We will use the PCAC language  \cite{treiman} to
introduce this term.
\begin{figure}[t!]
\begin{center}
\epsfig{bbllx=128pt,bblly=303pt,bburx=466pt,bbury=706pt,height=12truecm,
        figure=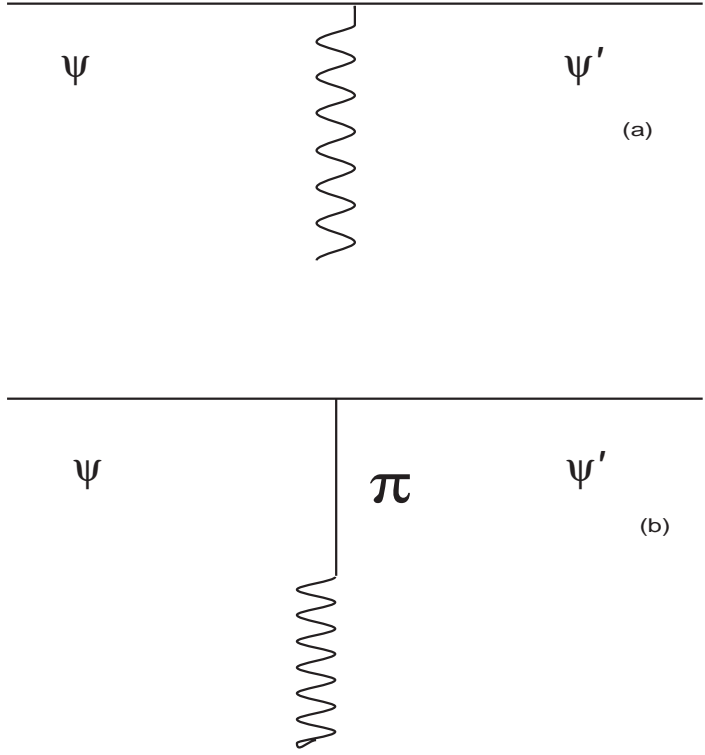}
\caption{\label{fig:pione} \footnotesize
          (b) represents the pion pole contribution to the diagram
          in (a).}
\end{center}
\end{figure}
The derivative operator of the axial current, $\partial A$, is a
pseudoscalar operator with odd G-parity, isospin one and
hypercharge zero. It has therefore all quantum numbers of the
$i$-component of pion triplet. Let us consider the matrix element:
\begin{equation}
\langle \psi^\prime | \partial A |\psi \rangle,
\end{equation}
where $\psi$ is a light quark field, see fig. \ref{fig:pione}(a).
Many diagrams contribute to diagram in fig. \ref{fig:pione}(a); we
choose that in fig. \ref{fig:pione}(b). This  is equivalent to:
\begin{equation}
\langle \psi^\prime | \partial A |\psi \rangle=i\frac{\langle {\rm
VAC}| \partial A |\pi \rangle \langle \pi
\psi^\prime|\psi\rangle}{q^2-m_\pi^2}=i\frac{f_\pi
m_\pi^2}{q^2-m_\pi^2} {\rm Amp}(\psi\to \psi^\prime \pi),
\end{equation}
where  ${\rm Amp}(\psi\to \psi^\prime \pi)$ denotes the amplitude
for the process $\psi\to \psi^\prime \pi$ (the $i$ comes from the
propagator). Clearly:
\begin{equation} {\rm Amp}(\psi\to \psi^\prime
\pi)=\frac{q^2-m_\pi^2}{i f_\pi m_\pi^2} \langle \psi^\prime
|\partial A |\psi \rangle,
\end{equation}
in the  $q^2\to m_\pi^2$ limit. PCAC hypothesis consists
essentially in defining the following off-mass-shell amplitude:
\begin{equation}
\widetilde{\rm Amp}(\psi\to \psi^\prime \pi)=\frac{q^2-m_\pi^2}{i
f_\pi q^2} \langle \psi^\prime |\partial A |\psi \rangle,
\end{equation}
where $q$ is the pion momentum. At this point one must suppose
that this off-mass shell  amplitude varies smoothly within the
$(0,m_\pi^2)$ $q^2$-range. The chiral limit of $\widetilde{\rm
Amp}$ is defined taking the $m_\pi \to 0$ limit and then the
mass-shell-limit $q^2\to 0$  (the two procedures do not commute)
hoping that the hypothesis on the smooth variability with respect
to $q^2$ holds well (as is confirmed by the Goldberger-Treiman
relation and by its physical consequences). The chiral off-mass
shell amplitude is therefore:
\begin{equation}
\widetilde{\rm Amp}(\psi\to \psi^\prime \pi)\to -\frac{i}{f_\pi}
\langle \psi^\prime |\partial A |\psi \rangle =
\frac{q_\mu}{f_\pi} \langle \psi^\prime | A^\mu |\psi \rangle.
\end{equation}
We can therefore have a Yukawa type coupling in the Lagrangian:
\begin{equation}
-\frac{i}{f_\pi}\bar{\psi}\gamma_\mu \gamma_5 \psi q^\mu \pi,
\end{equation}
and generalize it, see (\ref{eq:aa}), in the following way:
\begin{equation}
\bar{\psi} \gamma \cdot {\cal A} \gamma_5 \psi.
\end{equation}
The Manhoar-Georgi Lagrangian to lowest oder, including also
colour in the covariant derivative, is:
\begin{eqnarray}
\label{eq:mangeolag} {\cal L}_{\rm eff}&=&\bar{\psi}[\gamma \cdot
(iD + {\cal V})]\psi +g_A \bar{\psi} \gamma \cdot {\cal A}
\gamma_5 \psi -m\bar{\psi}\psi \nonumber\\ &+& \frac{f_\pi^2}{4}
{\rm Tr}[\partial^\mu \Sigma \partial_\mu \Sigma^+]-\frac{1}{2}
{\rm Tr}[G_{\mu\nu}G^{\mu\nu}],
\end{eqnarray}
where $G_{\mu\nu}$ is the well known non-abelian strenght tensor
of the gluon field $G_\mu=G_\mu^a T^a$. $g_A$ must be computed
through a matching with QCD or extracted by a comparison with
experimental data, or, as will be our case, with some more
fundamental effective model.

\subsubsection{\label{sec:mesoni} Heavy mesons and chiral
Lagrangians}

We are going to discuss of an effective chiral Lagrangian
describing the interaction of soft pions (and kaons) with mesons
containing an heavy quark. As we saw in section {\bf
\ref{sec:hqet}}, heavy meson fields should be described through
the HQET formalism. In order to implement the flavor-spin
symmetry, the field describing an heavy meson has to be
independent on heavy quark mass and spin. On the contrary, it can
be characterized by the total angular momentum $s_\ell$ of light
degrees of freedom. To each $s_\ell$ value, there corresponds a
doublet of states degenerate in mass with total angular momentum
$J=s_\ell \pm \frac{1}{2}$. In correspondence of
$s_\ell=\frac{1}{2}$, for example, we have the $P$ and $P^*$
mesons being respectively the pseudoscalar and vector components
of the spin symmetry doublet. If the heavy quark is $c$, $P$ and
$P^*$ correspond to $D$ and $D^*$, while if the heavy quark is
$b$, they are the states $B$ and $B^*$ respectively.

Let us consider the negative parity doublet $(P,P^*)$. We can
associate a unique {\it super-field} $H$ \cite{rep} describing
both states. This super-field must have two spinor indices: one
connected to the heavy quark and the other to the light quark. The
structure of $H$ is that of a $4\times 4$ Dirac matrix. If one
performs a Lorentz transformation, $H$ behaves like:
\begin{equation}
H\to D(\Lambda) H D(\Lambda)^{-1},
\end{equation}
where $\Lambda$ is the usual $4\times 4 $ representation of the
Lorentz group. An explicit matrix representation is the following:
\begin{equation}
\label{eq:acca} H=\frac{1+\gamma\cdot v}{2}[P^*_\mu \gamma^\mu - P
\gamma_5],
\end{equation}
and we define:
\begin{equation}
\bar{H}=\gamma_0 H^\dag \gamma_0.
\end{equation}
$v$ is the velocity of the heavy meson and the following
transversality condition holds: $v^\mu P^*_\mu=0$ and
$M_H=M_P=M_{P^*}$. We mention also the following useful relations:
 $\gamma\cdot v
H=-H\gamma \cdot v $, $\bar{H}\gamma\cdot v=-\gamma\cdot v
\bar{H}=\bar{H}$. $P$ and $P^{*\mu}$ are the annihilation
operators normalized in the following way:
\begin{eqnarray}
\label{eq:norm1} \langle {\rm VAC}|P|Q\bar{q}(0^-)\rangle &=&
\sqrt{M_H}\\ \label{eq:norm2} \langle {\rm
VAC}|P^{*\mu}|Q\bar{q}(1^-)\rangle &=& \epsilon^\mu \sqrt{M_H}.
\end{eqnarray}
The formalism apt to describe higher spin meson states and the
effective Lagrangian terms associated to them, is extensively
developed in  \cite{luca}. We are interested in considering the
$p$-wave ($\ell=1$) of the $Q\bar{q}$ system. HQET predicts the
existence of two degenerate doublets: $(0^+,1^+)$ and $(1^+,2^+)$
for each heavy quark $c$ or $b$. The related superfields are:
\begin{eqnarray}
\label{eq:esse} S&=&\frac{1+\gamma\cdot v}{2}[P^{*\prime}_{1\mu}
\gamma^\mu\gamma_5 - P_0]\\ \label{eq:ti} T^\mu&=&
\frac{1+\gamma\cdot v}{2}\left[P^{*\mu\nu}_2\gamma_\nu -
\sqrt{\frac{3}{2}} P^*_{1\nu}\gamma_5\left(g^{\mu\nu}-\frac{1}{3}
\gamma^\nu(\gamma^\mu-v^\mu) \right)\right].
\end{eqnarray}
These two doublets have respectively  $s_\ell=\frac{1}{2}$ and
$s_\ell=\frac{3}{2}$. This classification with respect to $s_\ell$
is the more reasonable one since we know that the dynamics is
completely independent on the spin and on the mass of the heavy
quark $Q$. Observe that in the limit of infinite heavy quark mass,
$s_\ell$ and $s_Q$ are separately conserved. We can introduce the
total spin ${\bf J}$, defined as the total angular momentum of the
heavy  and light quark in the rest frame of the heavy quark:
\begin{equation}
{\bf J}={\bf s_\ell}+{\bf s}_Q.
\end{equation}

Heavy mesons can interact with $\pi$ fields through their light
degrees of freedom. The interaction terms of the NG boson octet
with heavy meson fields must be written in an effective Lagrangian
including the essential symmetry properties: chiral symmetry and
heavy flavor-spin symmetry. This heavy-light chiral effective
Lagrangian  can be expanded with respect to:
\begin{itemize}
\item  NG bosons momenta \\
\item  $\frac{1}{m_Q}$ powers
\end{itemize}

An heavy-light Lagrangian has been introduced almost
simultaneously by different groups \cite{hlwise}:
\begin{equation}
\label{eq:wise} {\cal L}=\frac{f_\pi^2}{4}{\rm Tr}[\partial^\mu
\Sigma \partial_\mu \Sigma^+] - {\rm Tr}[\bar{H} iv\cdot D H] +
g{\rm Tr}[\bar{H}H\gamma\cdot {\cal A} \gamma_5] +.....,
\end{equation}
where ellipses indicate the presence of an infinite number of
operators having higher dimensionality, including those
responsible for explicit chiral symmetry breaking, {\it i.e.},
terms containing light hadron masses, and those of order
$\Lambda_{\rm QCD}/m_Q$, violating the flavor-spin symmetry (such
as the color magnetic moment operator). The covariant derivative
has been defined in (\ref{eq:dcov}).

Let us consider for example the term having the factor of $g$ in
eq. (\ref{eq:wise}): it describes the coupling of $H$-type mesons
with NG bosons, see eq. (\ref{eq:aa}). We will come back on terms
of this kind many times. Including also $S$ and $T$ mesons we
have:
\begin{equation}
\label{eq:forti} {\cal L}= ig{\rm Tr}[\bar{H}H\gamma\cdot {\cal A}
\gamma_5]+ ig^\prime {\rm Tr}[\bar{S}S\gamma\cdot {\cal A}
\gamma_5]+ig^{\prime\prime} {\rm Tr}[\bar{T}^\mu T_\mu\gamma\cdot
{\cal A} \gamma_5].
\end{equation}
At the lower order of the derivative expansion we can also write
the interaction terms describing transitions between different
doublets; for example:
\begin{equation}
\label{eq:effes} {\cal L}= i f {\rm Tr}[\bar{S}T^\mu {\cal A}_\mu
\gamma_5]+ i f^\prime {\rm Tr}[\bar{H}S\gamma \cdot {\cal A}
\gamma_5] + h.c.
\end{equation}
A particular case is that of transitions between $T$ and $H$
states. Let's suppose to consider the following s-wave interaction
Lagrangian:
\begin{equation}
\label{eq:esempio} {\cal L}=ir{\rm Tr}[\bar{H}T^\mu {\cal A}_\mu
\gamma_5].
\end{equation}
We can therefore write the $S$-matrix element:
\begin{equation}
\langle {\rm out} D \pi|D_2 {\rm in}\rangle,
\end{equation}
as:
\begin{eqnarray}
\langle D \pi|i{\cal L}|D_2 \rangle &=& -r\langle D|P|{\rm
VAC}\rangle \langle \pi|\frac{\partial_\mu \pi}{f_\pi}|{\rm
VAC}\rangle \langle {\rm VAC}|P^{*\mu\nu}_2|D_2\rangle {\rm
Tr}\left[\frac{1+\gamma \cdot v}{2} \gamma_\nu\right]
\\
&=&ir \sqrt{m_D m_{D_2}} \frac{q_\mu}{f_\pi}
2\epsilon^{\mu\nu}v_\nu,
\end{eqnarray}
where the first order in the expansion (\ref{eq:aa}) has been
used. Due to transversality, this term is certainly zero, since
$\epsilon^{\mu\nu}v_\nu=0$. Since the process we are discussing is
entirely due to the strong interaction, the velocity conservation
rule, introduced in {\bf \ref{sec:hqet}},  holds, {\it i.e.}, the
velocity of $D$ and $D_2$ are the same.

We conclude that the Lagrangian (\ref{eq:esempio}) cannot be the
right one  to describe the $T\to H\pi$ transition. We need one
more Lorentz index coming from the insertion of another derivative
under the trace sign. This derivative should be accompained with a
negative power of $\Lambda_\chi$, giving the right mass dimension
to the interaction term and controlling the expansion in NG bosons
momenta.

The $d$-wave Lagrangian is:
\begin{equation}
\label{eq:tih} {\cal L}=\frac{h_1}{\Lambda_\chi} {\rm
Tr}[\bar{H}T^\mu(i D_\mu \gamma\cdot {\cal
A})\gamma_5]+\frac{h_2}{\Lambda_\chi} {\rm Tr}[\bar{H}T^\mu(i
\gamma \cdot D {\cal A}_\mu)\gamma_5]+{\rm h.c.}
\end{equation}
With an analogous approach super-fields having higher spins may be
constructed.
\section{\label{chap:secondo}CQM}

        \subsection{\label{sec:modello}The CQM model}
CQM is a Constituent-Quark-Meson-Model that has been introduced
and discussed in a number of recent research papers
\cite{art1,art2,art3,art4,art5}. The model, based on an effective
Lagrangian describing quark-meson interactions, is relativistic,
incorporates the flavor-spin symmetry in the heavy sector and the
chiral symmetry in the light quark sector. In the following
sections CQM will be reviewed in detail. Section {\bf
\ref{sec:unpi}} and subsequent sections are instead devoted to the
CQM phenomenological applications to heavy meson physics.

In the well known old paper \cite{njl}, Y. Nambu and   G.
Jona-Lasinio discuss the possibility that the nucleon mass can be
due to an unknown primary interaction bounding hypothetical
massless primary fermions. In their model, the same interaction
bounds nucleon pairs giving rise to pions and the mass of the
Dirac particle emerges as a result of the primary interaction in
the same way as the energy gap in a BCS superconductor
\cite{strocchi} is connected to the formation of correlated Cooper
pairs of electrons, as a result of a phonon-mediated ``primary''
interaction (the finite energy that is needed to break a Cooper
pair is proportional to the BCS gap).

The primary interaction in the  Nambu-Jona-Lasinio model is a non
linear four fermion interaction, already discussed in an older
paper by Heisenberg et al.:
\begin{equation}
\label{eq:int1}
G[(\bar{\psi}\gamma_\mu\psi)^2-(\bar{\psi}\gamma_\mu
\gamma_5\psi)^2].
\end{equation}

In some recent papers \cite{ebert,bije}, the problem of the {\it
bosonization} of a Nambu-Jona-Lasinio (NJL) like Lagrangian, where
the fundamental fields are quarks, has been extensively studied.
The aim is to derive an effective theory of mesons starting from a
model (the NJL model)  incorporating global chiral symmetry and
its spontaneous breakdown: order parameter of the chiral symmetry
breaking is the light {\it constituent} quark mass which is
proportional to the chiral condensate and can be calculated
through a gap equation \cite{strocchi}, as we will briefly see.

Let us consider a  $U(N_f)$ multiplet $\psi=\psi_{f,c}$, the
$\lambda$ matrices $\lambda\in U(N_f)$, normalized according to
 ${\rm Tr}_f \lambda^\alpha\lambda^\beta=2\delta^{\alpha\beta}$, and let's write:
\begin{equation}
\label{eq:exemplum} {\cal L}_{\rm NJL}= i\bar{\psi}
\gamma\cdot\partial \psi + G \sum_{\alpha=0}^{N_f^2 - 1} \left[
\left(\bar{\psi}\frac{\lambda^\alpha}{2}\psi\right)^2
+\left(\bar{\psi}\frac{\lambda^\alpha}{2}i\gamma_5\psi\right)^2\right].
\end{equation}
(Interactions (\ref{eq:int1},\ref{eq:exemplum}) are interconnected
by Fierz theorem). Taking the derivative of (\ref{eq:exemplum})
with respect to $\bar{\psi}$, we can write the following equation
of motion:
\begin{equation}
i\gamma\cdot\partial\psi+\frac{G}{2}\sum_{\alpha}
(\bar{\psi}\lambda^\alpha\psi)\lambda^\alpha\psi=0,
\end{equation}
where we require that  $(\bar{\psi}\lambda^\alpha i
\gamma_5\psi)=0$ \cite{miranski}. If we also require that
$(\bar{\psi}\lambda^\alpha\psi)=0$ for each $\alpha \neq 0$,
following an analogy with the approximations made to solve the BCS
equation of motion \cite{miranski,strocchi}, and remind that
$\lambda^0=\sqrt{\frac{2}{N_f}}{\bf 1}$, defining:
\begin{equation}
\frac{G}{N_f}\langle {\rm VAC}|\bar{\psi}\psi|{\rm VAC}\rangle=-
m_{\rm dyn},
\end{equation}
the equation of motion becomes:
\begin{equation}
(i\gamma\cdot\partial-m_{\rm dyn})\psi=0.
\end{equation}
Let us now consider that:
\begin{equation}
-\frac{G}{N_f}\langle {\rm VAC}|\bar{\psi}\psi|{\rm VAC}\rangle=
\frac{G}{N_f} \lim_{x\to 0} {\rm Tr}_f {\rm Tr}_c \langle {\rm
VAC}|\psi_{f,c}(x)\bar{\psi}_{f,c}(0)|{\rm VAC} \rangle=iG
N_c\Delta_F(x),
\end{equation}
where $\Delta_F(x)$ is the Feynman propagator:
\begin{equation}
\Delta_F(x)=\frac{1}{(2\pi)^4}\int d^4p e^{-ipx} \frac{\gamma\cdot
p+m_{\rm dyn}}{p^2-m_{\rm dyn}^2+i\epsilon}.
\end{equation}
Using the Dirac equation, we obtain the following expression for
the dynamical mass:
\begin{equation}
\label{eq:dinamica} m_{\rm dyn}= 2G \frac{iN_c}{(2\pi)^4}\int^{\rm
reg} d^4p \frac{m_{\rm dyn}}{p^2-m_{\rm dyn}^2+i\epsilon}.
\end{equation}
The non-trivial solution, ${\rm m_{\rm dyn}}\neq 0$, is connected
to the spontaneous breaking of chiral symmetry.

In the papers by D. Ebert {\it et al.}  \cite{ebert}, the NJL
model is extended to include two light quark fields and an heavy
one. In such a way the bosonization  produces collective meson
fields having a light-light or heavy-light constituent quark
content.

\subsubsection{\label{sec:boson} Bosonization}

The basic idea of the bosonization technique is that of
re-formulating a field theory written in terms of microscopic
degrees of freedom, such as quarks and gluons, as a field theory
in which  meson fields are on the same footing of elementary
fields. Many attempts to bosonize the QCD Lagrangian, {\it i.e.},
to yield a meson theory mathematically derived from first
principles, have been unsuccessfully performed. Some progress in
this direction has been made in two-dimensional QCD \cite{witten}.

In the path integral language, this is how bosonization works:
\begin{equation}
\int Dq D\bar{q} e^{i\int_{x} {\cal L}_{\rm NJL}} \to \int D\sigma
D\pi D\rho...e^{i\int_{x} {\cal L}_{\rm bos}}, \label{eq:intermed}
\end{equation}
where $D\sigma$, $D\pi$, $D\rho$...are the integration measures
associated to the meson fields. The effective Lagrangian ${\cal
L}_{\rm bos}$ is written as a function of these fields.

The Hubbard-Stratonovich transform is the first step in
(\ref{eq:intermed}): the four quark NJL interaction is substituted
by Yukawa couplings of the quark fields with meson fields:
\begin{equation}
e^{i\int_{x} {\cal L}_{\rm NJL}}\to\int D\sigma D\pi...e^{i\int_x
{\cal L}_{\rm NJL}^\prime}.
\end{equation}
${\cal L}_{\rm NJL}^\prime$ is the semi-bosonized  ${\cal L}_{\rm
NJL}$. ${\cal L}_{\rm NJL}^\prime$ is Gaussian with respect to
functional integration over microscopic fields. Therefore,
integrating over $Dq$ and $D\bar{q}$ one obtains a determinant
containing the meson fields. This can be loop expanded and the
Feynman diagrams coming in this expansion can be evaluated in the
region of small meson momenta, the most interesting for our
purposes.

The NJL Lagrangian studied in  \cite{ebert} is:
\begin{equation}
\label{eq:due} {\cal L}_{\rm
NJL}=-\frac{G}{2}\left(\bar{\psi}\gamma_\mu
\frac{\lambda^\alpha}{2}\psi\right)\left(\bar{\psi}\gamma^\mu
\frac{\lambda^\alpha}{2}\psi\right),
\end{equation}
where $q=(u,d,s)^T$, $Q_v=b$ or $Q_v=c$, $\psi=(q,Q)^T$ and $G$ is
a coupling having dimension of $(mass)^{-2}$, while $\lambda$ are
matrices of $SU(N_c)$. The free Lagrangian is therefore the sum of
the familiar Dirac Lagrangian for light quarks and the free
effective Lagrangian for heavy quarks:
\begin{equation}
{\cal L}_0=\bar{q}(i\gamma\cdot \partial - \tilde{m})q+\bar{Q}_v
(iv\cdot\partial)Q_v.
\end{equation}

The bosonization is then performed on:
\begin{equation}
\label{eq:ebermodel} {\cal L}= {\cal L}_0+{\cal L}_{\rm NJL}.
\end{equation}
Fierz theorem allows to rearrange the bosonized Lagrangian ${\cal
L}_{\rm bos}$ as a sum of three pieces. We are interested only in
two of them: ${\cal L}_{\rm bos}={\cal L}^{ll}+{\cal L}^{hl}$. The
former is related only to the light degrees of freedom, the latter
includes also the heavy quark and heavy meson fields. The third
term, ${\cal L}^{hh}$, is not relevant when one is interested in
studying the physics of mesons containing only one heavy quark, as
is the case here.

${\cal L}$ in eq. (\ref{eq:ebermodel}), has the global colour
symmetry, the chiral $SU(3)\times SU(3)$ symmetry (as the current
light quark masses go to zero) and the flavor-spin symmetry of
HQET (observe that also the interaction term is independent on the
heavy quark mass and spin). These properties are preserved in
${\cal L}_{\rm bos}$.

Technical details about the bosonization method are far beyond the
scopes of this work. The interested reader is referred to
\cite{ebert} and references therein.

The CQM Lagrangian is a phenomenological extension of ${\cal
L}_{\rm bos}$.

\subsubsection{\label{sec:cqmlag} The CQM effective Lagrangian}

As discussed in the last section, the CQM Lagrangian is made up of
two terms, like  ${\cal L}_{\rm bos}$, but does not exactly
coincide with it (${\cal L}_{\rm CQM} \neq {\cal L}_{\rm bos}$)
for reasons that will be clear soon:
\begin{equation}
{\cal L}_{\rm CQM}={\cal L}^{ll}+{\cal L}^{hl}.
\end{equation}
The first term describes the light degrees of freedom and it is
very similar to the Georgi-Manohar Lagrangian given in
(\ref{eq:mangeolag}). The differences are that, in the CQM
Lagrangian, there are no gluons and the light fields are defined
differently:
\begin{equation}
\label{eq:cqmllag} {\cal L}^{ll}=\bar{\chi}[\gamma \cdot
(i\partial + {\cal V})]\chi+ \bar{\chi} \gamma \cdot {\cal A}
\gamma_5 \chi -m\bar{\chi}\chi + \frac{f_\pi^2}{8} {\rm
Tr}[\partial^\mu \Sigma
\partial_\mu \Sigma^+].
\end{equation}
where now we define $f_\pi=130$ MeV. The absence of gluons is
rather plausible since this Lagrangian originates from the
bosonization of an underlying NJL interaction Lagrangian, where
gluons are absent from the start.

The light $\chi$ fields are also a consequence of the bosonization
of an underlying NJL. What emerges is that $\chi=\xi q$, where $q$
are the familiar light quark fields and $\xi=e^{i\frac{\pi}{f}}$.
We will always consider the expansion of $\xi$ to be truncated at
the zero-th order in the pion field. We will need the first order
of this expansion only in section {\bf \ref{sec:btopi}}.

From a detailed comparison with (\ref{eq:mangeolag}), it is also
evident that in (\ref{eq:cqmllag}) $g_A=1$, again as a result of
bosonization. The mass $m$ in (\ref{eq:cqmllag}) is dynamically
generated according to the mechanism explained in section {\bf
\ref{sec:modello}}.

Let us now focus on ${\cal L}^{hl}$.  Here we have the Yukawa type
interactions between quark and meson fields emerging from
bosonization, plus two phenomenological terms put by hand:
\begin{eqnarray}
{\cal L}^{hl}&=&\bar{Q}_v (iv\cdot\partial)Q_v-
\left[\bar{\chi}\left(\bar{H}+\bar{S}+i\bar{T}^\mu\frac{\partial_\mu}{\Lambda}
\right)Q_v+h.c.\right]\nonumber\\ \label{eq:meson}
&+&\frac{1}{2G_3}{\rm Tr}[\bar{H}H]+ \frac{1}{2G_3^\prime}{\rm
Tr}[\bar{S}S] +\frac{1}{2G_4}{\rm Tr}[\bar{T}^\mu T_\mu].
\label{eq:russa}
\end{eqnarray}
The first term is the well known heavy quark kinetic term of HQET,
see {\bf \ref{sec:hqet}}.

\begin{figure}[t!]
\begin{center}
\epsfig{bbllx=5cm,bblly=15cm,bburx=15cm,bbury=30cm,height=12truecm,
        figure=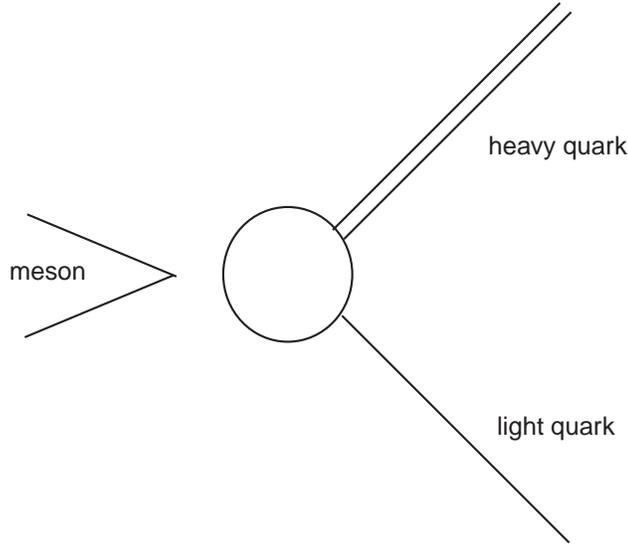}
\caption{\label{fig:vertex} \footnotesize
          The CQM  $Q-{\rm Meson}-q$ vertex.}
\end{center}
\end{figure}

The second term is responsible for the $Q-{\rm Meson}-q$ vertices
shown in fig. \ref{fig:vertex}: these are the most relevant aspect
of the CQM model. The meson fields $H$, $S$ and $T$ have been
introduced in  {\bf \ref{sec:mesoni}}.

The vertices for $H$ and $S$ mesons have been derived from
bosonization. The vertex involving the $T$ field is instead a
phenomenological term, introduced according to the philosophy of
effective theories ($\Lambda=1$ GeV in eq. (\ref{eq:russa})). This
value of $\Lambda$ should also be assumed as the ultaviolet cutoff
in the regularization of the model $\Lambda=\Lambda_\chi$, see the
discussion in {\bf \ref{sec:lambdachi}}.

Following \cite{ebert} we will adopt as UV cutoff $\Lambda=1.25$
GeV. This is a value quite close to the $\Lambda_\chi$ discussed
in {\bf \ref{sec:lambdachi}}. On the other hand one finds that the
CQM phenomenological predictions are not very sensitive to the
variation of the UV cutoff, at least if one varies it within
$10-15\%$.

Last three terms in (\ref{eq:meson}) are the kinetic terms for
$H$, $S$ and $T$ fields. The first two terms come from
bosonization, while the third is inserted by hand as a
phenomenological term resembling the first two. Bosonization also
predicts that $\frac{1}{2G_3}=-\frac{1}{2G_3^\prime}$ and we can
therefore write:
\begin{eqnarray}
\label{eq:cqmhl} {\cal L}^{hl}&=&\bar{Q}_v (iv\cdot\partial)Q_v-
\left[\bar{\chi}\left(\bar{H}+\bar{S}+i\bar{T}^\mu\frac{\partial^\mu}{\Lambda_\chi}Q_v
\right)+h.c.\right]\nonumber\\ &+&\frac{1}{2G_3}{\rm
Tr}[(\bar{H}+\bar{S})(H-S)] +\frac{1}{2G_4}{\rm Tr}[\bar{T}^\mu
T_\mu].
\end{eqnarray}
The dynamical information $\frac{1}{2G_3}=-\frac{1}{2G_3^\prime}$
is crucial for the CQM calculation of the coupling constants:
there are not sufficient experimental data to determine two
different constants $G_3$ and $G_3^\prime$.

The kinetic terms will be rewritten, in the form discussed in
\cite{rep}, in the next section, where we will also discuss the
problem of determining the mass difference between the $H$ and $S$
multiplets. Again the dynamical information
$\frac{1}{2G_3}=-\frac{1}{2G_3^\prime}$ helps this evaluation. We
will call $\Delta_H$, $\Delta_S$ and $\Delta_T$ the mass
differences between the masses of $H$, $S$ and $T$ multiplets and
the heavy quark there contained. The mass difference between $S$
and $H$ will be simply $\Delta_S-\Delta_H$ and it will be zero as
soon as the light constituent quark mass $m\to 0$ ({\it i.e.}, in
the chiral unbroken phase).

\subsubsection{\label{sec:reg} Regularization}

As we pointed out in  {\bf \ref{sec:cqmlag}}, CQM is the fusion of
a Manohar-Georgi like Lagrangian for the light quark sector with a
quark-meson Lagrangian for the heavy quark sector. We  therefore
know that the upper energy scale, {\it i.e.}, the energy scale
over which the effective theory should be substituted by a more
fundamental theory, is $\Lambda_\chi$. It could seem strange that
the heavy quark mass is itself higher than the UV cutoff, but we
have to remind that in HQET the on shell momentum of the heavy
quark, $m_Q v$, is not a dynamical quantity since, due to the
velocity superselection rule, $v$ is not dynamical, see {\bf
\ref{sec:hqet}}. The dynamical quantity due to the interaction
between the heavy quark and the light degrees of freedom is the
residual momentum  $k^\mu$, which is necessarily $k\simeq
\Lambda_{\rm QCD}<\Lambda_\chi$.

CQM does not include the gluon fields, as it is obvious
considering that it results from a path integral bosonization of a
NJL model, and does not incorporate confinement of quarks. This
may appear as a strong limitation of the model but, according to a
common opinion,  it is physically much more important to work with
non confining models possessing chiral symmetry and its
spontaneous breakdown than with confining models where chiral
symmetry and its breakdown are not properly incorporated. In the
former case one is describing a world which is essentially the
same as the real one for what concernes the hadronic spectra; the
only difference would be that of the theoretical admissibility of
asymptotic quark states. The latter case presents an hadronic
spectrum completely messed up with respect to the observed one. We
show here how one can face the problem that CQM is not a confining
model: introducing an infrared cutoff $\mu$.

The kinematical condition for an heavy meson having mass $M$ to
decay into its free constituent quarks is:
\begin{equation}
\label{eq:precedente} M>m_Q+m.
\end{equation}
Since the meson momentum is $P=m_Qv+k$, where $P=Mv$, eq.
(\ref{eq:precedente}) is equivalent to the condition $v\cdot k>m$.
In the frame where the heavy meson is at rest, the latter
condition means $k_0>m$, {\it i.e.}, ${\rm inf}(k)=m$. Therefore
one should consider residual momenta $k$ larger than $m$ to be
sure that the unphysical threshold condition (\ref{eq:precedente})
holds, as it should in a not-confining model. For lower $k$ values
one is in the energy region where confinement must be necessarily
taken into account.

On the other hand, the value of the constituent light mass is
determined by a gap equation (see also (\ref{eq:dinamica}))
\cite{ebert}:
\begin{equation}
\langle \Sigma \rangle =m=\tilde{m}+8mI_1(m^2),
\label{eq:gapequation}
\end{equation}
where the chiral ray $\Sigma$ has been defined in
(\ref{eq:chiralradius}), while the $I_1$ integral is given in the
Appendix together with other integrals met in  CQM applications.
$I_1$ is calculated with an UV and an IR  cutoff introduced
according to the Shwinger's regularization method, as we will
discuss in a while. As the infrared cutoff varies, the $m$ value
varies accordingly and, following what we have observed before, we
can choice as an infrared cutoff $\mu \simeq m$ (the running
momenta in the CQM loops we will deal with, are of the size of the
heavy quark residual momenta). In the second paper in reference
\cite{ebert}, it is shown the $m$ vs. $\mu$ plot obtained from
(\ref{eq:gapequation}) for a fixed $\Lambda$. This plot has the
typical shape of a second order phase transition order parameter
with a critical $\mu$ at $\mu_{\rm c}\simeq 550$ MeV. For
$\mu>\mu_{\rm c}$, $m$ is zero, {\it i.e.}, the chiral symmetry is
unbroken. For $\mu=m=300$ MeV, one is in the broken (physical)
phase at the edge of a plateau.

Therefore the boundary energy values of the effective theory are
chosen to be $\mu=300$ MeV, $\Lambda\simeq\Lambda_\chi\simeq 1.25$
GeV, and the light constituent mass is dynamically generated by a
NJL gap equation: $m=300$ MeV (which represents the degenerate $u$
and $d$ masses. We will not consider $s$ quarks).

The last step is the choice of the prescription to implement the
cutoffs in the calculations. For a non renormalizable model, this
step is part of the definition of the model itself. The proper
time Shwinger regularization has shown to be the most adequate for
our purposes.

After a continuation of the light propagator in the Euclidean
domain, the following prescription is used:
\begin{equation}
\label{eq:regint} \int d^4 \ell_E \frac{1}{\ell^2_E + m^2}\to \int
d^4 \ell_E\int_{1/\Lambda^2}^{1/\mu^2} ds e^{-s(\ell_E^2+m^2)}.
\end{equation}
All the CQM calculations are performed applying this receipt. If
one tries to insert the cutoffs as the bounds of the Euclidean
integral measure, besides the problem that Euclidean translation
invariance is then lost, one also has to face the problem that the
choice of the infrared cutoff is not only conditioned by $m$ (see
the discussion made above), but also by $\Delta_H$, that is the
free parameter of our model. The regularization receipt
(\ref{eq:regint}) acts in the sense of modifying the Euclidean
light propagator through a factor depending on the difference of
two exponential functions of $(\ell^2_E+m^2)$. This could affect
the Ward-Takahashi relation for, {\it e.g.}, the vertex of the
axial current $A$ with the light quarks, causing the emergence of
a mass term for the pion, even if  one considers the chiral limit
$\partial A=0$ from the beginning. Anyway, due to the structure of
the regularization receipt, this should be a very soft effect.

        \subsection{\label{sec:renorm} Renormalization constants and masses}

\begin{figure}[t!]
\begin{center}
\epsfig{bbllx=5cm,bblly=18cm,bburx=15cm,bbury=30cm,height=8truecm,
        figure=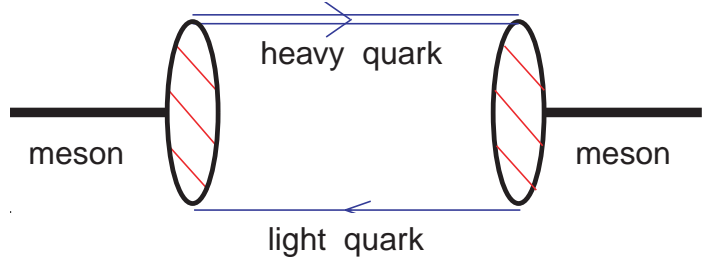}
\caption{\label{fig:autoenergia} \footnotesize
          CQM meson self energy diagram. }
\end{center}
\end{figure}

The simplest CQM loop diagram that one can obtain contracting two
vertices $Q-{\rm Meson}-q$, see fig. \ref{fig:vertex}, is the
meson self energy diagram shown in fig. \ref{fig:autoenergia}.

For in and out $H$ fields, we can write down the following loop
integral:
\begin{equation}
\label{eq:auto} iN_c\int^{\rm reg} \frac{d^4\ell}{(2\pi)^4}
\frac{{\rm
Tr}[H\gamma\cdot(\ell-k+m)\bar{H}]}{[(\ell-k)^2-m^2+i\epsilon]
[v\cdot\ell+i\epsilon]}={\rm Tr}[\bar{H}\Pi_H(v\cdot k)H].
\end{equation}
The rules applied are the standard ones for loop integrals. The
expressions of the usual Dirac propagator and of the heavy quark
propagator, defined in {\bf \ref{sec:hqet}}, have been inserted in
the integral together with the vertex prescriptions derived from
the heavy-light Lagrangian ${\cal L}^{hl}$. The regularization
procedure is that of the Shwinger's proper time, see {\bf
\ref{sec:reg}}.

First of all let us observe that we can perform the expansion:
\begin{equation}
\Pi(v\cdot k)\simeq \Pi(\Delta)+\Pi^{\prime}(\Delta) (v\cdot k
-\Delta),
\end{equation}
since we know that $k$ smoothly fluctuates around $(M-m_Q)v$, {\it
i.e.}:
\begin{equation}
k^\mu= \Delta v^\mu -q^\mu,
\end{equation}
where $q$ parameterizes this small fluctuation, see eq.
(\ref{eq:smallfluctu}), and $\Delta$ is defined as $\Delta=M-m_Q$
modulo $1/m_Q$ corrections. This expansion of  $\Pi$ can now be
inserted in the self energy expression for the $H$ field (for $S$
and $T$ fields the procedure is exactly the same) and subtracting
from ${\cal L}^{hl}$ the counter-terms ${\rm Tr}[\bar{H}\Pi(v\cdot
k)H]$, $-{\rm Tr}[\bar{S}\Pi(v\cdot k)S]$ and $-{\rm
Tr}[\bar{T}^\mu\Pi(v\cdot k)T_\mu]$, one obtains a modified
kinetic part of ${\cal L}^{hl}$ that can be written as follows
\cite{art1}:
\begin{eqnarray}
{\cal L}^{hl}_{\rm ren}&=& -{\rm Tr}[\bar{H}_{\rm ren}(iv\cdot
\partial-\Delta_H)H_{\rm ren}]+{\rm Tr}[\bar{S}_{\rm ren}(iv\cdot
\partial-\Delta_S)S_{\rm ren}]\nonumber\\
\label{eq:kin} &+&{\rm Tr}[\bar{T}_{\rm ren}^\mu(iv\cdot
\partial-\Delta_T)T_{\rm ren \mu}],
\end{eqnarray}
provided that:
\begin{eqnarray}
\label{eq:deltas}
\frac{1}{2G_3}&=&\Pi_H(\Delta_H)=\Pi_S(\Delta_S)\\
\label{eq:deltat} \frac{1}{2G_4}&=&\Pi_T(\Delta_T)\\
\label{eq:hren} H_{\rm ren}&=&\frac{H}{\sqrt{Z_H}}
\\
S_{\rm ren}&=&\frac{S}{\sqrt{Z_S}}
\\
T_{\rm ren}&=&\frac{T}{\sqrt{Z_T}},
\end{eqnarray}
where  the renormalization constants $Z$ are defined as follows:
\begin{equation}
Z_j^{-1}=\left(\frac{d}{dx}\Pi(x)\right)_{x\to \Delta_j},
\end{equation}
with $j=H,S,T$. As showed, the kinetic part of ${\cal L}^{ hl}$,
that originally was written as:
\begin{equation}
\frac{1}{2G_3}{\rm Tr}[\bar{H}H]- \frac{1}{2G_3}{\rm Tr}[\bar{S}S]
+\frac{1}{2G_4}{\rm Tr}[\bar{T}^\mu T_\mu],
\end{equation}
it is substituted by the  form given in (\ref{eq:kin}), see
\cite{rep} (see also (\ref{eq:wise})). If compared with
(\ref{eq:wise}) (the expression (\ref{eq:kin}) is extended to
include the meson fields $S$ and $T$), (\ref{eq:kin}) does not
contain the fields ${\cal V}$, since in the CQM model the pions
are not coupling directly to the meson fields and includes mass
terms such as $\Delta_{H,S,T}$. In the chiral Lagrangian approach
for heavy meson states, see section {\bf \ref{sec:mesoni}}, the
fundamental fields of the Lagrangian are the meson fields. CQM is
a somehow more fundamental approach since it includes, together
with meson fields, also the quark fields. When  one adds in
(\ref{eq:wise}) the kinetic terms related to the $S$ and $T$
fields following a chiral Lagrangian approach, see \cite{luca},
one must also subtract two mass shifting terms: $\delta m_S {\rm
Tr}[\bar{S}S]$ and $\delta m_T {\rm Tr}[\bar{T}^\mu T_\mu]$, where
$\delta m_S=m_S-m_H=\Delta_S-\Delta_H$ and $\delta
m_T=m_T-m_H=\Delta_T-\Delta_H$ are defined in the $m_Q\to \infty$
limit. In the CQM model, that contains explicitly the heavy and
light quark fields,  $\delta m_S$ and $\delta m_T$ are substituted
by $\Delta_S$ and $\Delta_T$, {\it i.e.}, the mass differences
between the heavy meson masses $M_{S,T}$ and the mass of the heavy
quark involved $m_Q$; $\Delta_H$ comes in the $H$ kinetic term.

$\Delta_H$ is the free CQM parameter. We cannot deduce  it from
the model, but we can fix it by reasonable numerical values. On
the other hand, $\Delta_S$ and $\Delta_T$ can be computed once
$\Delta_H$ is fixed. In the case of $\Delta_S$, one only needs to
solve (\ref{eq:deltas}). In the case of $\Delta_T$, we will use
some experimental information and a $1/m_Q$ correction to the
meson mass formula (we will discuss this point later on).

The CQM expressions for  $\Pi_H(\Delta_H)$, $\Pi_S(\Delta_S)$ and
$\Pi_T(\Delta_T)$ are here given. They allow to calculate the
renormalization constants $Z_{H,S,T}$:
\begin{eqnarray}
\Pi_H (\Delta_H) &=& I_1 + ( \Delta_H + m) I_3(\Delta_H)\\
\label{eq:donata} \Pi_S (\Delta_S) &=& I_1 + (\Delta_S -
m)I_3(\Delta_S)\\ \label{eq:antonio} \Pi_T (\Delta_T) &=& {1\over
{\Lambda_{\chi}^2}} \left[ -{I'_1\over 4}+ \frac{m+\Delta_T}{3}
[I_0(\Delta_T) + \Delta_T I_1 + (\Delta_T^2-m^2) I_3(\Delta_T)]
\right]\\
\label{eq:llaaxxii} Z_H^{-1} &=& (\Delta_H+m) {\frac{\partial
I_3(\Delta_H)}{\partial \Delta_H}}
 +I_3(\Delta_H)\\
Z_S^{-1} &=& (\Delta_S-m) {\frac{\partial I_3(\Delta_S)}{\partial
\Delta_S}} +I_3(\Delta_S),
\end{eqnarray}
and finally:
\begin{eqnarray}
Z_T^{-1} &=&\frac{1}{3 \Lambda^2_\chi} \Big[ (\Delta_T^2-m^2)
\left[(m+\Delta_T) {\frac{\partial I_3(\Delta_T)}{\partial
\Delta_T}} +I_3(\Delta_T)\right]\nonumber \\ &+& (m+\Delta_T)
\left[{\frac{\partial I_0(\Delta_T)}{\partial \Delta_T}}+ I_1 +2
\Delta_T I_3(\Delta_T)\right] +I_0 +\Delta_T I_1 \Big],
\end{eqnarray}
where $\Lambda_\chi=1$ GeV. The $I$ integrals are given in the
Appendix.

At this point let us fix $\Delta_H$ and compute numerically
$\Delta_{S,T}$, the couplings $G_3$, $G_4$ and the renormalization
constants $Z_{H,S,T}$.

We will consider everywhere in this work $H_{\rm ren}$, $S_{\rm
ren}$ and $T_{\rm ren}$, but often, when notation is evident, we
will drop the `ren'.

The  $\Delta_H$ values will be taken in the range $\Delta_H=0.3,
0.4,0.5$ GeV.  $\Delta_S$ and $G_3$ follow directly from eq.
(\ref{eq:deltas}). From eqs. (\ref{eq:donata}) and
(\ref{eq:antonio}) it is evident that $\Delta_S-\Delta_H=0$ if
$m\to 0$. Finally, $\Delta_T$ is obtained as follows.

Take $M_H$ and $M_T$ to be the spin averaged masses related to the
$H$ and $T$ multiplets respectively (a weighted average of the
experimental masses of the particles in each doublet is taken. The
weights are given by the number of polarization states that each
particle can assume according to its spin). We can write the
following two $O(\frac{1}{m_Q})$ equations:
\begin{eqnarray}
\label{eq:arg1} M_H&=&m_Q+\Delta_H+\frac{\Delta_H^\prime}{m_Q}
\\
\label{eq:arg2} M_T&=&m_Q+\Delta_T+\frac{\Delta_T^\prime}{m_Q}.
\end{eqnarray}
This couple of equations can be written both if $m_Q=m_c$ or
$m_Q=m_b$.

In the case of $m_Q=m_c$ we must use  experimental information
about the  $D_2^*$ and $D_1^*$ states. As for  $D_2^*$, we have
$m_{D^{*}_2(2460)^0}=2458.9\pm 2.0$ MeV,
$\Gamma_{D^{*}_2(2460)^0}=23\pm 5$ MeV and
$m_{D^{*}_2(2460)^{\pm}}=2459\pm 4$ MeV,
$\Gamma_{D^{*}_2(2460)^\pm}=25^{+8}_{-7}$  MeV. These particles
are identified with the $2^+$ states of the $T$ multiplet
$\frac{3}{2}^+$. As  for $D^{*}_1(2420)$, experimentally it is
found $m_{D^{*}_1(2420)^0}=2422.2\pm 1.8$ MeV,
$\Gamma_{D^{*}_1(2420)^0}=18.9^{+4.6}_{-3.5}$ MeV; this particle
can be identified with the $1^+$ state of the $T$ multiplet. We
are  ignoring a possible small mixing between the  $1^+$ states
belonging respectively to the $S$ and $T$ multiplets  \cite{rep}.
$D_2^*$ and $D_1^*$ states are quite broad states since their
strong decays proceed in $d$ wave, as pointed out in  {\bf
\ref{sec:mesoni}}.

From this analysis we obtain that:
\begin{equation}
\label{eq:one} \Delta_T - \Delta_H +\frac{(\Delta^\prime_T
-\Delta^\prime_H)}{m_c}\simeq 470 {\rm MeV}.
\end{equation}

If on the other hand we consider the $m_Q=m_b$ case, experimental
data on positive parity resonances show a bunch of states, not
easily resolvable, having a mass $M_{B^{**}}=5698\pm 12$ MeV and a
width  $\Gamma=128\pm 18$ MeV \cite{pdg}. If we identify this mass
with the $T$ multiplet narrow states mass, we obtain:
\begin{equation}
\label{eq:two} \Delta_T - \Delta_H +\frac{(\Delta^\prime_T
-\Delta^\prime_H)}{m_b}\simeq 380 {\rm MeV}.
\end{equation}
Reasonable values of the heavy constituent masses are
$m_b=m_B-300$ MeV and $m_c=m_D-300$ MeV, where  300 MeV is the
constituent mass discussed in {\bf \ref{sec:reg}}, while $m_B$ and
$m_D$ are the experimental masses of the $B$ and $D$ mesons
respectively. Solving simultaneously (\ref{eq:one}) and
(\ref{eq:two}), one gets:
\begin{equation}
\Delta_T - \Delta_H \simeq 335 \;{\rm {MeV}}.
\end{equation}
The results for $\Delta_S$ and $\Delta_T$ as functions of
$\Delta_H$, are shown in Table \ref{t:senza1};
\begin{table} [htb]
\hfil \vbox{\offinterlineskip \halign{&#&
\strut\quad#\hfil\quad\cr \hline \hline &$\Delta_H$ && $\Delta_S$
&& $\Delta_T$& \cr \hline &$0.3$&& $0.545$&& $0.635$ &\cr &$0.4$&&
$0.590$&& $0.735$ &\cr &$0.5$&& $0.641$&& $0.835$ &\cr \hline
\hline }} \caption{$\Delta$ values in (in GeV)} \label{t:senza1}
\end{table}
in Table \ref{t:senza2} we list the CQM $G_j$ and $Z_j$ values.
\begin{table} [htb]
\hfil \vbox{\offinterlineskip \halign{&#&
\strut\quad#\hfil\quad\cr \hline \hline &$\Delta_H $&& $1/G_3 $&&
$Z_H $&& $Z_S $&& $Z_T $&& $1/G_4$ &\cr \hline &$0.3$&& $0.16$&&
$4.17$&& $1.84$&& $2.95$&&  $0.15$ & \cr &$0.4$&& $0.22$&&
$2.36$&& $1.14$&& $1.07$&&  $0.26$ & \cr &$0.5$&& $0.345$&&
$1.142$&& $0.63$&& $0.27$&& $0.66$ & \cr \hline \hline }}
\caption{Renormalization constants and couplings. $\Delta_H$ in
GeV; $G_3$, $G_4$ in GeV$^{-2}$, $Z_j$ in GeV$^{-1}$.}
\label{t:senza2}
\end{table}

Through $\Delta_S$, CQM predicts the following $m_S$ mass value
(in literature these are the $D_0,D_1^{*\prime}$ states):
\begin{equation}
m_S=2165\pm 50 {\rm MeV},
\end{equation}
where the central value is given in correspondence of
$\Delta_H=0.4$ GeV, while the upper and lower values are related
to the remaining two $\Delta_H$ values. This determination does
not account for $1/m_c$ corrections (see eqs. (\ref{eq:arg1}) and
(\ref{eq:arg2})). These states are very difficult to be observed
experimentally since of their large width: from \cite{col1} one
expects $\Gamma(D_0\to D^+\pi^-)\simeq 180$ MeV and
$\Gamma(D_1^{*\prime}\to D^{*+}\pi^-)\simeq 165$ MeV. Theoretical
predictions of $m_S$ available in literature are anyway larger, a
typical value is $m_S=2350$ MeV. Very recent CLEO data
\cite{zoeller} indicate, as the mass of the $S$ multiplet,
$m_S=2461$ MeV. This discrepancy of about $300$ MeV between CQM
and experimental data can be attributable to the absence of
$O(1/m_c)$ corrections in CQM calculations. Anytime we will need
to use  $m_S$ in applications, we will use both the CQM predicted
value, for consistency, and the experimental one.

        \subsection{\label{sec:reson} ${\cal L}^{ll}$ extension to
        include $\rho$ and $a_1$ resonances.}

The effective Lagrangian (\ref{eq:cqmllag}) for the CQM light
sector can be extended to include $\rho$ and $a_1$ resonances. The
operative hypothesis needed is that of Vector-Meson-Dominance
(VMD) (and of Axial-Meson-Dominance for $a_1$). We will briefly
sketch the VMD hypothesis: then the insertion of $\rho$ in ${\cal
L}^{ll}$ (analogously for $a_1$) will turn out to be a simple
step.

Let us consider the electromagnetic form factor for $\pi^+$:
\begin{equation}
\langle \pi^+(p^\prime)|J_\mu|\pi^+(p)\rangle=(p+p^\prime)_\mu
F_\pi(t).
\end{equation}
where $t=(p^\prime-p)^2$. The best way to determine this form
factor is to consider the process $\gamma p\to n \pi^+$ where
$\gamma$ is a virtual photon coming from the scattering of an
electron \cite{fey-phint}. Now one does the hypothesis that
$F_\pi(t)$ is an analytic function in the variable $t$ with a
branching cut on the real positive axis. This hypothesis allows to
write the following dispersion relation for $F_\pi$:
\begin{equation}
F_\pi(t) = \frac{1}{\pi}\int_{4m_\pi^2}^{\infty}dt^\prime
\frac{{\cal I}m(F_\pi(t^\prime))}{t^\prime-t}.
\end{equation}
The lower integration bound is the threshold above which the form
factor is different from zero. Observe that for $t\to \infty$ one
has $F_\pi(t)\to 0$, since the probability of producing only two
pions, when an enormous number of higher energy states are
accessible, is extremely low.

Let us suppose that  $F_\pi(t)$ is {\it dominated} by the $\rho^0$
resonance (VMD-hypothesis) in such a way that one can write, for
the absorbitive part \cite{marshak}:
\begin{equation}
{\cal I}m(F_\pi(t^\prime))\propto \delta(t^\prime-m_\rho^2).
\end{equation}
Then we have that:
\begin{equation}
\label{eq:alluno} F_\pi(t)=\frac{m_\rho^2}{m_\rho^2-t}.
\end{equation}
\begin{figure}[t!]
\begin{center}
\epsfig{bbllx=5cm,bblly=15cm,bburx=15cm,bbury=26cm,height=8truecm,
        figure=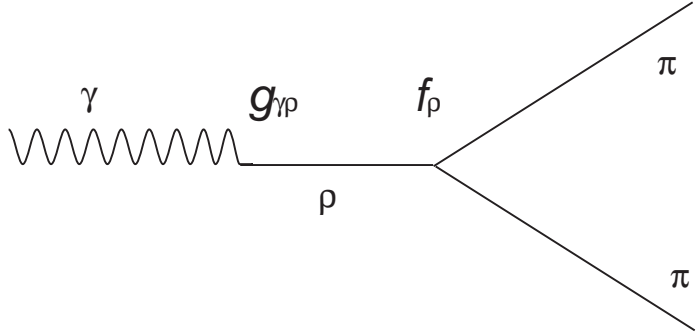}
\caption{\label{fig:rho} \footnotesize
          $\gamma-\rho$ coupling diagram. }
\end{center}
\end{figure}
On the other hand, if we consider the diagram in fig.
\ref{fig:rho} \cite{sakurai}, where the $\gamma-\rho$ coupling is
shown, we can write:
\begin{equation}
\label{eq:alldue} eF_\pi(t)=\frac{g_{\gamma
\rho}f_\rho}{m_\rho^2-t},
\end{equation}
where the $f_\rho$ coupling with pions is an universal coupling,
{\it i.e.}, is the same with two nucleons, two $\rho$'s, etc.
$f_\rho$'s universality is a consequence of electric charge
conservation and of the complete $\rho$ dominance.

From eqs. (\ref{eq:alluno},\ref{eq:alldue}) we conclude that the
$\gamma-\rho$ coupling constant is given by:
\begin{equation}
g_{\gamma \rho}=e\frac{m_\rho^2}{f_\rho}.
\end{equation}
In a paper by  N.M. Kroll, T.D. Lee and B. Zumino \cite{kroll}, it
is shown that, at lower order in $e$, it makes sense to consider
the interaction of an interpolating $\rho$ field,
$\frac{m_\rho^2}{f_\rho}\rho^\mu$, with the photon:
\begin{equation}
\label{eq:mado} e\frac{m_\rho^2}{f_\rho} \rho^\mu A_\mu,
\end{equation}
even if this term shows to be manifestly not gauge-invariant (one
can prove that introducing an $e^2$ term in the interaction
Lagrangian, this  problem  is solved). Equation (\ref{eq:mado})
can be considered  as the coupling of the gauge field with the
current:
\begin{equation}
J_\mu=\frac{m_\rho^2}{f_\rho} \rho_\mu.
\end{equation}
This identity between the  $\rho$ interpolating field and the
$J_\mu$ current makes sense only if $\rho^\mu$  is coupled to a
conserved external current, as one can easily show writing the
Lagrangian for the $\rho^\mu$ field as:
\begin{equation}
\label{eq:star} {\cal
L}^\rho=-\frac{1}{4}(F_{\mu\nu})^2-\frac{1}{2}m_\rho^2
\rho^\mu\rho_\mu+j\cdot \rho,
\end{equation}
where an interaction $j\cdot \rho$ with an external current $j$
(the $\rho$ source) is included. If one takes the divergence of
the equation of motion derivable from (\ref{eq:star}), and if one
considers that $\partial\cdot J=0$, then one has $\partial\cdot
j=0$. The equation of motion is:
\begin{equation}
(\partial^2+m^2)\rho^\mu=j^\mu. \label{eq:essa}
\end{equation}
Eq. (\ref{eq:essa}) produces the following relation between matrix
elements:
\begin{equation}
\langle B|J^\mu|A\rangle=\frac{m_\rho^2}{f_\rho} \frac{\langle
B|j^\mu|A\rangle}{m^2_\rho-t}.
\end{equation}
Therefore, at the level of matrix elements, we can see that the
$\rho$ and the photon sources coincide, modulo a factor
$\frac{1}{f_\rho}$. A problem arises if one observes that $f_\rho$
is not directly measurable at $t=0$, due to the finite $\rho$
mass: in  $\rho\to \pi\pi$ one measures  $f_\rho$ for
$t=m^2_\rho\neq 0$. This means that one has to formulate the
problematic hypothesis that $f_\rho$ is essentially constant in
the $(0,m_\rho^2)$ interval (in this respect VMD is similar to
PCAC). This problem is even stronger for $a_1$, that has a mass
larger than the $\rho$ one by a factor of $3/2$.

Let us go back to CQM. We can couple the $\rho$ interpolating
field to the vector fermion light quark current:
\begin{equation}
\frac{m_\rho^2}{f_\rho} \rho_\mu \bar{\chi}\gamma^\mu\chi.
\end{equation}
The same thing can be made for $a_1$, writing an analog
interaction for the $a_1$ interpolating field. These interaction
terms give Feynman rules for CQM vertices between light quark
current -$\rho$, -$a_1$:
\begin{eqnarray}
\label{eq:feyrho} \rho~{\rm vertex} &=& i
\frac{m_\rho^2}{f_\rho}\gamma^\mu \epsilon_\mu^*
\\
\label{eq:feya1} a_1~{\rm vertex} &=& i
\frac{m_{a_1}^2}{f_{a_1}}\gamma^\mu \gamma_5 \epsilon_\mu^{\prime
*},
\end{eqnarray}
where $\epsilon$ and $\epsilon^\prime$ are the polarizations of
$\rho$ and $a_1$ respectively.

The expression for the CQM effective Lagrangian ${\cal L}^{ll}$,
must incorporate these results. Let us write ${\cal L}^{ll}$ using
a notation mediated by the Hidden-Symmetry approach, which
incorporates VMD \cite{art2}:
\begin{eqnarray}
\label{eq:feyrholag} {\cal L}^{ll}&=&{f_{\pi}^2\over 8}
\partial_{\mu} \Sigma^{\dagger} \partial^{\mu} \Sigma +\frac{1}{2
g_V^2} {\rm Tr}[{\cal F(\rho)}_{\mu \nu} {\cal F(\rho)}^{\mu \nu}]
+\frac{1}{2 g_A^2} {\rm Tr}[{\cal F}(a)_{\mu \nu} {\cal F}(a)^{\mu
\nu}] \nonumber\\ &+& {\bar \chi} (i  D^\mu \gamma_\mu -m) \chi
\nonumber\\
 &+& {\bar \chi} (
{\cal A}^\mu \gamma_\mu \gamma_5- i h_\rho \rho^{\mu} \gamma_\mu -
i h_a a^{\mu} \gamma_\mu \gamma_5) \chi,
\end{eqnarray}
where:
\begin{equation}
{\cal F}(x)_{\mu \nu} = \partial_\mu x_\nu - \partial_\nu x_\mu +
[x_\mu,x_\nu],
\end{equation}
describes the strength tensor for the $\rho$ and $a_1$ fields.
Moreover:
\begin{equation}
\rho_\mu = i \frac{g_V}{\sqrt{2}} {\hat \rho}_\mu  ~~~~~~~~~~~~g_V
= \frac{m_\rho}{f_\pi} \simeq 5.8
\end{equation}
and, in an analogous way, we also write ($m_a\simeq 1.26$ GeV):
\begin{equation}
a_\mu = i \frac{g_A}{\sqrt{2}} {\hat a}_\mu
 ~~~~~~~~~~~~g_A = \frac{m_a}{f_\pi} \simeq 9.5, \label{GA}
\end{equation}
where $\hat \rho$ and $\hat a$ are Hermitian  $3\times 3$ matrices
related to positive and negative parity light mesons. If we
consider:
\begin{eqnarray}
h_\rho &=&\frac{\sqrt{2} m^2_\rho}{g_V f_\rho}
\\
h_a &=&\frac{\sqrt{2} m^2_a}{g_A f_a},
\end{eqnarray}
then we are implementing VMD and AMD hypotheses and we are
recovering (\ref{eq:feyrho},\ref{eq:feya1}) vertices. Numerically
one finds:
\begin{equation}
h_\rho\simeq h_a\simeq 0.95.
\end{equation}

For us $f_{\rho}=0.152~{\rm GeV}^2$, as emerges from $\rho^0$ and
$\omega$ decays in  $e^+ e^-$, and $f_a= 0.25 \pm 0.02~{\rm
GeV}^2$, as it comes out from $\tau \to \nu_\tau \pi \pi \pi$
decay \cite{9art2}. This result agrees with a determination made
for $f_a$ by QCD sum rules  \cite{10art2}. Lattice QCD predicts
$f_a=0.30 \pm 0.03~{\rm GeV}^2$, \cite{11art2}. Since $1/f_a$
multiplies all amplitudes containing the $a_1$ meson, the
uncertainty on $f_a$ will induce an uncertainty on  normalizations
for all the amplitudes involving the light axial meson.
\section{\label{chap:terzo}Strong Couplings}

        \subsection{\label{sec:unpi}Processes with one $\pi$ in the final state}

\begin{figure}[t!]
\begin{center}
\epsfig{bbllx=128pt,bblly=12cm,bburx=466pt,bbury=706pt,height=8truecm,
        figure=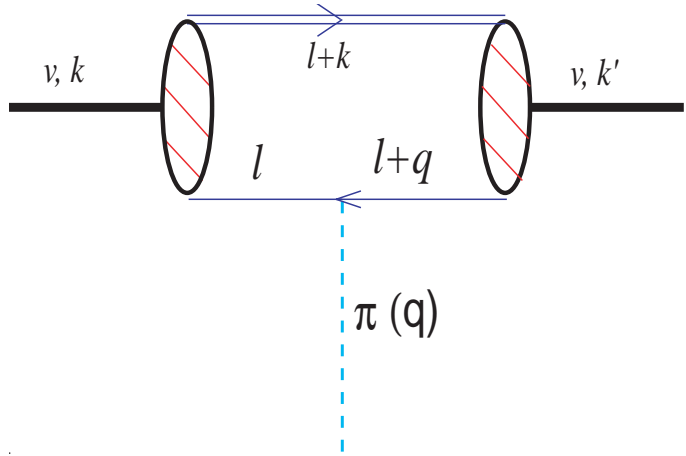}
\caption{\label{fig:1pione} \footnotesize
          CQM diagram for transitions meson $\to$ meson $+$ pion.}
\end{center}
\end{figure}

This is the first of two sections where the CQM model will be used
to compute the coupling constants for the strong decays $H\to
H\pi$, $S\to H\pi$, $S\to S\pi$, $T\to H\pi$, $T\to S\pi$ and $H
\to H(\rho,a_1)$, $H \to S(\rho,a_1)$. A technique allowing to go
beyond the soft pion limit hypothesis is also introduced. As
showed previously, CQM incorporates a direct coupling of the pion
to the light quark current. For strong processes with one pion in
the final state, there is only one CQM diagram describing the
decay of an heavy meson to another heavy meson and a pion. This
diagram is represented in fig. \ref{fig:1pione}. Different
processes have different in and out heavy meson states and the
soft pion limit hypothesis must be discussed case by case. The
soft pion limit allows to simplify calculations, but it is a rough
approximation if, {\it e.g.}, transitions $S\to H \pi$ or $T\to H
\pi$ are considered. In the exact chiral limit we can write, for
the pion momentum, $q^\mu=(q_\pi,0,0,q_\pi)$ . In the heavy meson
rest frame, $v=(1,0,0,0)$, we have $v\cdot q=q_\pi$. Moreover
$v\cdot q=v\cdot (p-p^\prime)=\Delta_S-\Delta_H\neq 0$, where $p$
and $p^\prime$ denote the momenta of the in and out mesons
respectively and we have used the relation  $v\cdot k=\Delta$, $k$
being the residual momentum, (see {\bf \ref{sec:hqet}}), and
$\Delta$ the mass difference between the heavy meson and the heavy
quark there contained. Therefore, the soft pion limit is not very
reliable for transitions such as $S \to H \pi$, where
$\Delta_S-\Delta_H\simeq 140-190$ MeV.
We should observe that, if we adopt the soft pion limit
hypothesis, the CQM diagram of fig. \ref{fig:1pione} shows a very
soft NG boson emitted from an internal line of a diagram, while we
should have expeced  an Adler zero of the emitting amplitude in
this case. Anyway the CQM regularization scheme forces $\ell^2$ in
the loop to be quite close to $m^2$ and this saves the soft pion
approximation (see discussion in \cite{weinbook}, pp. 175,176).

In \cite{art1} the calculation of CQM amplitudes for the
transitions $H\to H\pi$ and $S\to H\pi$ has been performed using
in both cases the soft pion limit, which works as a good
approximation only in the first case. In \cite{art3}, a technique
allowing to avoid the soft pion limit has been introduced for the
$T\to H\pi$ process. The same technique gives the possibility of
improving also the $S\to H\pi$ calculation. Evidently, the soft
pion limit is a good approximation for the $S\to S \pi$ process
\cite{art5}. Recent CLEO data \cite{zoeller} indicate that
$m_S\simeq m_T\simeq 2460$ MeV so that the soft pion limit may be
used also for the $T\to S\pi$ process \cite{art5}. In the next two
subsections we will show how to compute the mentioned processes in
the soft and not-soft pion hypothesis.

\subsubsection{\label{sec:hhpi} $H\to H\pi$, the soft pion limit}
\begin{figure}[t!]
\begin{center}
\epsfig{bbllx=128pt,bblly=12cm,bburx=466pt,bbury=706pt,height=8truecm,
        figure=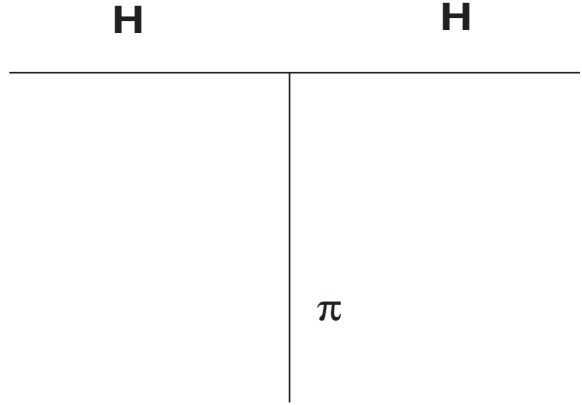}
\caption{\label{fig:fig6} \footnotesize
          In the low energy language of chiral Lagrangians, the pion is directly
          coupled to the meson fields.}
\end{center}
\end{figure}

Let's consider the first term in the Lagrangian (\ref{eq:forti}):
\begin{equation} \label{eq:prec} {\cal L}=ig{\rm
Tr}[\bar{H}H\gamma\cdot {\cal A}\gamma_5],
\end{equation}
where the meson field  $H$ has been defined in (\ref{eq:acca}).
The transition $1^- \to 0^- \pi$ is allowed. We can therefore
consider $\langle D\pi|i{\cal L}|D^*\rangle$ and, using
(\ref{eq:prec}), we obtain:
\begin{equation} \label{eq:raffa} \langle D\pi|i{\cal
L}|D^*\rangle= g\left(-\frac{iq^\mu}{f_\pi}\right){\rm
Tr}[\tilde{P}\gamma_5\frac{1+\gamma \cdot v}{2}\gamma^\sigma
\tilde{P}^*_\sigma\gamma_\mu\gamma_5],
\end{equation}
where ${\cal A}$  has been expanded up to the first order in $\pi$
and the zeroth order in the expansion has been neglected.
Observing that $\tilde{P}=\langle H|P|{\rm VAC}\rangle$ and
$\tilde{P}^*_\sigma=\langle {\rm VAC}|P^*_\sigma|H\rangle$ can be
made explicit using (\ref{eq:norm1},\ref{eq:norm2}), the
interaction term (\ref{eq:raffa}) reduces to:

\begin{equation}
\label{eq:chiamala} -ig\frac{2m_H}{f_\pi}(\epsilon\cdot q),
\end{equation}
where $\epsilon$ represents the polarization of the $1^-$ state.
As already observed, this interaction effective Lagrangian
describes the coupling of the pion to the meson states, the
fundamental fields at low energy, see fig. \ref{fig:fig6}.

In the CQM model, where the fundamental fields are the heavy and
light constituent quarks, the same coupling is {\it modeled} as a
coupling of the pion to the light quark current. Figure
\ref{fig:1pione} shows a one loop CQM diagram containing two
vertices  meson-$Q$-$\chi$, see fig.  \ref{fig:vertex}, described
by ${\cal L}^{hl}$ introduced in (\ref{eq:cqmhl}) and one vertex
pion-$\chi\chi$. This diagram can be computed as a standard loop
diagram in the following way:

\begin{equation}
\label{eq:looopy} (-1)i^3 i^3 Z_H m_H\frac{N_c}{16\pi^4}\int
d^4\ell \frac{{\rm Tr }[(\gamma\cdot
\ell+m)(-\frac{q^\mu}{f_\pi}\gamma_\mu\gamma_5) (\gamma\cdot \ell
+ m)\gamma_5\frac{1+\gamma\cdot v}{2}\gamma^\sigma
\epsilon_\sigma]}{(\ell^2-m^2)^2(v\cdot \ell+\Delta_H)}.
\end{equation}
Here is a legenda for the factors appearing in the preceding
expression:

\begin{itemize}
\item $(-1)$ from the fermion loop\\
\item $i^3$ from the three quark propagators\\
\item $i^3$ from the Feynman rules for the $\chi\chi\pi$ vertex,
described in ${\cal L}^{ll}$, and for  the two $\chi H  Q$
vertices, described in ${\cal L}^{hl}$; both carry a factor of
$(-i)$\\
\item $Z_H$ is due to the fact that the two meson fields coming in
the loop integral must be the renormalized fields, $\sqrt{Z_H}$
being the renormalization constant\\
\item $m_H$ comes from the normalization conditions
(\ref{eq:norm1},\ref{eq:norm2})\\
\item $N_c$ is the number of colours running in the loop ($N_c=3$)\\
\item $(-\frac{q^\mu}{f_\pi}\gamma_\mu\gamma_5)$ is the one pion
expansion of the $\gamma\cdot{\cal A}\gamma_5$ term contained in
${\cal L}^{ll}$.\\
\end{itemize}
At this point one should calculate the trace in (\ref{eq:looopy})
and the following integrals:
\begin{eqnarray} \label{eq:zetasoff} Z&=&\frac{iN_c}{16\pi^4}\int
d^4\ell \frac{1}{(\ell^2-m^2)^2(v\cdot \ell+\Delta_H)}
\\
Z_\mu&=&\frac{iN_c}{16\pi^4}\int d^4\ell
\frac{\ell_\mu}{(\ell^2-m^2)^2(v\cdot \ell+\Delta_H)}
\\
Z_{\mu\nu}&=&\frac{iN_c}{16\pi^4}\int d^4\ell \frac{\ell_{\mu}
\ell_{\nu}}{ (\ell^2-m^2)^2(v\cdot \ell+\Delta_H)}.
\end{eqnarray}

The technique for calculating $Z$ with the proper time
regularization procedure will be explained in detail in the next
section, where the general case in which the two light propagators
carry different momenta ($\ell$ and $\ell+q$ respectively) is
analyzed. The soft pion approximation is achieved performing the
$q_\pi\to 0$ limit. The expression for  $Z$ is then equivalent to
the expression for  $I_4(\Delta_H)$ given in the Appendix. The
Lorentz structures must be treated as in the following example:
\begin{equation} Z_{\mu\nu}=\frac{iN_c}{16\pi^4}\int d^4\ell
\frac{\ell_{\mu}\ell_{\nu}}{ (\ell^2-m^2)^2(v\cdot
\ell+\Delta_H)}=Qv_\mu v_\nu +R g_{\mu\nu}.
\end{equation}
In order to obtain $Q$ and $R$, the contraction with $v^\mu v^\nu$
and $g^{\mu\nu}$ in needed. In such a way one obtains integrals of
the type described in the Appendix.

A comparison between (\ref{eq:chiamala}) and (\ref{eq:looopy})
shows that, in order to obtain $g$, {\it one has to extract from}
(\ref{eq:looopy}) {\it the coefficient of} $\epsilon\cdot q$. The
CQM expression for $g$ is then the following:
\begin{equation} g=Z_H\left[\frac{1}{3} I_3(\Delta_H)
-2\left(m+\frac{1}{3}\Delta_H \right)(I_2 + \Delta_H
I_4(\Delta_H))-\frac{4}{3}m^2 I_4(\Delta_H)\right],
\end{equation}
where all the $I$ integrals are listed in the Appendix.
Numerically:
\begin{equation}
g=0.456\pm0.040,
\end{equation}
where the central  value corresponds to $\Delta_H=0.4$ GeV and the
lower (higher) values, correspond to $\Delta_H=0.3$ GeV
($\Delta_H=0.5$ GeV). These values agree with what is found using
the QCD sum rules method, $g=0.44\pm0.16$ \cite{rep,32art1}. A
good aggreement is also found with relativistic quark models
giving $g\simeq 0.40$ \cite{31art1} and $g=0.34$ \cite{33art1}.
CQM disagrees with the determination by Le Yaouanc and Becirevic
\cite{becirevic} according to which $g=1$.

The coupling constant $g$ allows the determination of the hadronic
width:

\begin{equation} \Gamma(D^{*+} \to D^0 \pi^+ ) = \frac{g^2}{6\pi
f_\pi^2} |{\vec p_\pi}|^3.\label{strong}
\end{equation}
Numerical values are described in Table \ref{t:dipi}.
\begin{table} [htb]
\hfil \vbox{\offinterlineskip \halign{&#&
\strut\quad#\hfil\quad\cr \hline \hline &Decay&& $\Delta_H=0.4$
GeV && $\Delta_H=0.5$ GeV && Exp.&\cr \hline &${D^*}^0\to D^0
\pi^0$&& $65.5$ && $70.1$ && $61.9\pm 2.9$&\cr &${D^*}^0\to D^0
\gamma$&& $34.5$ && $29.9$ &&$38.1\pm 2.9$&\cr &${D^*}^+\to D^0
\pi^+$&& $71.6$ && $71.7$ && $68.3\pm 1.4$&\cr &${D^*}^+\to D^+
\pi^0$&& $28.0$ && $28.1$ && $30.6\pm 2.5$&\cr &${D^*}^+\to D^+
\gamma$&& $0.4$ &&  $0.24$ &&$1.1^{+2.1}_{-0.7}$ &\cr \hline
\hline }} \caption{Theoretical and experimental B.R.'s for $D^*$.
The theoretical values are computed for $\Delta_H=0.4,0.5$ GeV.}
\label{t:dipi}
\end{table}

Table \ref{t:dipi}, following \cite{art1}, includes also the
B.R.'s predicted by CQM for the radiative decays. The CQM
calculation of radiative processes proceeds without qualitatively
new elements with respect to what explained so far.

The soft pion approximation is also suitable for the $S\to S \pi$
process \cite{art5}. The meson interaction Lagrangian is given by
the second term in  eq. (\ref{eq:forti}):
\begin{equation}
{\cal L}=ig^\prime{\rm Tr}[\bar{S}S\gamma\cdot{\cal A}\gamma_5].
\end{equation}
Once calculated the CQM loop diagram, a comparison with the
mesonic amplitude  $\langle S\pi|i{\cal L}|S\rangle$ gives:

\begin{equation} g^\prime=Z_S\left[\frac{1}{3} I_3(\Delta_S)
-2\left(-m+\frac{1}{3}\Delta_S\right)(I_2 + \Delta_S
I_4(\Delta_S))-\frac{4}{3}m^2 I_4(\Delta_S)\right],
\end{equation}
to which correspond the numerical values:
\begin{equation}
g^\prime=-0.13 \pm 0.04.
\end{equation}
Observe that $g^\prime$ can be obtained by $g$ with the simple
exchange $m\to -m$, $\Delta_H \to \Delta_S$ and $Z_H \to Z_S$.
Indeed the loop integral for the process $1^+\to 0^+ \pi$ can be
obtained by the loop integral for $H\to H\pi$, shifting the
$\gamma_5$ matrix contained in (\ref{eq:looopy}) on the right side
of $\gamma_\sigma$. In so doing, one writes  the expression that
would have written for the $S\to S\pi$ process but having
$-\gamma\cdot v$, instead of $\gamma\cdot v$, in the projector.
Expanding the trace in (\ref{eq:looopy}) in it's four terms
different from zero, one realizes that this substitution in the
projector is equivalent to an $m\to-m$ exchange.

According to the CLEO data, $m_S\simeq m_T$, therefore the soft
pion limit is a good approximation also for $T\to S\pi$. The CQM
approach gives \cite{art5}:
\begin{equation} f=2\sqrt{Z_TZ_S}[mV-T],
\end{equation}
where $V$ and $T$ are linear combinations of the $I_i$ integrals
discussed in the Appendix, while $f$ is defined in
(\ref{eq:effes}). Numerically we find:
\begin{equation}
f=0.91^{+0.56}_{-0.27}.
\end{equation}

\subsubsection{\label{sec:thpi} $T\to H\pi$, $q_\pi\neq0$}

Let's go back to the integral $Z$ given in  eq.
(\ref{eq:zetasoff}) and evaluate it in the general case of pions
bringing a momentum $q_\pi\neq 0$:
\begin{equation}
\label{eq:beyondi} Z=\frac{iN_c}{16\pi^4}\int
\frac{d^4\ell}{(\ell^2-m^2)((\ell+q)^2-m^2)(v\cdot
\ell+\Delta+i\epsilon)},
\end{equation}
where the Feynman contour prescription is made explicit in the
expression for the heavy quark propagator. Since the pion is the
NG boson of the broken chiral symmetry, we put:
\begin{equation}
q^\mu=(q_\pi,0,0,q_\pi).
\end{equation}

Let's write (\ref{eq:beyondi}) in the following way:
\begin{eqnarray}
Z=H_1+H_2&=& \int^{\rm reg}
\frac{d^4\ell}{(\ell^2-m^2)((\ell+q)^2-m^2)(v\cdot
\ell+\Delta-i\epsilon)}\nonumber\\ &-&2\pi i \int^{\rm reg}
d^4\ell \frac{\delta(v\cdot
\ell+\Delta)}{(\ell^2-m^2)((\ell+q)^2-m^2)},
\end{eqnarray}
where the proper time regularization is concerned and the Plemelij
identity has been invoked \cite{bernard}:
\begin{equation}
\frac{1}{x-(x_0 \pm i\epsilon)}={\rm P.V.} \frac{1}{x-x_0} \pm
i\pi \delta (x-x_0),
\end{equation}
in such a way that poles with negative real part lie in the upper
complex $\ell_0$ plane.

As a first step we compute $H_1$:
\begin{eqnarray}
\label{eq:primopezz} H_1&=&\frac{d}{dm^2}\int^{\rm reg}
d^4\ell\int_0^1 \frac{dx}{((\ell+qx)^2-m^2)(v\cdot\ell
+\Delta)}\nonumber\\ &=&-i\frac{d}{dm^2}\int d^4\ell_E \int_0^1
\frac{dx}{((\ell+qx)^2_E+m^2)(i\ell_4 +\Delta)}\nonumber
\\
&=&-\frac{d}{dm^2}\int
d^4\ell_E\int_0^1\frac{dx}{(\ell^2_E+m^2)(\ell_4 -i\Delta(x))}.
\end{eqnarray}
In the first equation the Feynman integral trick has been used
while, in the second, the Euclidean rotation is performed. At this
stage  $q_4=-iq_\pi$ and we can shift
$\ell_4\to\ell^\prime_4+iq_\pi x$ as shown in the last equation in
(\ref{eq:primopezz}). Here $\Delta(x)=\Delta-q_\pi x$. Now we must
exponentiate the light quark propagator and cutoff the integrals
according to the Shwinger's proper time prescription described in
{\bf \ref{sec:reg}}. We recall also the numerical factor
$\frac{iN_c}{16\pi^4}$ and write ($N_c=3$):
\begin{eqnarray}
H_1&=& -\frac{iN_c}{16\pi^4}\frac{d}{dm^2}
\int_{1/\Lambda^2}^{1/\mu^2} ds e^{-sm^2}\int d^3{\bf {\bf
\ell}}e^{-s{\bf \ell}^2} \int d\ell_4
\frac{e^{-s\ell_4^2}}{\ell_4-i\Delta(x)}\nonumber\\ &=&
-\frac{3}{16\pi^{3/2}}\int_{1/\Lambda^2}^{1/\mu^2}ds
\frac{e^{-sm^2}}{s^{1/2}}\int_{0}^{1}dx e^{s\Delta^2(x)}(1-{\rm
erf}(\sqrt{s}\Delta(x))).
\end{eqnarray}
In the second equation we have used the following formula:
\begin{equation}
\frac{1}{i\pi}\int dt \frac{e^{-t^2}}{t-iz}=e^{z^2}(1-{\rm
erf}(z)),
\end{equation}
where:
\begin{equation}
{\rm erf}(z)=\frac{2}{\sqrt{\pi}}\int_{0}^{z}dt e^{-t^2}.
\end{equation}
In a similar way let's work out $H_2$:
\begin{eqnarray}
H_2&=&-2\pi i \frac{d}{dm^2}\int^{\rm reg} d^4\ell
\int_{0}^{1}dx\frac{\delta(\ell_0+\Delta)}{((\ell+qx)^2-m^2)}\nonumber\\
&=& -2\pi  \frac{d}{dm^2}\int d^4\ell_E
\int_{0}^{1}dx\frac{\delta(i\ell_4+\Delta)}{((\ell+qx)^2_E+m^2)}\nonumber\\
&=& -2\pi  \frac{d}{dm^2}\int d^4\ell_E
\int_{0}^{1}dx\frac{\delta(i\ell_4+\Delta(x))}{(\ell^2_E+m^2)}\nonumber\\
&=& 2\pi i \frac{d}{dm^2} \int d^3\ell \int_{0}^{1}dx
\int_{1/\Lambda^2}^{1/\mu^2}ds e^{-s(m^2+{\bf {\bf
\ell}}^2-\Delta^2(x))}\nonumber\\
&=&2\frac{3}{16\pi^{3/2}}\int_{1/\Lambda^2}^{1/\mu^2}ds
\frac{e^{-sm^2}}{s^{1/2}}\int_{0}^{1}dx e^{s\Delta^2(x)},
\end{eqnarray}
where the same steps as for $H_1$ have been followed. In the last
equation $\frac{iN_c}{16\pi^4}$ is taken into account. Summing the
expressions obtained for $H_1$ and $H_2$:
\begin{equation}
Z=\frac{3}{16\pi^{3/2}}\int_{1/\Lambda^2}^{1/\mu^2}ds
\frac{e^{-sm^2}}{s^{1/2}}\int_{0}^{1}dx e^{s\Delta^2(x)}(1+{\rm
erf}(\sqrt{s}\Delta(x))).
\end{equation}
The soft pion limit amounts to consider $q_\pi\to 0$,
$\Delta(x)\to \Delta$ and therefore $Z\to I_4(\Delta)$, where
$I_4(\Delta)$, given in the Appendix, comes from the integration
of (\ref{eq:zetasoff}). The CQM calculation of the process $T\to
H\pi$ \cite{art3}, requires not only the calculation of $Z$, but
also of $Z_\mu$, $Z_{\mu\nu}$ and $Z_{\mu\nu\lambda}$ since the
CQM loop integral for $T\to H\pi$ is:
\begin{equation}
-\frac{i}{2f_\pi}\sqrt{Z_H Z_T m_H m_T} q^\mu \eta^{\sigma\nu}
\frac{iN_c}{16\pi^4} \int^{\mathrm {reg}}d^4\ell \frac{{\rm
Tr}[(\gamma\cdot\ell + m)\gamma_\mu \gamma_5 (\gamma\cdot (\ell+q)
+ m) \gamma_5 (1+\gamma\cdot v) \gamma_\nu k_\sigma]
}{(\ell^2-m^2)[(\ell+q)^2-m^2](v\cdot \ell + \Delta)}.
\label{eq:mostro}
\end{equation}
Let's consider, for example, $Z_{\mu\nu}=Cv_\mu v_\nu+ Dq_\mu
q_\nu+E(v_\mu q_\nu+q_\mu v_\nu)+Og_{\mu\nu}$. If we contract both
members of this equation with the tensors $v_\mu v_\nu$, $q_\mu
q_\nu$,..we obtain a linear system of four simultaneous equations
to be solved with respect to  $C,D,E,O$, knowing the (integral)
expressions $v^\mu v^\nu Z_{\mu\nu}$, $q^\mu q^\nu Z_{\mu\nu}$,
$v^\mu q^\nu Z_{\mu\nu}$ and $g^{\mu\nu} Z_{\mu\nu}$. The matrix
${\bf A}$, acting on the vector $(C,D,E,O)$, contains powers of
$q_\pi$ multiplied by numerical factors. The hadronic matrix
element that one wants to compute, informs about the powers of
$q_\pi$ that should be eliminated in ${\bf A}$.

Let' recall now the  eq. (\ref{eq:tih}):
\begin{equation}
\label{eq:tih} {\cal L}=\frac{h_1}{\Lambda_\chi} {\rm
Tr}[\bar{H}T^\mu(i D_\mu \gamma\cdot {\cal
A})\gamma_5]+\frac{h_2}{\Lambda_\chi} {\rm Tr}[\bar{H}T^\mu(i
\gamma \cdot D {\cal A}_\mu)\gamma_5]+{\rm h.c.}
\end{equation}
Using the same strategy followed in the calculation of $g$   in
{\bf \ref{sec:unpi}}, we find that $h^\prime=h_1+h_2$ is given by
\cite{art3}:
\begin{eqnarray}
h^\prime &=& \sqrt{Z_T Z_H} \Big\{ \frac{m^2}{q_\pi}
\Big[I_2+\Delta_T Z(\Delta_T )+ \frac{1}{2 q_\pi}
(I_3(\Delta_H)-I_3(\Delta_T))\Big] \nonumber \\ &+&
P(R_i(\Delta_T),S_i(\Delta_T),q_\pi)\Big\}. \label{eq:acc1}
\end{eqnarray}
The polynomial $P(R_i,S_i,q_\pi)$, given in the Appendix, is a sum
of $q_\pi$ powers multiplied by some linear combinations of the
integrals $I_i$ and  $Z$ that we called $R_i$ and $S_i$. If we
write down the hadronic matrix elements for the processes $T\to
H\pi$ (the following three), $S\to H\pi$ and $H\to H\pi$ (the last
two), we understand that, {\it e.g.}, in the CQM calculation of
$h^\prime$ one should take account only of two powers of $q_\pi$,
therefore we neglect terms $O(q_\pi^3)$ ({\it e.g.}, in ${\bf
A}$).
\begin{eqnarray}
\langle D^{+}(p^{\prime}) \pi^{-}(q)| D_2^{*0}(p,\eta) \rangle &=&
ig_1 \eta^{\mu\nu} q_\mu q_\nu \label{eq:mat1}\\ \langle
D^{*+}(p^{\prime},\epsilon) \pi^{-}(q)| D_2^{*0}(p,\eta) \rangle
&=& ig_2 \eta^{\mu\nu} q_\mu
\epsilon_{\lambda\sigma\nu\tau}\epsilon^{*\lambda}\frac{p^{\sigma}}{m_T}
q^{\tau} \label{eq:mat2}\\ \langle D^{*+}(p^{\prime},\epsilon)
\pi^{-}(q)| D_1^{*0}(p,\eta) \rangle &=&
ig_3\eta_{\nu}\epsilon^{*}_{\sigma}
[3q^{\nu}q^{\sigma}+g^{\nu\sigma} \left(q^2-\frac{(p\cdot
q)^2}{m_T}\right)] \label{eq:mat3}\\ \langle D^{0}(p^{\prime})
\pi^{+}(q)| D_0^+(p) \rangle &=& ig_4\frac{m_S^2-m_H^2}{m_S}
\label{eq:mat4}\\ \langle D^{0}(p^{\prime}) \pi^{+}(q)|
D^{*+}(p,\epsilon) \rangle &=& ig_5q^{\mu}\epsilon_{\mu}.
\label{eq:mat5}
\end{eqnarray}
The processes involving the $c$ quark are the most interesting
since the charmed states, even the positive parity ones, are
object of wide experimental investigation. The coupling constants
 $g_1,..,g_5$ are connected to the coupling constants appearing at
 the level of meson Lagrangians by the following relations:
\begin{eqnarray}
g_1& = & g_2 = 2\sqrt{m_H m_T} \frac{h^{\prime}}{\Lambda_{\chi}
f_{\pi}}\\ g_3 & = & \sqrt{\frac{2 m_H
m_T}{3}}\frac{h^{\prime}}{\Lambda_{\chi} f_{\pi}}\\ g_4 & = &
\sqrt{m_H m_S}\frac{f^\prime}{ f_{\pi}}\\ g_5 & = & \frac{2
m_H}{f_\pi} g,
\end{eqnarray}
where $f^\prime$ has been introduced in (\ref{eq:effes}).

Using recent data on the masses \cite{pdg} and the CQM value for
$h^\prime=0.65$, one obtaines the following widths:
\begin{eqnarray}
\Gamma (D_2^{*0} \to D^+ \pi^-) &=& 4.59 \times 10^7
\frac{h^{\prime 2}}{\Lambda_\chi^2}~{\rm MeV} = 19.4~{\rm MeV}\\
\Gamma (D_2^{*0} \to D^{*+} \pi^-) &=& 1.33 \times 10^7
\frac{h^{\prime 2}}{\Lambda_\chi^2}~{\rm MeV} = 5.6~{\rm MeV}\\
\Gamma (D_1^{*0} \to D^{*+} \pi^-) &=& 1.47 \times 10^7
\frac{h^{\prime 2}}{\Lambda_\chi^2}~{\rm MeV} = 6.2~{\rm MeV}.
\label{eq:width21}
\end{eqnarray}
Due to the neutral pion decay channel, these widths should be
multiplied by a factor of $1.5$.

According to the chiral Lagrangian for heavy mesons, the total
width of the state $D_2^{*0}$ is dominated by decays with only one
pion in the final state. Therefore we can use the experimental
value for this width, $23 \pm 5$ MeV, to obtain $h^\prime_{\rm
exp}=0.51$, in good agreement with $h^\prime=0.65$ predicted by
CQM.

Equation (\ref{eq:width21}) gives the total width for $D_1^{*0}$
decaying to one pion, {\it i.e.}, $\Gamma=9.3$ MeV. This is only
one half of the measured total width  $18.9^{+4.6}_{-3.5}$ MeV.
This effect could be attributed to a mixing of $T(1^+)$ with
$S(1^+)$ or to strong $O(1/m_c)$ corrections \cite{14art3}.

\subsubsection{\label{sec:shpi} $S\to H\pi$, $q_\pi\neq0$}
Applying the technique developed in {\bf \ref{sec:thpi}}, we can
compute the strong coupling $f^\prime$ describing  $S\to H\pi$.
One finds:
\begin{equation}
f^\prime =\sqrt{Z_S Z_H}\left[R_1(\Delta_S)-
R_2(\Delta_S)-\frac{R_4(\Delta_S)} {q_\pi}+ m^2
Z(\Delta_S)\right]. \label{eq:accap}
\end{equation}
In  this computation only terms up to order $q_\pi^1$ have been
considered. Numerically we find:
\begin{equation}
f^\prime=-0.76\pm 0.13,
\end{equation}
where the error is induced by the variation of $\Delta_H$ in the
range $0.3,0.4,0.5$~GeV. In this computation the CQM value
$m_S=2.165\pm 0.05$ GeV has been used (the error in this last
determination doesn't affect much the $f^\prime$ numerical value).
Anyway, recent CLEO data \cite{zoeller}  give $m_S\simeq 2.461$
GeV for the broad charmed state  $S(1^+)$. This discrepancy
between the experimental value and the CQM determination, is most
likely due to $O(1/m_c)$ corrections. Since  $m_S$ determines the
$q_\pi$ value, the experimental value gives a different numerical
result for $f^\prime$:
\begin{equation}
f^\prime=-0.56\pm 0.11, \label{eq:newh}
\end{equation}
consistent with a QCD sum rules determination \cite{col1}:
\begin{equation}
f^\prime=-0.52\pm 0.17.
\end{equation}
If one computes the process  $S\to H\pi$ applying the soft pion
limit hypothesis, the following numerical determination for
$f^\prime$ is obtained \cite{art1}:
\begin{equation}
f^\prime=-0.85\pm 0.02.
\end{equation}
Neglecting mixing effects $S(1^+)-T(1^+)$ we can evaluate the
width of $D_1^\prime$ using the CLEO results for $m_S$ and the CQM
value for $f^\prime$. We get:
\begin{equation}
\Gamma(D_1^\prime \to D^* \pi)=240\; {\rm MeV},
\end{equation}
which probably accounts for the entire $D_1^\prime$ width, due to
the limited phase space. This result can be compared with the CLEO
total width for the $1^+$ state, $290^{+101}_{-79}\pm 26 \pm 36$
MeV.

        \subsection{\label{sec:rhoa}Processes with $\rho$ and  $a_1$ in the final state}

In this section we will discuss the strong coupling constants
 $HH\rho$, $HS\rho$, $HHa_1$ and $HSa_1$.
These will turn out to be essential when the semileptonic decays
$B\to (\rho,a_1)\ell\nu$ \cite{art2} are examined. According to
the notations introduced in \cite{rep}, these couplings are
parametrized in the following way:
\begin{eqnarray}
\label{eq:strong1} {\cal L}_{H H \rho}&=&i\lambda {\mathrm
{Tr}}(\overline{H} H \sigma^{\mu \nu} {\cal F(\rho)}_{\mu \nu})
-i \beta {\mathrm {Tr}}(\overline{H} H \gamma^\mu \rho_\mu)
\\
\label{eq:strong2} {\cal L}_{H S \rho}&=&-i \zeta {\mathrm
{Tr}}(\overline{S} H \gamma^\mu \rho_\mu)+ i\mu {\mathrm
{Tr}}(\overline{S} H \sigma^{\mu \nu} {\cal F(\rho)}_{\mu \nu})
\\
\label{eq:strong3} {\cal L}_{H H a_1}&=& -i \zeta_A {\mathrm
{Tr}}(\overline{H} H \gamma^\mu a_\mu )+ i\mu_A {\mathrm
{Tr}}(\overline{H} H \sigma^{\mu \nu} {\cal F}(a)_{\mu \nu} )
\\
\label{eq:strong4} {\cal L}_{H S a_1}&=&i\lambda_A {\mathrm
{Tr}}(\overline{S} H \sigma^{\mu \nu} {\cal F}(a)_{\mu \nu} )  -i
\beta_A {\mathrm {Tr}}(\overline{S} H \gamma^\mu a_\mu ).
\end{eqnarray}

The CQM approach for the determination of these constants amounts
to a loop calculation of diagrams like the one in fig.
\ref{fig:1pione}, where the pion is substituted by a  $\rho$ or
$a_1$  according to the Feynman rules obtained in  {\bf
\ref{sec:reson}}, followed by a comparison with the hadronic
matrix elements determined by the interaction Lagrangians above
listed.

Even if the strategy is the same as before, we have some new
technical difficulties in dealing with $\rho$ and $a_1$, partly
because of the polarizations of these states, partly because of
the fact that $\rho$ and $a_1$ are not massless.

The first mentioned problem has influence on the expressions for
the hadronic matrix elements. Consider, for example, the
transition $H(1^-)\to S(0^+) \rho$. This process has contributions
in $s$ and $d$ wave. The $d$ wave contribution comes only from the
Lagrangian term containing the factor $\mu$ in (\ref{eq:strong2})
while, the $s$ wave contributions, come also from the Lagrangian
term containing the $\zeta$ factor. With obvious notation, the
matrix element is written in the following way:
\begin{equation}
\langle\rho^+(\epsilon,q)S(p^\prime)|H(\eta,p)\rangle=-i\epsilon^{*}_{\mu}
\eta_\lambda(Sg^{\mu\lambda}+Dv^\mu q^\lambda).
\end{equation}
Applying the notations introduced in {\bf \ref{sec:reson}}, we can
compute explicitly the traces in (\ref{eq:strong2}) obtaining:
\begin{eqnarray}
S&=&\frac{g_V}{\sqrt{2}}\sqrt{m_H m_S}(2\zeta-4\mu(v\cdot q))
\\
D&=&\frac{g_V}{\sqrt{2}}\sqrt{m_H m_S}(4\mu).
\end{eqnarray}
Therefore to obtain $S$ and $D$, a CQM calculation of $\zeta$ and
$\mu$ is required. Similar considerations apply for matrix
elements containing the $T$ state.

Since $\rho$ and $a_1$ are not massless particles, the loop
integral $Z$, given in (\ref{eq:beyond}), must be examined in the
case of a general $q$ ($q^2\neq 0$) momentum. Once that $Z$ is
computed, the problem of determining $Z_\mu$, $Z_{\mu\nu}$,..is
consequential.

As we know, $Z$ is given by:
\begin{equation}
\label{eq:beyond} Z=\frac{iN_c}{16\pi^4}\int
\frac{d^4\ell}{(\ell^2-m^2)((\ell+q)^2-m^2)(v\cdot
\ell+\Delta+i\epsilon)}.
\end{equation}
Let's consider the following identity:
\begin{equation}
\frac{1}{(\ell^2-m^2)}\frac{1}{((\ell+q)^2-m^2)}=\frac{1}{q^2+2\ell\cdot
q} \left[\frac{1}{(\ell^2-m^2)}-\frac{1}{((\ell+q)^2-m^2)}\right],
\end{equation}
where $q$ is, {\it e.g.}, the $\rho$ momentum. Then we can write:
\begin{equation}
q^\mu=m_\rho v^{\prime\mu}.
\end{equation}
Therefore, considering transitions $(1)\to (2)\rho$ with (1) and
(2) generic  in and out meson states in the CQM diagram and
calling $x$ the mass value of $\rho$ or $a_1$, the following
expression for $Z$ easily follows:
\begin{equation}
Z=\frac{iN_c}{32x\pi^4}\int
\frac{d^4\ell}{(\ell^2-m^2)}\left[\frac{1}{(v\cdot\ell+\Delta_1)
(v^\prime\cdot\ell+\frac{x}{2})}-\frac{1}{(v\cdot\ell+\Delta_2)
(v^\prime\cdot\ell-\frac{x}{2})}\right],
\end{equation}
where:
\begin{equation}
v\cdot v^\prime=\frac{v\cdot q}{x}=\frac{\Delta_1-\Delta_2}{x}.
\end{equation}
Applying the notations given in the Appendix, $Z$ can be also be
written as:
\begin{equation}
Z=\frac{I_5(\Delta_1,x/2,\omega)-I_5(\Delta_2,-x/2,\omega)}{2x},
\end{equation}
where  $\omega=v\cdot v^\prime$ and this equation is well defined
for all $\omega$ values. At this point, we must care of integrals
with several Lorentz indices:
\begin{eqnarray}
Z^{\mu}&=&\Omega_1 v^\mu +\Omega_2 v^{\prime \mu}
\\
Z^{\mu\nu}&=&\Omega_3g^{\mu\nu} +\Omega_4 v^\mu v^\nu+ \Omega_5
v^{\prime\mu}v^{\prime\nu}+\Omega_6[v^\mu
v^{\prime\nu}+v^{\prime\mu}v^\nu],
\end{eqnarray}
where the $\Omega_i$ are reported in the Appendix. The CQM results
for $\lambda$ and $\beta$ are:
\begin{eqnarray}
\lambda &=&\frac{m^2_{\rho}}{ \sqrt{2} g_V f_{\rho}} Z_H
(-\Omega_1+ m Z)\\ \beta &=&\sqrt{2}\frac{m^2_{\rho}}{ g_V
f_{\rho}}   Z_H [2 m \Omega_1+
 m_\rho \Omega_2 + 2 \Omega_3 - \Omega_4 + \Omega_5-
 m^2 Z ].
\end{eqnarray}
Here the functions $Z$, $\Omega_j$ are computed with
$\Delta_1=\Delta_2=\Delta_H$, $x=m_\rho$, $\omega=m_\rho/(2m_B)$
where one takes the first $1/m_Q$ correction to $\omega=0$.
Moreover:
\begin{eqnarray}
\mu &=& \frac{m^2_\rho}{\sqrt{2} g_V f_{\rho}} \sqrt{Z_H
Z_S}\left( -\Omega_1- 2 \frac{\Omega_6}{m_\rho}+ m Z\right) \\
\zeta &=& \frac{\sqrt{2} m^2_\rho}{g_V f_{\rho}} \sqrt{Z_H Z_S}
\left(m_\rho \Omega_2 +2 \Omega_3 +\Omega_4 +\Omega_5 -m^2 Z
\right),
\end{eqnarray}
where $\Delta_1=\Delta_H$, $\Delta_2=\Delta_S$, $x=m_\rho$  and
$\omega=(\Delta_1-\Delta_2)/{m_\rho}$.

The axial couplings of $a_1$ to $H$ and $S$ are here listed:
\begin{eqnarray}
\lambda_A &= & \frac{m^2_a}{\sqrt{2} g_A f_a}   \sqrt{Z_H Z_S}
\left(-\Omega_1 +2\Omega_2\frac{m}{m_a} + m Z \right)\\ \beta_A &
= & \sqrt{2}\frac{m^2_a}{ g_A f_a}   \sqrt{Z_H Z_S} (m_a \Omega_2
+2\Omega_3   -\Omega_4 + \Omega_5
  + m^2 Z),
\end{eqnarray}
where now $\Delta_1=\Delta_H$, $\Delta_2=\Delta_S$, $x=m_a$ and
$\omega=(\Delta_1-\Delta_2)/{m_a}$. Moreover:
\begin{eqnarray}
\mu_A & = & \frac{m^2_a}{\sqrt{2} g_A f_a} Z_H \left( m \left(Z+
2\frac{\Omega_2}{m_a}\right) - \Omega_1
-2\frac{\Omega_6}{m_a}\right) \\ \zeta_A & = & \frac{\sqrt{2}
m^2_a}{g_A f_a} Z_H ( -2m\Omega_1+ m_a\Omega_2 +
2\Omega_3+\Omega_4+\Omega_5+m^2 Z),
\end{eqnarray}
where $\Delta_1=\Delta_H$, $x=m_a$, $\omega=m_a/(2m_B)$.
Numerically the results are:

$$
\begin{array}{cclcccl}
\lambda &=& 0.60~{\rm {GeV}}^{-1} &\hspace{0.5truecm}& \lambda_A
&=& 0.85 \times (0.25~ {\rm GeV}^2/f_a)~{\rm {GeV}}^{-1}\nonumber
\\ \beta &=& -0.86 & & \beta_A &=& -0.81 \times (0.25~ {\rm
GeV}^2/f_a) \nonumber \\ \mu &=& 0.16~{\rm {GeV}}^{-1} & & \mu_A
&=& 0.23 \times (0.25~ {\rm GeV}^2/f_a)~{\rm {GeV}}^{-1}
\\ \zeta &=& 0.01 & &\zeta_A &=&  0.15 \times (0.25~ {\rm
GeV}^2/f_a).
\end{array}
$$

As already mentioned, $f_a$ is the principal source of theoretical
error on these results. Another source of uncertainty is the
variation of $\Delta_H$ in the range $\Delta_H=0.3-0.5$ GeV (above
we have used only the $\Delta_H=0.4$ GeV value). The latter
produces a significative uncertainty only in the
$\zeta,\beta_A,\zeta_A$ determination since $\zeta = 0.01\pm
0.19$, $\beta_A = -0.81^{+0.45}_{-0.24}$ and $\zeta_A =
0.15^{+0.16}_{-0.14}$ while, in the other cases, only a few
percent variation is observed. A theoretical uncertainty of $\pm
15 \%$ can be added to the constants $\lambda$, $\mu$,
$\lambda_A$, $\mu_A$. This follows, for example, from the
calculation of $\lambda$ performed in \cite{art1} for the
processes $B^*\to B \gamma$ and $D^*\to D\gamma$. Other
evaluations of the of $\lambda$ can be found in  \cite{18art2} and
\cite{19art2}.
\section{\label{chap:quarto}Semileptonic decays}

        \subsection{\label{sec:clep}Semileptonic decays: leptonic constants}

CQM, as all constituent quark models, can only give
model-dependent  predictions that do not have the features of
theoretical solidity achievable by the SVZ sum rules approach
\cite{shif} or by lattice QCD. Anyway, effective constituent quark
models are sometimes easy to use tools allowing to evaluate the
hadronic matrix elements.

One of the main goals of the next future experimental physics is
the high precision measurement of CKM matrix elements $V_{cb}$ and
$V_{ub}$; $B$ physics experiments, such as Belle \cite{1art2} and
BaBar \cite{2art2}, are already running in this direction. The
semileptonic decays play a central role for the determination of
CKM elements, therefore, a wide set of theoretical predictions for
these processes, is a very actual need.

In the next sections we will consider how CQM can help this need.
The greatest part of the results we will obtain have already been
determined through other approaches.

The first target will be the determination of $s$ and $p$ wave
semileptonic $B$ decays to charmed mesons. The Isgur-Wise form
factors governing the hadronic matrix elements relative to these
processes will be calculated through CQM loop diagrams where an
external weak current  interacts with the heavy quark internal
line, inducing a boost on the velocity of the heavy quark and
possibly a change in its flavor.

Next we will consider the semileptonic decays $B\to \rho \ell \nu$
and $B\to a_1 \ell\nu$. An experimental determination by CLEO
\cite{4art2,pdg} gives a branching ratio for the former amounting
to:
\begin{equation}
{\cal B}(B^0 \to \rho^- \ell^+ \nu)=(2.5 \pm
0.4^{+0.5}_{-0.7}\pm0.5) \; \times \; 10^{-4}. \label{eq:cleo}
\end{equation}
This process offers an important test for CQM.

Experimental information are instead missing in the $B\to a_1
\ell\nu$ channel. CQM has given the first prediction for the
branching ratio of this process. A QCD sum rule prediction for the
same process has been recently carried out in a paper by Aliev and
Savci \cite{aliev}. Discrepancies with this work will be discussed
later on.

A role of prominent importance in the determination of $V_{ub}$ is
covered by the semileptonic channel $B\to \pi \ell \nu$ and a
precise measure of this process is one of the major goals of
$B$-factories. CQM predicts a new contribution to the form factors
governing $B\to \pi \ell \nu$. This new contributions are due to
CQM diagrams in which the $\pi$ is directly attached to the $B$
loop. They influence the form factors inducing corrections between
$10 \%$ and $30 \%$, according to the moment carried away by the
weak current.

\begin{figure}[t!]
\begin{center}
\epsfig{bbllx=128pt,bblly=12cm,bburx=466pt,bbury=706pt,height=8truecm,
        figure=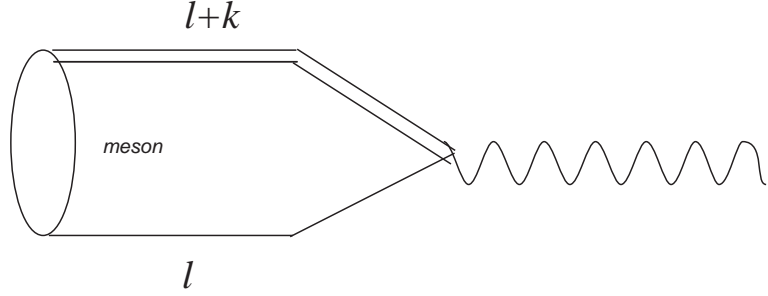}
\caption{\label{fig:lepto} \footnotesize
          CQM diagram for the lepton decay constants.
}
\end{center}
\end{figure}

Let's start with the evaluation of the leptonic constants
$\hat{F}$ and $\hat{F}^+$ through diagrams of the kind of that in
fig. \ref{fig:lepto}. These constants will be useful to calculate
the so called polar contribution to $B\to \rho$, $B\to a_1$. Their
definition is:
\begin{eqnarray}
\label{eq:matf1} \langle {\rm VAC}|{\bar q} \gamma^\mu \gamma_5 Q
|H( 0^-, v)\rangle  & =&i \sqrt{m_H} v^\mu {\hat F}\\
\label{eq:matf2} \langle {\rm VAC}|{\bar q} \gamma^\mu  Q |S( 0^+,
v)\rangle & =&i \sqrt{m_S} v^\mu {\hat F}^+.
\end{eqnarray}

Computing the loop diagram given in fig. \ref{fig:lepto}, with the
CQM rules one easily finds:
\begin{eqnarray}
{\hat F}&=&\frac{\sqrt{Z_H}}{G_3}\\ {\hat F}^+ &=&
\frac{\sqrt{Z_S}}{G_3},
\end{eqnarray}
where the renormalization constants $Z$ and the coupling constant
$G_3$ have been discussed in {\bf \ref{sec:renorm}}. The
calculation proceeds following the usual CQM strategy: from the
loop integral one extracts the $v^\mu$ contribution and a
comparison with the matrix elements (\ref{eq:matf1}) and
(\ref{eq:matf2}) gives $\hat{F}$ and $\hat{F}^+$. The numerical
values are given in Table  \ref{t:fhat}.

\begin{table} [htb]
\hfil \vbox{\offinterlineskip \halign{&#&
\strut\quad#\hfil\quad\cr \hline \hline &$\Delta_H$&& ${\hat F}$&&
${\hat F}^+$&\cr \hline &$0.3$&& $0.33$&& $0.22$&\cr &$0.4$&&
$0.34$&& $0.24$&\cr &$0.5$&& $0.37$&& $0.27$&\cr \hline \hline }}
\caption{${\hat F}$ and ${\hat F}^+$ for three $\Delta_H $ values.
$\Delta_H$ is expressed in GeV; leptonic constants are in
 GeV$^{3/2}$.} \label{t:fhat}
\end{table}

Neglecting logarithmic corrections, ${\hat F}$ and ${\hat F}^+$
are connected, in the $m_Q\to \infty$ limit, to the leptonic decay
constants $f_B$ and $f^+$:
\begin{eqnarray}
\label{eq:effeb} \langle 0|{\bar q} \gamma^\mu \gamma_5 b
|B(p)\rangle &=& i  p^\mu f_B \\ \langle 0|{\bar q} \gamma^\mu  b
|B_0(p)\rangle &=& i  p^\mu f^+,
\end{eqnarray}
through the relations $f_B={\hat F}/\sqrt{m_B}$ and $f^+={\hat
F}^+/\sqrt{m_{B_0}}$. For example, if $\Delta_H=400$ MeV, one
obtains:
\begin{eqnarray}
f_B &\simeq& 150 \;{\rm {MeV}}\\ f^+ &\simeq& 100 \;{\rm {MeV}}.
\end{eqnarray}

The QCD sum rules analysis  \cite{16art1} gives ${\hat F}=0.30 \pm
0.05$ GeV$^{3/2}$, neglecting radiative corrections and even
higher values, $0.4-0.5$ GeV$^{3/2}$, including $\alpha_s$
corrections. Another QCD sum rules analysis \cite{17art1} suggests
${\hat F}^+=0.46\pm 0.06$ GeV$^{3/2}$, which is somehow higher
with respect to the CQM values. Lattice QCD \cite{3art1} gives
$f_B=170\pm 35$ MeV.

        \subsection{\label{sec:btoc}$b\to c$ transitions}

We are interested in studying some weak heavy meson decays, {\it
i.e.}, decays with flavor changing of the heavy quark due to the
presence of the charged weak current $W$. Let's focus on the $b\to
c \ell \bar{\nu}$ transitions. Such processes are described by the
following four fermion operator:
\begin{equation}
{\cal O}=\frac{G_F
V_{cb}}{\sqrt{2}}\bar{c}\gamma^\mu(1-\gamma_5)b\bar{\nu}\gamma_\mu
(1-\gamma_5)\ell.
\end{equation}
At the quark level one can calculate the matrix element:
\begin{equation}
{\rm Amp}_q=\langle c\ell\bar{\nu}|{\cal O}|b\rangle=\frac{G_F
V_{cb}}{\sqrt{2}} \bar{u}_c(p_c)\gamma^\mu(1-\gamma_5)u_b(p_b)
\bar{u}_\nu(p_\nu)\gamma_\mu(1-\gamma_5)u_\ell(p_\ell),
\end{equation}
that is relevant only at very small distances and fails when, as
is the case in the hadron world, quarks cannot be treated as
asymptotic states. The hadronic amplitude ${\rm Amp}_h$ is:
\begin{equation}
{\rm Amp}_h=\langle D^* \ell\bar{\nu}|{\cal O}|B\rangle= \frac{G_F
V_{cb}}{\sqrt{2}} \langle
D^*|\bar{c}\gamma^\mu(1-\gamma_5)b|B\rangle \langle
\ell\bar{\nu}|\bar{\nu}\gamma_\mu(1-\gamma_5)\ell|{\rm
VAC}\rangle.
\end{equation}

The hadronic matrix term  in ${\rm Amp}_h$  accounts for non
perturbative QCD effects. It is possible to parameterize these
effects through form factors depending on $q^2$, where $q$ is the
momentum carried by the weak current. Strong interaction
symmetries, and in particular HQET symmetries (in the limit
$m_Q\to \infty$), have influence on the properties of these form
factors. Many times this influence acts in the sense of
simplification, reducing the number of form factors needed to
parameterize the matrix element.

\subsubsection{\label{sec:xiis} The Isgur-Wise function
$\xi(\omega)$}

Let's consider an elastic scattering process $B(v)\to B(v^\prime)$
mediated by the weak current $V_\mu$. A model of what happens in
the hadron is the following: due to the action of the weak
current, the heavy quark is suddenly substituted by another heavy
quark having the same flavor but different velocity $v^\prime$.
Light degrees of freedom must rearrange themselves in order to
give rise to a $B$ meson moving with velocity $v^\prime$. This
rearranging process is mediated by an exchange of soft gluons
(having momenta of order $\simeq \Lambda_{\rm QCD}$) with the
heavy quark acting as the source of chromoelectric field. The
larger is $v^\prime-v$, the smaller is the probability to have an
elastic transition: we have a suppression of the elastic form
factor. In the $m_b\to \infty$ limit, the form factor can only
depend on $v$ and $v^\prime$ in the scalar combination $v\cdot
v^\prime=\omega$. We can introduce a dimensionless probability
function, $\xi(\omega)$, that works as the form factor of the
transition. This function is known, in HQET, as the Isgur-Wise
function:
\begin{equation}
\label{eq:elast} \langle
\bar{B}(v^\prime)|\bar{h}_{v^\prime}\gamma^\mu h_{v}
|\bar{B}(v)\rangle=m_B \xi(\omega) (v+v^\prime)^\mu,
\end{equation}
where $m_B$ is due to a particular choice of the normalization of
the external meson states. To convince oneself that there are no
terms proportional to $(v-v^\prime)$, it is sufficient to multiply
the matrix element in (\ref{eq:elast}) by $(v-v^\prime)_\mu$  and
remind from section {\bf \ref{sec:hqet}} that $\gamma\cdot v
h_v=h_v$ and $h_v \gamma\cdot v=h_v$.

The interpretation of $\xi(\omega)$, as the probability for the
elastic transition $B(v)\to B(v^\prime)$, suggests to assign a
value of $\xi=1$ when $\omega=1$, {\it i.e.}, when there is no
change of velocity; smaller probability values (form factor
suppression) are assigned when $v\neq v^\prime$.

We can examine this point in greater detail. The $\xi(\omega)$
function describes the response of light degrees of freedom to the
change of velocity of the static source of colour. This change can
also be accompanied by a flavor change of the heavy quark. If
$v=v^\prime$, the current causing the transition is therefore a
symmetry current since light degrees of freedom do not resolve the
flavor of the heavy quark (this is the flavor symmetry of HQET).
This symmetry current comes together with conserved charges,
generators of  the flavor symmetry, which are connected to the
current by the well known relation:
\begin{equation}
Q^{f^\prime f}=\int d^3x h^{+\prime}_v(x)h_v(x),
\end{equation}
where $f^\prime=f$ means that the current hasn't produced heavy
quark  flavor change. One can verify explicitly  that the weak
current responsible for the $\omega=1$ transition is a conserved
current:
\begin{equation}
\partial_\mu \bar{h}^\prime_v v^\mu h_v=0,
\end{equation}
where the property $\gamma\cdot vh_v=h_v$  and the equation of
motion derived by (\ref{eq:hqet}) have been used.

Meson states must be eigenstates of charge operators in such a way
that:
\begin{eqnarray}
\label{eq:uso} Q^{f^\prime f} |P(v)\rangle&=&|P^\prime(v)\rangle
\\
Q^{f f}|P(v)\rangle&=&|P(v)\rangle,
\end{eqnarray}
where $P$ denotes a pseudoscalar meson (let's simplify the
discussion  taking the case of $J=0$ mesons). If we write:
\begin{equation}
\langle P^\prime(v^\prime)|\bar{h}^\prime_{v^\prime}\gamma^\mu
h_v|P(v)\rangle =m_P\xi(\omega)(v+v^\prime)^\mu,
\end{equation}
and if we consider  $v=v^\prime$, take the $\mu=0$ component,
integrate with respect to ${\bf x}$ and use the (\ref{eq:uso}), we
get:
\begin{equation}
\xi(1)=1,
\end{equation}
assuming the following meson state normalization:
\begin{equation}
\langle M(p^\prime)|M(p)\rangle= 2m_M v^0(2\pi)^3 \delta^3({\bf
p}-{\bf p}^\prime).
\end{equation}

The normalization condition $\xi(\omega=1)=1$ is particularly
relevant not only because it is connected to the flavor symmetry
of HQET, but also because it is not affected by $1/m_Q$
corrections. The relevant corrections are $O(1/m_Q^2)$. This is
consequence of the Luke theorem \cite{luke}, generalization of the
Ademollo-Gatto theorem \cite{agatto}, according to which, at
$\omega=1$, there are no corrections to the hadronic matrix
elements responsible for the semileptonic decays $B\to
D\ell\bar{\nu}$ and $B\to D^*\ell\bar{\nu}$. The leading
corrections to the normalization of these matrix elements are, at
the zero recoil point, of order $1/m_c^2$. Since $\Lambda_{\rm
QCD}^2/m_c^2 \simeq 10\%$, zeroth order predictions in the $1/m_c$
expansion are, for $\omega=1$, very accurate. Far from the zero
recoil point $\omega=1$, corrections of order $1/m_c$ are
suppressed by $(\omega-1)$ factors.

The normalization condition $\xi(1)=1$ is therefore a good table
test to understand if CQM allows for a correct calculation of the
Isgur-Wise function.

Let's go back to the weak transitions  $b\to c$. The decay $b\to
c$ is mediated by the left-handed current
$J^\mu=\bar{c}\gamma^\mu(1-\gamma_5)b$. This operator not only
carries momentum, but  also rotates the orientation of the the
spin $s_Q$ of the heavy quark during its decay. For an assigned
value of total angular momentum of the light degrees of freedom
$s_\ell$, the relative orientation of  $s_Q$ determines if the
hadron in its final state is a $D$ or $D^*$. The heavy quark spin
symmetry induces relations connecting the matrix elements for the
decays $B\to D\ell \nu$, $B\to D^* \ell \nu$ in such a way that
they could be expressed in terms of a unique universal form factor
$\xi(\omega)$.

Let's consider for example the following matrix element:
\begin{equation}
\label{eq:hdr} \langle
D(p^\prime)|V^\mu|B(p)\rangle=f_+(q^2)(p+p^\prime)^\mu+
f_-(q^2)(p-p^\prime)^\mu,
\end{equation}
where  $V^\mu$ is the vectorial component of $J^\mu$. In HQET one
has :
\begin{equation}
\label{eq:hdrhqet} \langle D(p^\prime)|V^\mu|B(p)\rangle=\sqrt{m_B
m_D} \xi(\omega)(v+v^\prime)^\mu.
\end{equation}
Therefore we can see a relation between  (\ref{eq:hdr}) and
(\ref{eq:hdrhqet}):
\begin{equation}
f_{\pm}=C_{cb}\frac{m_D\pm m_B}{2\sqrt{m_B m_D}}\xi(\omega),
\end{equation}
where $C_{cb}$ is the correction that emerges from the matching
with QCD, see {\bf \ref{sec:1/mQQCD}}. In the following we will
consider $C_{cb}=1$ for simplicity. In an analogous way one can
consider the matrix elements describing the processes $B\to D^*$,
where $V^\mu$ and  $A^\mu$ are both relevant. One finds:
\begin{eqnarray}
\langle D(v^\prime)|{\bar c} \gamma_\mu (1-\gamma_5) b|
B(v)\rangle  & =& \sqrt{m_B m_D}  \; \xi(\omega) (v_\mu +
v^{\prime }_{\mu}) \\ \langle D^*(v^\prime, \epsilon)|{\bar c}
\gamma_\mu (1- \gamma_5) b| B(v) \rangle  & =& \sqrt{m_B m_{D^*}}
\; \xi(\omega)[i \epsilon_{\mu \nu \alpha \beta}
\epsilon^{*\nu}v^{\prime \alpha}v^\beta  \nonumber \\ &-&
(1+\omega)\epsilon^*_\mu+ (\epsilon^*\cdot v)v^\prime_\mu].
\end{eqnarray}

\begin{figure}[t!]
\begin{center}
\epsfig{bbllx=128pt,bblly=12cm,bburx=466pt,bbury=706pt,height=8truecm,
        figure=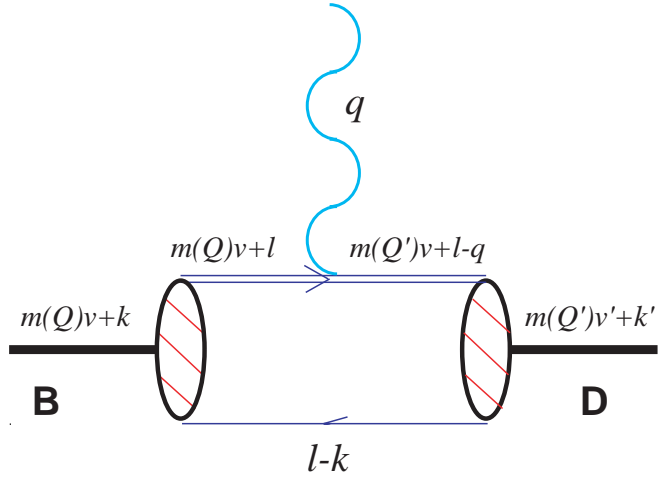}
\caption{\label{fig:btoc} \footnotesize
          Weak current insertion on the heavy quark propagator
          line.}
\end{center}
\end{figure}

In order to determine $\xi(\omega)$ with CQM, one needs to
calculate the loop diagram in fig. \ref{fig:btoc}.
We will address the simple case in which $Q=Q^\prime=b$. Let's
observe that the momentum $q$ is given by:
\begin{equation}
\label{eq:stupida} q=(m_b v+k)-(m_b v^\prime+k^\prime),
\end{equation}
and the momentum carried by the heavy quark having $v^\prime$
velocity is then:
\begin{equation}
m_b v+\ell-q=m_b v^\prime +\ell+k^\prime-k,
\end{equation}
where we have used the relation (\ref{eq:stupida}). Using the CQM
approach developed in so far, one can write the loop integral
describing the diagram of fig. \ref{fig:btoc} in the following
way:
\begin{equation}
\label{eq:aaa} m_H Z_H (-1)i^3i^2 \frac{N_c}{16\pi^4}\int d^4\ell
\frac{{\rm Tr}\left[(\gamma\cdot
(\ell-k)+m)(-\gamma_5)\frac{1+\gamma\cdot v^\prime}{2} \gamma^\mu
\frac{1+\gamma\cdot
v}{2}\gamma_5\right]}{((\ell-k)^2-m^2)(v^\prime\cdot(\ell+k^\prime-k))(v\cdot
\ell)},
\end{equation}
where for simplicity we are considering as in and out states the
pseudoscalar mesons of the $H$ multiplet. We can perform the shift
$\ell-k\to \ell$ and, observing that $v\cdot k=v^\prime \cdot
k^\prime=\Delta_H$, we can rewrite (\ref{eq:aaa}) in the following
way:
\begin{equation}
m_H Z_H (-1)i^3i^2 \frac{N_c}{16\pi^4}\int d^4\ell \frac{{\rm
Tr}\left[(\gamma\cdot\ell+m)(-\gamma_5)\frac{1+\gamma\cdot
v^\prime}{2} \gamma^\mu \frac{1+\gamma\cdot
v}{2}\gamma_5\right]}{(\ell^2-m^2)(v^\prime\cdot\ell+\Delta_H)(v\cdot
\ell+\Delta_H)}.
\end{equation}
Once computed the trace, $\xi(\omega)$  is extracted by comparison
with the transition amplitude  (\ref{eq:elast}).

The CQM expression for $\xi(\omega)$ is:
\begin{equation}
\xi(\omega)=Z_H \left[ \frac{2}{1+\omega} I_3(\Delta_H)+\left(
m+\frac{2 \Delta_H}{1+\omega} \right) I_5(\Delta_H,
\Delta_H,\omega) \right],\label{eq:laxi}
\end{equation}
where the $I_i$ integrals are listed in the Appendix. Let's
observe that $I_5(\Delta,\Delta,1)=\frac{\partial}{\partial
\Delta}I_3(\Delta)$. Recalling the equation (\ref{eq:llaaxxii}),
we then have \cite{art1}:
\begin{equation}
\xi(1)=1,
\end{equation}
as expected.

A very accurate determination  of $V_{cb}$ can be obtained
measuring the recoil spectrum of $D^*$ produced in semileptonic
decays of  $B$. In particular, measuring the differential decay
rate  for $\bar{B}\to D^*\ell\nu$, one obtains an indirect measure
of the product $V_{cb}\xi(\omega)$. If we expand $\xi(\omega)$
around $\omega=1$  and if we suppose  $\xi(1)=1$ (taking into
account $1/m_Q^2$ corrections and bound state effects gives a
value of $\xi(1)\simeq 0.91$ \cite{rass8}), then near $\omega=1$:
\begin{equation}
(1-\rho^2_{IW}(\omega-1))V_{cb},
\end{equation}
where $\rho^2_{IW}$, defined by:
\begin{equation}
\rho^2_{IW}=-\frac{d\xi}{d\omega}(1),
\end{equation}
works as a fit parameter allowing to extrapolate back to the
$V_{cb}$ value that one could measure at zero recoil.

The CQM numerical values for $\rho^2_{IW}$ are given in Table
\ref{t:ffeslopes}. What emerges is that CQM results are in a quite
good agreement with those obtained with QCD sum rules:
$\rho^2_{IW}= 0.54-1.0$ \cite{18art1}. After an overall
examination of the results obtained for the Isgur-Wise function in
a series of quark models \cite{19art1,20art1,21art1,22art1},
authors in \cite{23art1} find $\rho^2_{IW}=0.97-1.28$. Lattice QCD
indicates lower results, around $\rho^2_{IW}=0.64$ \cite{25art1}.

\begin{figure}[t!]
\begin{center}
\epsfig{bbllx=128pt,bblly=303pt,bburx=466pt,bbury=706pt,height=12truecm,
        figure=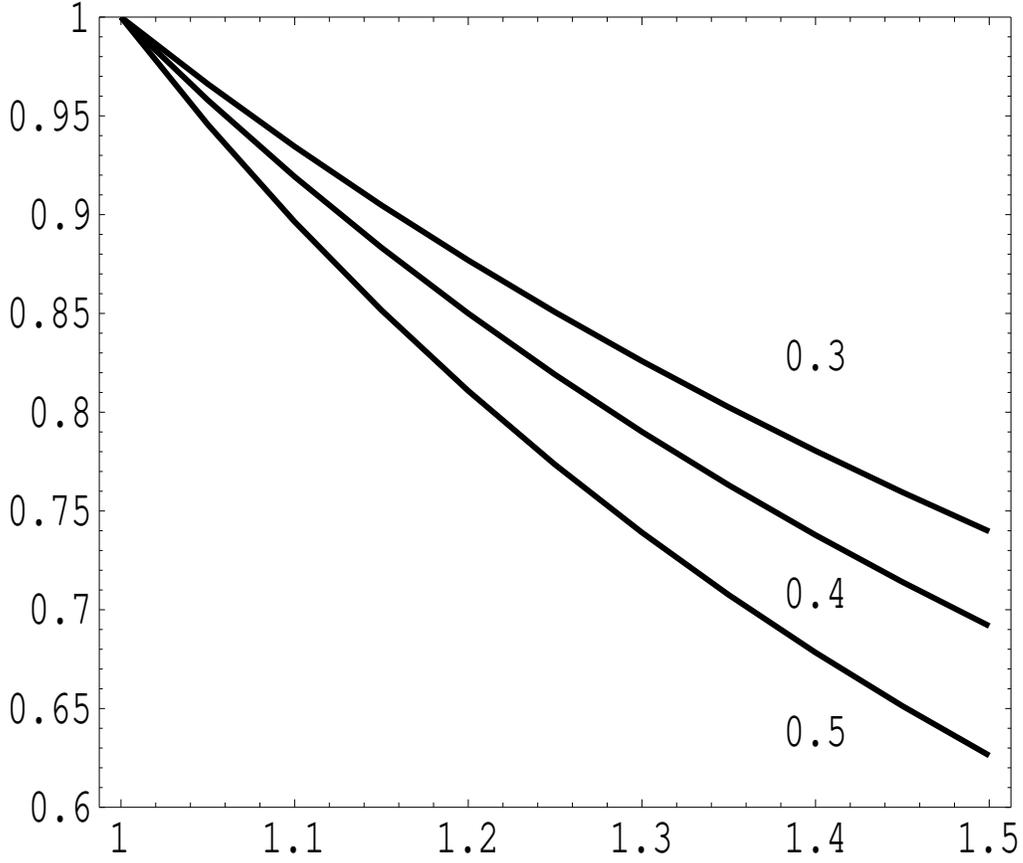}
\caption{\label{fig:plotxi} \footnotesize
          Isgur-Wise form factor vs. $\omega$ for different $\Delta_H$ values.}
\end{center}
\end{figure}

In fig. \ref{fig:plotxi} it is shown a plot of the $\xi(\omega)$
given in (\ref{eq:laxi}).

\subsubsection{\label{sec:tauis}$\tau_{1/2}(\omega)$ and
$\tau_{3/2}(\omega)$  form factors}

In this section we will consider the semileptonic decays of an $H$
meson to $S$ and $T$ mesons. These decays are written as:
\begin{equation}
B\to D^{**}\ell \nu,
\end{equation}
where $D^{**}$ can be either an $S$ state, {\it i.e.}, a charmed
meson $0^+$ or $1^+$ with $s_\ell=1/2$, or a $T$ state, {\it
i.e.}, a charmed meson $2^+$ or $1^+$ with $s_\ell=3/2$. There are
two form factors describing respectively these decays: they are
known as $\tau_{1/2}$ and  $\tau_{3/2}$ \cite{26art1}. The
relevant matrix elements are:
\begin{eqnarray}
&&<D_2^*(v^{\prime},\epsilon)| {\bar c} \gamma_\mu (1- \gamma_5)
b| B(v)> =\sqrt{ 3 m_B m_{D_2^*}} ~\tau_{3/2}(\omega)\times
\nonumber\\ &&\left[ i  \epsilon_{\mu  \alpha \beta \gamma }
\epsilon^{*\alpha \eta} v_\eta v^{\prime \beta}v^\gamma -
 [( \omega+1) \epsilon^*_{\mu \alpha} v^\alpha -
\epsilon^*_{\alpha \beta} v^\alpha v^\beta v^\prime_\mu]\right]
\label{tau32}
\end{eqnarray}
\begin{eqnarray}
&&<D_1^*(v^{\prime},\epsilon)| {\bar c} \gamma_\mu (1- \gamma_5)
b| B(v)> =\sqrt{\frac{m_B m_{D_1^*}}{2}} ~\tau_{3/2}(\omega)\times
\nonumber\\ &&\left\{ (\omega^2-1) \epsilon_{\mu}^* +  (
\epsilon^*\cdot v) [3~v_\mu -(\omega-2) v^\prime_\mu ]-  i
(\omega+1) \epsilon_{\mu \alpha \beta \gamma} \epsilon^{* \alpha}
v^{\prime \beta} v^\gamma \right\} \label{tau32bis}
\end{eqnarray}

\begin{eqnarray}
\langle D_0(v^\prime)|{\bar c} \gamma_\mu (1-\gamma_5) b|
B(v)\rangle & =& \sqrt{m_B m_{D_0}}~2~ \tau_{1/2}(\omega) (
v^{\prime }_{\mu}- v_\mu) \\ \langle D_1^{*\prime}(v^\prime,
\epsilon)|{\bar c} \gamma_\mu (1- \gamma_5) b|
 B(v)\rangle
& =& \sqrt{m_B m_{D_1^{*\prime}}}~\tau_{1/2}(\omega)\{
2~i~\epsilon_{\mu  \alpha \beta \gamma}
\epsilon^{*\alpha}v^{\prime \beta}v^\gamma \nonumber\\ &+&
2[(1-\omega)\epsilon^*_\mu+ (\epsilon^*\cdot v)v^\prime_\mu]\},
\end{eqnarray}
where $D_2^*$ and $D_1^*$ are, respectively, the $2^+$ and $1^+$
states of the $T$ multiplet, while $D_0$ and $D_1^{*\prime}$ are
the $0^+$ and $1^+$ $S$ states.

In the preceding equations we ignore logarithmic corrections. The
$\tau$ form factors can be calculated by means of CQM loop
diagrams, see again fig. \ref{fig:btoc}. The strategy for writing
down the loop integral is the same as before. We obtain the
following results:
\begin{equation}
\label{eq:tau12} \tau_{1/2}(\omega)=\frac{\sqrt{Z_H
Z_S}}{2(1-\omega)} \left[ I_3(\Delta_S)- I_3(\Delta_H)
+\left(\Delta_H - \Delta_S + m(1-\omega) \right) I_5(\Delta_H,
\Delta_S,\omega) \right]
\end{equation}
and:
\begin{eqnarray}
\tau_{3/2}(\omega)&=&-{{\sqrt{Z_H\,Z_T}}\over{{\sqrt{3}}}} \,
     \Big[m \Big( {{I_3(\Delta_H) - I_3(\Delta_T) -
              \left( \Delta_H - \Delta_T \right) \,
               I_5(\Delta_H,\Delta_T,\omega )}\over
            {2\,\left( 1 - \omega  \right) }} \nonumber \\
&-&{{I_3(\Delta_H) + I_3(\Delta_T)  +
              \left( \Delta_H + \Delta_T \right)  \,
               I_5(\Delta_H,\Delta_T,\omega )}\over
            {2\,\left( 1 + \omega  \right) }} \Big) \nonumber\\
&-& {{1}\over {2\,\left( -1 - \omega  + {{\omega }^2} +  {{\omega
}^3} \right) }} \Big( -3\,S(\Delta_H,\Delta_T, \omega ) -
           \left( 1 - 2\,\omega   \right) \,S(\Delta_T,\Delta_H,
\omega )\nonumber  \\ &+& (1-{{\omega }^2})
\,T(\Delta_H,\Delta_T,\omega  ) -
   2\, (1 - 2\,\omega ) \,U(\Delta_H,\Delta_T,\omega ) \Big)  \Big],
\end{eqnarray}
where $S,T,U$ are linear combinations of $I_i$ integrals, see the
Appendix.

At a first glance it may seem that the $\tau$ form factors are
diverging for $\omega\to 1$, see for example $\tau_{1/2}(\omega)$.
Let's take  $\omega\simeq 1$ and neglect  the term $m(1-\omega)$.
Since in the heavy quark propagator in $I_3(\Delta_S)$ appears the
velocity $v^\prime$, one can write the term in square brackets of
(\ref{eq:tau12}) in the following way:
\begin{equation}
-\frac{iN_c}{16\pi^4}\int d^4\ell
\frac{\ell_\mu(v^\mu-v^{\prime\mu})}{ (\ell^2-m^2)(v\cdot\ell
+\Delta_H)(v^\prime\cdot\ell +\Delta_S)}.
\end{equation}
Define  $\epsilon^\mu=v^\mu-v^{\prime\mu}$ and, consequently,
$\epsilon\cdot v=1-\omega$. The $\tau_{1/2}(\omega)$ limit for
$\omega\to 1$ is then:
\begin{equation}
\lim_{\omega\to 1}\tau_{1/2}(\omega)=\lim_{\epsilon\to 0}{\rm
Const.} \frac{1}{\epsilon\cdot v} A \epsilon\cdot v,
\end{equation}
where $A$, in the  $\epsilon\to 0$ limit, is given by:
\begin{equation}
-\frac{iN_c}{16\pi^4}\int d^4\ell \frac{\ell_\mu}{
(\ell^2-m^2)(v\cdot\ell +\Delta_H)(v\cdot\ell +\Delta_S)}=Av_\mu,
\end{equation}
and amounts to (just contracting by $v^\mu$ and using the
Appendix):
\begin{equation}
A=\frac{1}{2}I_3(\Delta_H)+\frac{1}{2}I_3(\Delta_S)+
\frac{1}{2}(\Delta_H+\Delta_S)I_5(\Delta_H,\Delta_S,1).
\end{equation}

Since the phase space allowed for $B$ decays to positive parity
states is small, ($\omega_{max}=1.33$ for $D_1^*,~D_2^*$ and
$\omega_{max}\simeq 1.215$ for $D_1^{*\prime},~D_0$), we can
consider the following approximation:
\begin{equation}
\tau_j(\omega)\simeq\tau_j(1)\times [ 1-\rho^2_j(\omega -1)].
\end{equation}
Numerically we find the results given in Table \ref{t:ffeslopes}
where we have a general scheme of all form factors calculated by
CQM in so far.
\begin{table} [htb]
\hfil \vbox{\offinterlineskip \halign{&#&
\strut\quad#\hfil\quad\cr \hline \hline &$\Delta_H$&& $\xi(1)$ &&
$\rho^2_{IW}$&& $\tau_{1/2}(1)$&& $\rho^2_{1/2}$&&
$\tau_{3/2}(1)$&& $\rho^2_{3/2}$&\cr \hline &$0.3$&&  $1$&&
$0.72$&& $0.08$&& $0.8$&& $0.48$&& $1.4$&\cr &$0.4$&&  $1$&&
$0.87$&& $0.09$&& $1.1$&& $0.56$&& $2.3$&\cr &$0.5$&&  $1$&&
$1.14$&& $0.09$&& $2.7$&& $0.67$&& $3.0$&\cr \hline \hline }}
\caption{Form factors and slopes. $\Delta_H$ in GeV.}
\label{t:ffeslopes}
\end{table}

Let's compare CQM results with those obtained by other methods,
see Table  \ref{t:letau}. As for $\tau_{3/2}$, we have a good
agreement with quark models. As for $\tau_{1/2}$, we find a good
agreement only with  \cite{23art1}.

\begin{table} [htb]
\hfil \vbox{\offinterlineskip \halign{&#&
\strut\quad#\hfil\quad\cr \hline \hline &$\tau_{1/2}(1)$&&
$\rho^2_{1/2}$&& $\tau_{3/2}(1)$&& $\rho^2_{3/2}$&& Ref. &\cr
\hline &$0.09$&& $1.1$&& $0.56$&& $2.3$&& CQM &\cr &$0.41$&&
$1.0$&& $0.41(input)$&& $1.5$&& \cite{27art1} &\cr &$0.25$&&
$0.4$&& $0.28$&& $0.9$&& \cite{17art1}&\cr &$0.31$&& $2.8$&&
$0.31$&& $2.8$&& \cite{22art1}&\cr &$0.41$&& $1.4$&& $0.66$&&
$1.9$&& \cite{28art1}&\cr &$0.059$&& $0.73$&& $0.515$&& $1.45$&&
\cite{23art1,21art1}&\cr &$0.225$&& $0.83$&& $0.54$&& $1.50$&&
\cite{23art1,19art1}&\cr \hline \hline }} \caption{Here $\Delta_H
= 0.4$ GeV. In the second paper in \cite{17art1} a slightly higher
determination for $\tau_{1/2}(1)$ is obtained using the SVZ method
up to the next to leading order.} \label{t:letau}
\end{table}

In Table  \ref{t:ibrs} we show the branching ratios for the $s$
and $d$ wave semileptonic decays of $B$ to charmed mesons. Here
$V_{cb}=0.038$ \cite{29art1}, $\tau_B=1.62$ psec. The results in
Table \ref{t:ibrs} seem to contradict recent experimental claims
that the broad resonances dominate and, therefore, they impose
some further understanding \cite{janne}. A possible direction to
look at, could be that of examining $\frac{1}{m_Q}$ corrections,
as pointed out in \cite{faustebe}. In the latter reference
\cite{faustebe}, it is shown how in the relativistic quark model
the $\frac{1}{m_Q}$ corrections act in the sense of an
approximately two-fold enhancement of the decay rates $B\to
D_0\ell\nu$ and $B\to D_1^{*\prime}\ell\nu$ which, at the leading
order of the heavy quark expansion, tend to be approximately one
order of magnitude smaller than the decay rates $B\to
D_1^*\ell\nu$ and $B\to D_2^*\ell\nu$ (due to the Lorentz
transformation properties of meson wave functions).

\begin{table} [htb]
\hfil \vbox{\offinterlineskip \halign{&#&
\strut\quad#\hfil\quad\cr \hline \hline &Decay&& $\Delta_H=0.3$&&
$\Delta_H=0.4$&& $\Delta_H=0.5$&& Exp. & \cr \hline &${B}\to
D\ell\nu $&& $3.0$&&$2.7$&&$2.2$&&$1.9 \pm 0.5$ \cite{pdg}& \cr
&${B}\to D^{*}\ell\nu$&& $7.6$&& $6.9$&& $5.9$&& $4.68\pm 0.25$
\cite{pdg} &\cr &${B}\to D_0\ell\nu$&& $0.03$&& $0.005$&&
$0.003$&& -- &\cr &${B}\to D^{*\prime}_{1}\ell\nu$&& $0.03$&&
$0.008$&& $0.0045$&&-- &\cr &${B}\to D_{1}^*\ell\nu$&& $0.27$&&
$0.18$&& $0.13$&& $0.74 \pm 0.16$ \cite{30art1}&\cr &${B}\to
D^{*}_{2}\ell\nu$&& $0.43$&& $0.34$&& $0.30$&& $<0.85$&\cr \hline
\hline }} \caption{Branching ratio (\%) for the $B$ semileptonic
decays in charmed states via CQM. } \label{t:ibrs}
\end{table}

\subsubsection{\label{sec:bjorken} The Bjorken sum rule}

A second important test table for CQM is the Bjorken sum rule.
Let's introduce it briefly considering the weak decay of an heavy
meson in which the quark $b(v)$ is substituted by a quark
$c(v^\prime)$. After the action of the weak current, light degrees
of freedom must rearrange themselves to build the new charmed
hadron state. There are many possible reconfigurations that can be
assumed and we can associate an amplitude to each of them; the
following sum rule holds:
\begin{equation}
{\rm Amp}(b(v)\to c(v^\prime))=\sum_{X_c} {\rm Amp}(B(v)\to
X_c(v^\prime)). \label{eq:aparti}
\end{equation}
It means that we must sum over all the possible charmed final
states. If there is a form factor suppression in the elastic
channel, as we move far from $\omega=1$, there is a compensating
growth of the amplitudes involving excited states (as the positive
parity ones). Indeed one can show that:
\begin{equation}
\rho^2_{IW}=\frac{1}{4}+\sum_k \left[|\tau_{1/2}^{(k)}(1)|^2~+~
2|\tau_{3/2}^{(k)}(1)|^2\right].
\end{equation}
This means that the compensation comes only from
$s_\ell=\frac{1}{2}^+, \frac{3}{2}^+$ states and $\rho^2_{IW}\geq
\frac{1}{4}$. Numerically we find that $S$ and $T$ states,
$(k=0)$, saturate almost completely the Bjorken sum rule for all
$\Delta_H$ values.

        \subsection{\label{sec:btora}$B\to \rho\ell\nu$, $B\to a_1\ell\nu$}

The form factors parameterizing the semileptonic decays $B\to \rho
\ell \nu$ and $B\to a_1\ell\nu$ are given by:
\begin{eqnarray}
\label{eq:monsterm}
<\rho^+(\epsilon(\lambda),p^\prime)|\overline{u}\gamma_\mu
(1-\gamma_5)b|\bar{B^0}(p)> & = & \frac{2
V(q^2)}{m_B+m_{\rho}}\epsilon_{\mu\nu\alpha\beta}
\epsilon^{*\nu}p^\alpha p^{\prime\beta}\nonumber\\ &-& i
\epsilon^{*}_{\mu}(m_{B} + m_{\rho})  A_{1}(q^{2}) \nonumber\\ &+&
i (\epsilon^{*}\cdot q) \frac{(p + p^\prime)_{\mu}}{m_B +
m_{\rho}}  A_{2}(q^{2})
\\
&+& i  (\epsilon^{*}\cdot  q) \frac{2  m_{\rho}}{q^{2}} q_{\mu}
[A_{3}(q^{2})  - A_{0}(q^{2})], \nonumber
\end{eqnarray}
where:
\begin{equation}
A_{3}(q^{2})  = \frac{m_{B} + m_{\rho}}{2  m_{\rho}} A_{1}(q^{2})
- \frac{m_{B} - m_{\rho}}{2  m_{\rho}} A_{2}(q^{2}),
\end{equation}
while for $a_1$ one has:
\begin{eqnarray}
<a^{+}_1(\epsilon(\lambda),p^\prime)|\overline{q^\prime}
\gamma_\mu(1-\gamma_5)b|\bar{B}^0(p)> & = & \frac{2
A(q^2)}{m_B+m_a}\epsilon_{\mu\nu\alpha\beta}
\epsilon^{*\nu}p^\alpha p^{\prime\beta}\nonumber\\ &-& i
\epsilon^{*}_{\mu}(m_{B} + m_{a})  V_{1}(q^{2}) \nonumber\\ &+& i
(\epsilon^{*}\cdot q) \frac{(p + p^\prime)_{\mu}}{m_B +  m_a}
V_{2}(q^{2})
\\
&+& i  (\epsilon^{*}\cdot  q) \frac{2  m_a}{q^{2}} q_{\mu}
[V_{3}(q^{2})  - V_{0}(q^{2})], \nonumber
\end{eqnarray}
where $m_a $ is the $a_1$ mass and:
\begin{equation}
V_{3}(q^{2})  = \frac{m_{B} + m_{a}}{2  m_{a}} V_{1}(q^{2}) -
\frac{m_{B} - m_{a}}{2  m_{a}} V_{2}(q^{2}).
\end{equation}
With this parameterization of the matrix elements  \cite{17art2},
the following relations hold for $q^2=0$:
\begin{eqnarray} A_{3}(0) &=& A_{0}(0)\label{eq:A3}\\ V_{3}(0) &=&
V_{0}(0), \label{V3}
\end{eqnarray}
where $q=(p-p^\prime)$.

\subsubsection{\label{sec:diretti} Direct contributions}

\begin{figure}[t!]
\begin{center}
\epsfig{bbllx=128pt,bblly=12cm,bburx=466pt,bbury=706pt,height=8truecm,
        figure=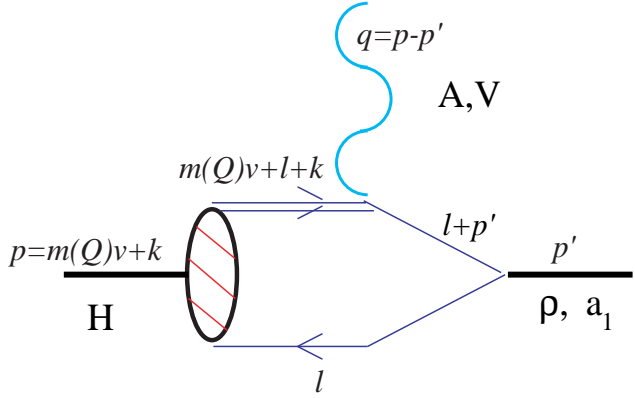}
\caption{\label{fig:diretti} \footnotesize
          The CQM direct diagram.}
\end{center}
\end{figure}

For direct contributions we mean the contributions to form factors
derived by CQM loop diagrams where the decaying meson couples
directly to $\rho$ or $a_1$, see fig. \ref{fig:diretti}. The
Feynman rules needed for the computation of these diagrams are the
usual ones, see also (\ref{eq:feyrho}) and (\ref{eq:feya1}).

The technique needed for computing the loop integrals has been
developed in {\bf \ref{sec:rhoa}}. The loop integral derived from
fig. \ref{fig:diretti} is:
\begin{equation}
(-1)(-i)\sqrt{Z_H m_H} \frac{N_c}{16\pi^4}\int^{\rm reg} d^4\ell
\frac{{\rm Tr}\left[ (\gamma\cdot\ell +m)(-\frac{m_\rho^2}{f_\rho}
\gamma\cdot\epsilon)(\gamma\cdot(\ell+q)+m)({\rm V,A})
\frac{1+\gamma\cdot v}{2}(-\gamma_5)
\right]}{(\ell^2-m^2)((\ell+q)^2-m^2)(v\cdot\ell+\Delta_H)}.
\end{equation}
In this expression:
\begin{itemize}
\item $(-1)$ comes from the fermion loop\\
\item $(-i)$ comes from the vertex $Q$-Meson-$\chi$.
The vertex with $\rho$ ($a_1$) doesn't introduce new $i$'s since
of (\ref{eq:feyrho})\\
\item $(-\frac{m_\rho^2}{f_\rho}
\gamma\cdot\epsilon)$, where $\epsilon$ is the  $\rho$ ($a_1$)
polarization, is the vertex described in (\ref{eq:feyrho}).
\end{itemize}
The CQM expressions for the form factors derived by the direct
diagrams calculations, are the following (here we can find the
$\Omega_i$ expressions introduced in {\bf \ref{sec:rhoa}}):
\begin{eqnarray}
V^{D}(q^{2}) &=& -\frac{m_{\rho}^2}{f_{\rho}} \sqrt{\frac{Z_H}
{m_B}} \left( \Omega_1 - m Z \right) (m_B + m_{\rho})\\ A^{D}_1
(q^{2}) &=& \frac{2 m_{\rho}^2}{f_{\rho}} \sqrt{Z_H m_B}
\frac{1}{m_B + m_{\rho}} \nonumber \\ && \left[ (m^2 + m m_{\rho}
{\bar{\omega}}) Z  -{\bar{\omega}} m_{\rho}\Omega_1 - m_\rho
\Omega_2 -2 \Omega_3 -\nonumber \right.\\ && \left. \Omega_4
-\Omega_5 -2 {\bar{\omega}} \Omega_6 \right]\\ A^{D}_2(q^{2}) &=&
\frac{m_{\rho}^2}{f_\rho }\sqrt{\frac{Z_H}{m_B}} \left( m Z
-\Omega_1 - 2 \frac{\Omega_6}{m_\rho} \right) (m_B + m_{\rho})\\
A^{D}_0 (q^{2}) &=& -\frac{m_\rho }{f_\rho} \sqrt{Z_H m_B}
\left[\Omega_1 \left(  m_\rho {\bar{\omega}} +2 m
\frac{q^2}{m_B^2} -
 \frac{r_1}{m_B}\right) +  m_\rho \Omega_2 + \nonumber \right.\\
&&\left. 2\Omega_3 + \Omega_4 \left(1- 2 \frac{q^2}{m_B^2}\right)
+ \Omega_5 + 2\Omega_6 \left(  \bar{\omega}- \frac {r_1}{m_B
m_\rho} \right)-\nonumber\right.\\ &&\left.  Z \left(m^2  - m
\frac{r_1}{m_B}  + m m_\rho {\bar{\omega}}\right) \right],
\end{eqnarray}
where:
\begin{equation}
{\bar{\omega}}=\frac{m_B^2+m_\rho^2-q^2}{2 m_B m_\rho},
\end{equation}
and:
\begin{equation}
r_1=\frac{m_B^2-q^2-m^2_\rho}{2}\label{r2}.
\end{equation}

$Z$ and $\Omega_j$ are given in the Appendix. In the above
expressions one must consider $\Delta_1=\Delta_H$,
$\Delta_2=\Delta_1 -m_\rho {\bar{\omega}}$, $x=m_\rho$.

The calculation for the $B\to a_1$ semileptonic transition
proceeds in a similar way. We find:
\begin{eqnarray}
A^{D}(q^2) &=&- \frac{m_{a}^2}{f_{a}} \sqrt{\frac{Z_H} {m_B}}
\left( \Omega_1 - m Z -\frac{2m}{m_a} \Omega_2 \right) (m_B + m_a)
\\ V^{D}_1 (q^{2})&=&\frac{2 m_{a}^2}{f_a} \sqrt{Z_H m_B}
\frac{1}{m_B + m_a} \nonumber \\ &&\left[(-m^2 + m m_a
\bar{\omega})Z + 2m\Omega_1 - \bar{\omega}m_a\Omega_1+ \nonumber
\right.\\ && \left.+(2m\bar{\omega} -m_a )\Omega_2-2\Omega_3
-\Omega_4-\Omega_5 -2\bar{\omega}\Omega_6  \right]\\
V^{D}_2(q^{2}) &=& \frac{m_a^2}{f_a}\sqrt{\frac{Z_H}{m_B}} \left(
m Z -\Omega_1 - 2 \frac{\Omega_6}{m_a}+2\frac{m}{m_a}\Omega_2
\right) (m_B + m_a)\\ V^{D}_0 (q^{2}) &=& -\frac{m_a }{f_a}
\sqrt{Z_H m_B} \left[\Omega_1 \left( m_a \bar{\omega}+2m
\frac{q^2}{m^{2}_B}- \frac{r_1^{\prime}}{m_B} -2m\right) +
\nonumber \right.\\ && \left. \Omega_2 \left( m_a + 2m \frac{
r_1^\prime}{m_B m_a} -
 2 m \bar{\omega}\right)+
2\Omega_3 + \Omega_4 \left(1- 2\frac{q^2}{m^{2}_B}
\right)+\Omega_5+ \nonumber \right.\\ &&\left. 2\Omega_6
\left(\bar{\omega} -   \frac{r_1^\prime}{m_B m_a} \right)+
Z\left(m^2+m\frac{r_1^\prime}{m_B}- m m_a \bar{\omega}
\right)\right],
\end{eqnarray}
where now:
\begin{eqnarray}
{\bar{\omega}}&=&\frac{m_B^2+m_a^2-q^2}{2 m_B m_a}\\ r_1^\prime &
= &\frac{m_B^2-q^2-m^2_a}{2}. \label{r2p}
\end{eqnarray}

\begin{table}
\begin{center}
\vbox{\offinterlineskip \halign{&#& \strut\quad#\hfil\quad\cr
\hline \hline &  && $V^D$ && $A^D_1$ && $A^D_2$ && $A^D_0$ &&
$A^D$ && $V^D_1$ && $V^D_2$&& $V^D_0$ &\cr \hline \hline &
$F^D(0)$ && $0.83$ && $0.69$ && $0.81$ && $0.33$ && $1.62$ &&
$1.13$ && $1.13$ && $1.13$ &\cr & $a_F$ && $0.93$ && $0$ && $0.87$
&& $2.9$ && $1.13$ && $0.18$ && $1.3$ && $1.9$ &\cr & $b_F$ &&
$0.02$ && $0$ && $-0.17$ && $2.6$ && $0.12$ && $0.04$ && $3.8$ &&
$0.93$ &\cr \hline \hline }} \caption{CQM direct diagrams
contributions to the form factors governing the semileptonic
decays $B\to \rho$ and $B\to a_1$. The  $F^D(0)$ values for the $B
\to a_1$ transition must be multiplied with the normalization
$0.25~{\rm GeV}^2 /f_a$. The theoretical uncertainty is around
$\pm 15 \%$. } \label{t:tab1}
\end{center}
\end{table}

These results are amenable to a numerical analysis. Let's consider
the following parametrization:
\begin{equation}
F^D(q^2)=\frac{F^D(0)}{1~-~a_F \left(\frac{q^2}{m_B^2}\right)
+~b_F \left(\frac{q^2}{m_B^2}\right)^2}, \label{eq:16b}
\end{equation}
for a generic direct form factor $F^D(q^2)$; $a_F, b_F$ have been
obtained through a numerical analysis of the $q^2$ region going
from $0$ to $q^2=16~{\rm GeV}^2$. Numerical values are listed in
Table \ref{t:tab1}.

The form factors describing the $B\to a_1$ transition at $q^2=0$
are proportional to the normalization factor $(0.25~{\rm
GeV}^2/f_a)$ (recall the problems in the determination of $f_a$
mentioned in {\bf \ref{sec:rhoa}}). These parameters are also
affected by a theoretical uncertainty of about $15\%$.

\subsubsection{\label{sec:polari} Polar contributions}

\begin{figure}[t!]
\begin{center}
\epsfig{bbllx=128pt,bblly=12cm,bburx=466pt,bbury=21cm,height=8truecm,
        figure=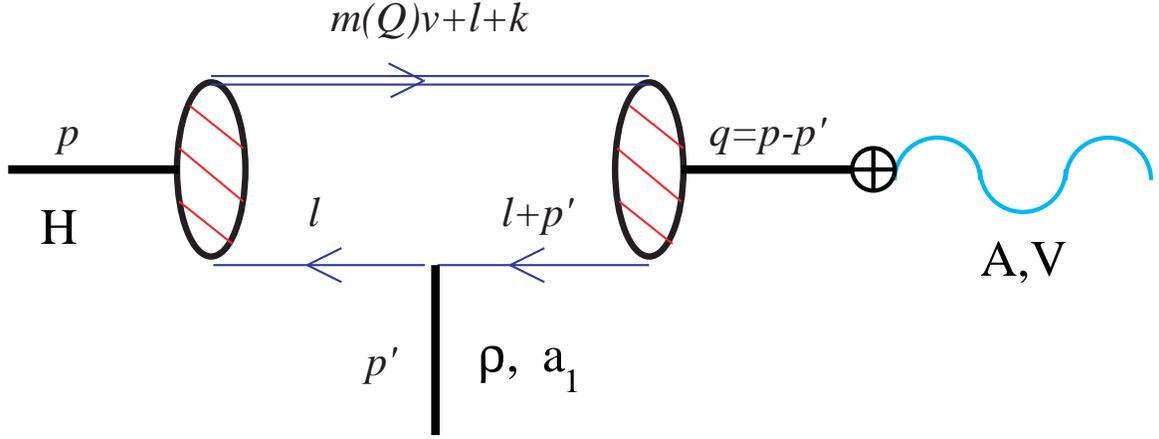}
\caption{\label{fig:polari} \footnotesize
          The CQM polar diagram.}
\end{center}
\end{figure}

The polar contributions to the form factors come from those CQM
diagrams in which the weak current is coupled to $B$ through an
heavy meson intermediate state.  In fig. \ref{fig:polari} it is
showed the related CQM loop diagram.
The form factor will then have a typical polar behaviour:
\begin{equation}
F^P(q^2)=\frac{F^P(0)}{1~-~ \frac{q^2}{m_P^2}}, \label{eq:fp}
\end{equation}
where $m_P$ is the mass of the intermediate virtual heavy meson
state. This behaviour is certainly valid nearby the pole. Let's
assume that it is valid all over the $q^2$ range that we want to
explore, {\it i.e.}, also for small $q^2$ values. This hypothesis
is a good one for the form factors $A_1^P, A_2^P$ (where the
superscript $P$ indicates that they are derived from the polar
diagram), since they are numerically small, with respect to
$A_1^D, A_2^D$, in the range of small $q^2$ values. Things are
different for $A_0^P(q^2)$ and $V_0^P(q^2)$, we shall come back on
this point later.

Using the strong couplings calculated in {\bf \ref{sec:rhoa}} and
the leptonic decay constants obtained in {\bf \ref{sec:clep}}, we
can calculate the different contributions to $F^P(0)$. As for the
semileptonic transition $B\to\rho$, they are \cite{art2}:
\begin{eqnarray}
V^P (0)&=&  -\sqrt{2} g_V \lambda {\hat F} \frac{m_B +m_\rho}{
m_B^{3/2}}\\ A^P_1 (0) &=&  \frac{\sqrt{2 m_B}g_V {\hat
F}^+}{m_{B_0} (m_B+m_{\rho})} (\zeta-2\mu {\bar \omega}
m_{\rho})\\ A^P_2 (0) &=& -\sqrt{2} g_V \mu {\hat F}^+
\frac{\sqrt{m_B} (m_B+m_\rho)} {m_{B_0}^2},
\end{eqnarray}
where ${\bar \omega}=m_B/(2 m_\rho)$, while $\lambda, \mu, \beta,
\zeta$ and $g_V$ have been defined and calculated in {\bf
\ref{sec:rhoa}}. As for $A^P_0 (q^2)$, we have to impose the
condition (\ref{eq:A3}); a possible choice is:
\begin{equation}
A^P_0 (q^2)= A^P_3(0) + g_V \beta {\hat F} \frac{1}{m_\rho\sqrt{2
m_B}} \frac{q^2}{m^2_B-q^2}. \label{eq:AP0}
\end{equation}
Let's discuss this equation in greater detail. The amplitude for
the semileptonic process $B\to \rho$ can be written, see  fig.
\ref{fig:ultima}, in the following way:
\begin{figure}[t!]
\begin{center}
\epsfig{bbllx=128pt,bblly=12cm,bburx=466pt,bbury=21cm,height=7truecm,
        figure=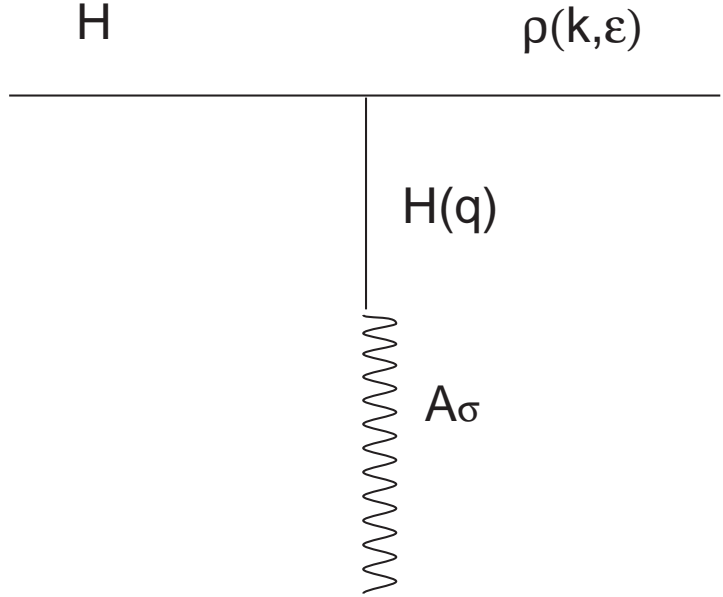}
\caption{\label{fig:ultima} \footnotesize
          The polar diagram at the level of meson interactions.}
\end{center}
\end{figure}
\begin{eqnarray}
{\rm Amp}(B\to\rho)&=&\langle {\rm VAC}|A^\sigma|H\rangle
\frac{i}{q^2-m_B^2} \langle H(q)\rho(k,\epsilon)|i{\cal
L}|H\rangle \nonumber\\ &=& i\frac{{\hat F}}{\sqrt{m_B}}q^\sigma
\frac{i}{q^2-m_B^2} C(\epsilon^*\cdot q),
\end{eqnarray}
where $H$ is the $0^-$ state. We have used eq. (\ref{eq:matf1})
and the interaction Lagrangian term multiplied by  $\beta$ in
(\ref{eq:strong1}). It is easy to find that:
\begin{equation}
C=i\sqrt{2}g_V\beta,
\end{equation}
therefore:
\begin{equation}
{\rm Amp}(B\to\rho)=i\sqrt{2}g_V\beta\frac{{\hat
F}}{\sqrt{m_B}}\frac{(\epsilon^*\cdot
q)}{q^2}\frac{q^2}{q^2-m_B^2}. \label{eq:this}
\end{equation}
Equation (\ref{eq:this}) should be compared with eq.
(\ref{eq:monsterm}):
\begin{equation}
i2m_\rho [A_3^P(q^2)-A_0^P(q^2)] \frac{(\epsilon^*\cdot q)}{q^2}.
\end{equation}
What then follows is (\ref{eq:AP0}).

As for the semileptonic transition $B\to a_1$, one has:
\begin{eqnarray}
A^P (0)&=&  -\sqrt{2} g_A \lambda_A {\hat F}^+ \frac{m_B +m_a}{
m_B^{3/2}}\\ V^P_1 (0) &=&  \frac{\sqrt{2 m_B}g_A {\hat F}}{m_B
(m_B+m_a)} (\zeta_A-2\mu_A {\bar \omega} m_a) \\ V^P_2 (0) &=&
-\sqrt{2} g_A \mu_A {\hat F} \frac{\sqrt{m_B} (m_B+m_a)} {m_B^2},
\end{eqnarray}
where ${\bar \omega}=m_B/(2 m_a)$. A similar reasoning as the one
made before can be applied for $V^P_0 (q^2)$:
\begin{equation}
V^P_0 (q^2)= V^P_3(0) + g_A \beta_A {\hat F}^+ \frac{1}{m_a\sqrt{2
m_B}} \frac{q^2}{m^2_{B_0}-q^2}. \label{eq:VP0}
\end{equation}

One should note that, if we do the hypothesis of massless leptons
in the final state, these $V_0^P$ and $A_0^P$ form factors do not
contribute to the process width and can therefore be ignored.

\begin{figure}[t!]
\begin{center}
\epsfig{bbllx=128pt,bblly=303pt,bburx=466pt,bbury=706pt,height=12truecm,
        figure=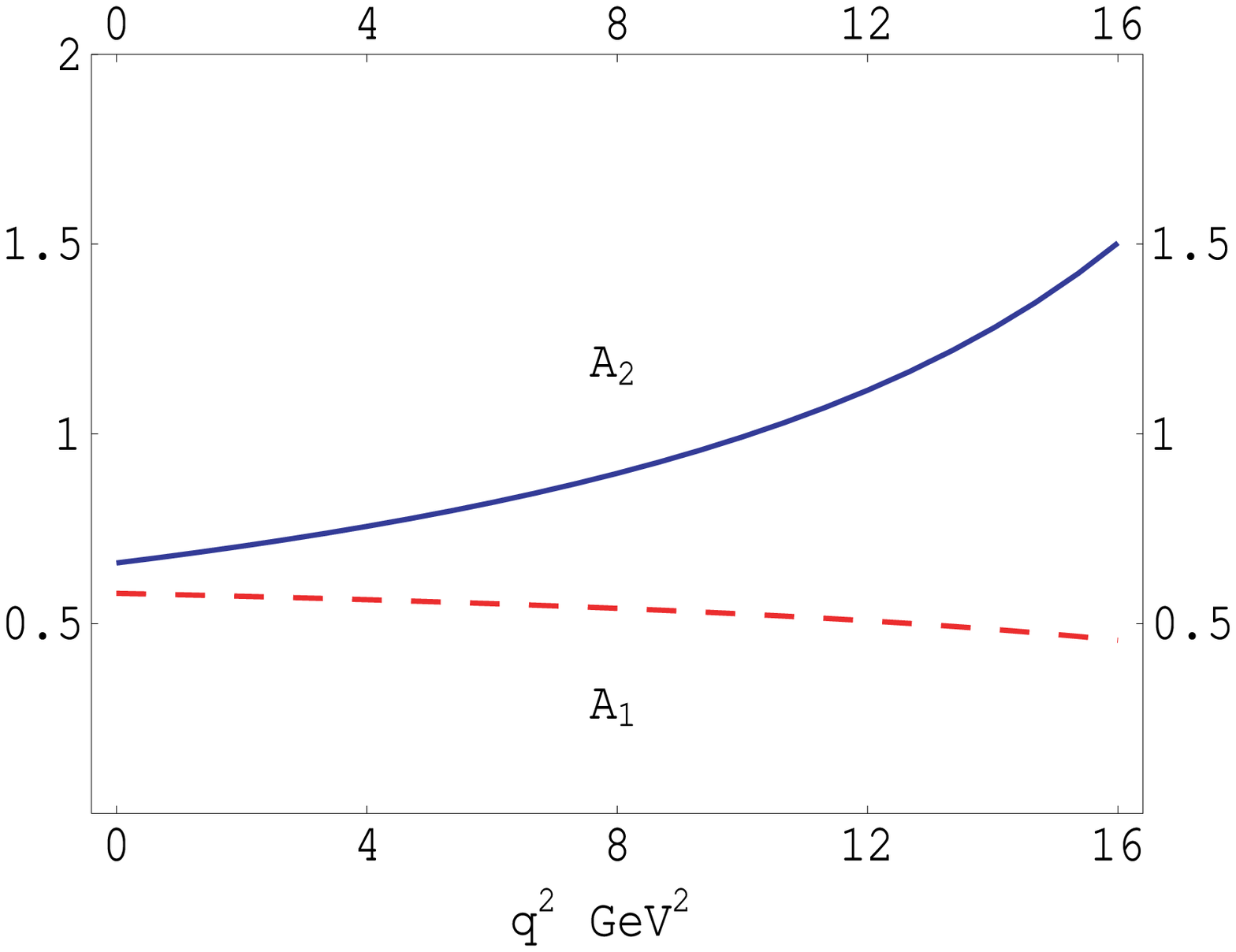}
\caption{\label{fig:plot1} \footnotesize
          Form factors $A_1$ and  $A_2$ for the semileptonic decay
          $B\to \rho$.}
\end{center}
\end{figure}

\begin{figure}[t!]
\begin{center}
\epsfig{bbllx=128pt,bblly=303pt,bburx=466pt,bbury=706pt,height=12truecm,
        figure=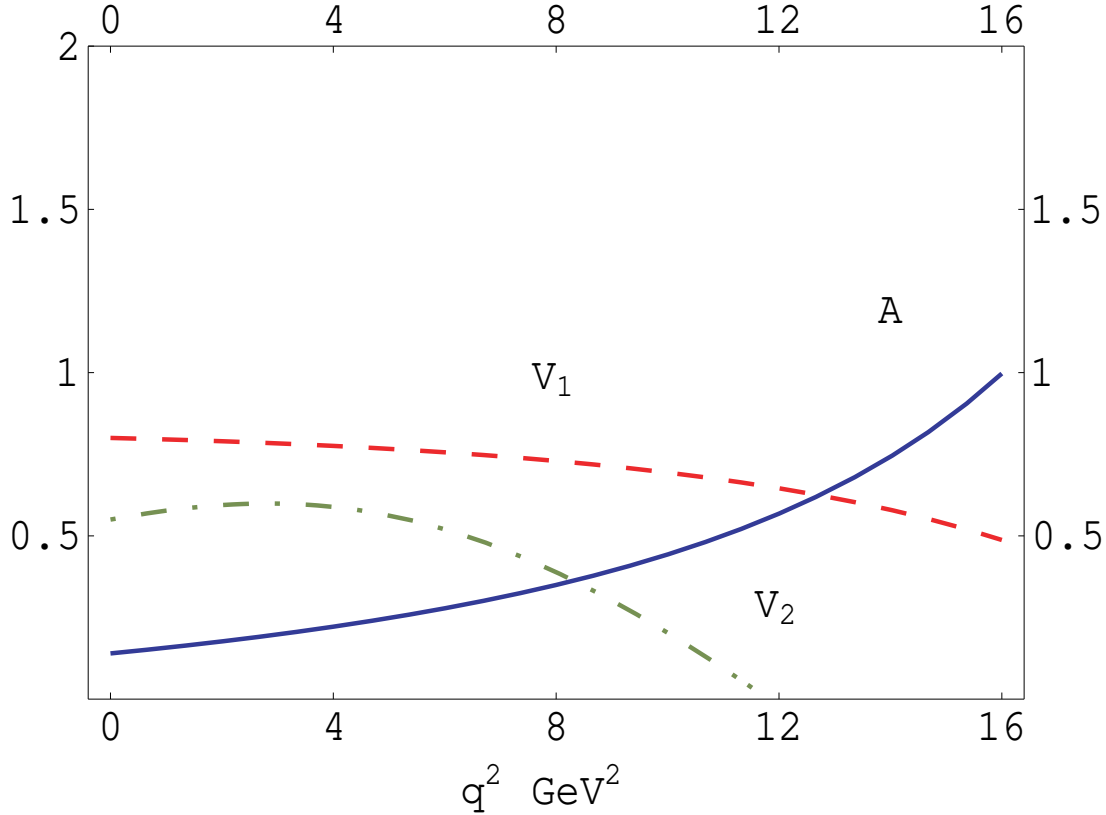}
\caption{\label{fig:plot2} \footnotesize
          Form factors $A$, $V_1$ and $V_2$ for the semileptonic
          decay $B\to a_1$.}
\end{center}
\end{figure}

Numerical results have been given in Table \ref{t:tabp} together
with the polar masses. A theoretical uncertainty of $\pm 15\%$ is
estimated. In fig. \ref{fig:plot1} we plot  $A_1$ and $A_2$, while
in fig. \ref{fig:plot2} the form factors $A$, $V_1$ and $V_2$ are
showed together. We don't plot $V(q^2)$ in fig. \ref{fig:plot1}
since the theoretical prediction for this value is affected by a
large uncertainty. Plots do not account of the errors given in
Tables.

\begin{table}
\hfil \vbox{\offinterlineskip \halign{&#&
\strut\quad#\hfil\quad\cr \hline \hline & && $V^P$ && $A^P_1$ &&
$A^P_2$ && $A^P_0$ && $A^P$ && $V^P_1$ && $V^P_2$ && $V^P_0$ &\cr
\hline \hline & $F^P(0)$ && $-0.84$ && $-0.11$ && $-0.15$ &&
$-0.019$ && $-1.48$ &&
 $-0.32$ &&
$-0.57$ && $0.07$ &\cr & $m_P$ && $5.3$ && $5.5$ && $5.5$ && $-$
&& $5.5$ && $5.3$ && $5.3$ && $-$ &\cr \hline \hline }} \caption{
Polar form factors for $B\to \rho$ and $B\to a_1$ semileptonic
decays. Polar masses have  GeV dimensions. The $F^P(0)$ values for
the $B \to a_1$ transition should be multiplied by the
normalization factor $(0.25~{\rm GeV}^2/f_a)$. The theoretical
uncertainty amounts to  $\pm 15 \%$.} \label{t:tabp}
\end{table}

\subsubsection{\label{sec:brw} Branching ratios and widths}

Using the numerical values given in the Tables, one can compute
semileptonic branching ratios and widths. First of all let's
compare CQM results for $B \to \rho$ with what is obtained by
other methods. In Table  \ref{t:tab2} we show the form factors
obtained summing up direct and polar contributions:
\begin{equation}
F(q^2)=F^D(q^2) + F^P(q^2).
\end{equation}
Here the CQM results are compared with those obtained by other
approaches. In particular, I would comment on the method followed
in \cite{20art2}. The idea is that of describing the heavy meson
field $H$ as an effective interpolating field given by the
operator product of the two constituent quark projectors times a
momentum space wave function, solution of the Bethe-Salpeter
equation for the Richardson potential (having the confining
behavior at large distances and the Coulombic one at short
distances). With such a description of the heavy field, the
calculation of the correlator $\langle {\rm VAC}|T \{ J_1 J J_2
\}|{\rm VAC} \rangle$, typical of potential models, where $J_1$
and $J_2$ are the effective interpolating currents of the light
mesons in the final state, $M_1$ and $M_2$, while $J$ is the
current of the process, is substituted by the calculation of the
amplitude $\langle M_1 M_2|J|H\rangle$. This approach gives
results for the semileptonic decays $B\to \rho \ell \nu$ and $B\to
K^* \ell^+ \ell^-$ that are in quite good agreement with those
obtained by lattice and QCD sum rules (in the $q^2$ region where
these latter methods are predictive).

As one can see, the CQM result obtained for $V^{\rho}(q^2)$ is
affected by a large theoretical uncertainty since it turns out to
be the sum of two numerical values almost identical in magnitude,
but opposite in sign. Aside from this problem, the CQM values are
generally in good agreement with those found by QCD sum rules;
with respect to other approaches they show to be systematically
higher.
\begin{table}
\hfil \vbox{\offinterlineskip \halign{&#&
\strut\quad#\hfil\quad\cr \hline \hline & && CQM &&
PM\cite{20art2} && LCSR \cite{21art2} && SR \cite{22art2} && LLCSR
\cite{23art2} &\cr \hline \hline & $V^{\rho}(0)$ && $-0.01 \pm
0.25$ && $0.45~\pm~0.11$ && $0.34~\pm~0.05$ && $0.6~\pm~0.2$ &&
$0.35^{+0.06}_{-0.05}$ &\cr & $A_1^{\rho}(0)$ && $0.58\pm 0.10$ &&
$0.27~\pm~0.06$ && $0.26~\pm~0.04$ && $0.5~\pm~0.1$ &&
$0.27^{+0.05}_{-0.04}$ &\cr & $A_2^{\rho}(0)$ && $0.66\pm 0.12$ &&
$ 0.26~\pm~0.05$ && $0.22~\pm~0.03$ && $0.4~\pm~0.2$ &&
$0.26^{+0.05}_{-0.03}$ &\cr & $A_0^{\rho}(0)$ && $0.33 \pm 0.05$
&& $0.29~\pm~0.09$ &&  && $ 0.24\pm 0.02$ &&
$0.30^{+0.06}_{-0.04}$ &\cr \hline \hline }} \caption{Form factors
for the  $B \to \rho$ transition at $q^2=0$. The CQM results are
compared  with those obtained by other theoretical approaches: a
potential model (PM), light cone sum rules (LCSR), QCD sum rules
(SR), lattice calculations with (SR), {\it i.e.} (LLCSR). The
large theoretical uncertainty on $V^\rho (0)$ is due to the strong
cancellation between direct and polar contributions.}
\label{t:tab2}
\end{table}

Using $V_{ub}=0.0032$ and $\tau_B=1.56 \times 10^{-12}$ s, we
obtain for $B\to \rho \ell\nu$:
\begin{eqnarray}
{\cal B}(\bar B^0 \to \rho^+ \ell \nu) &=& (2.5 \pm 0.8) \times
10^{-4} \nonumber \\ \Gamma_0(\bar B^0 \to \rho^+ \ell \nu) &=&
(4.4 \pm 1.3) \times 10^{7} \; s^{-1} \nonumber \\ \Gamma_+ (\bar
B^0 \to \rho^+ \ell \nu)&=& (7.1 \pm 4.5) \times 10^{7} \; s^{-1}
\nonumber \\ \Gamma_- (\bar B^0 \to \rho^+ \ell \nu)&=& (5.5 \pm
3.7) \times 10^{7} \; s^{-1} \nonumber \\ (\Gamma_+ + \Gamma_-)
(\bar B^0 \to \rho^+ \ell \nu)&=& (1.26 \pm 0.38) \times 10^8 \;
s^{-1},
\end{eqnarray}
where $\Gamma_0$, $\Gamma_+$, $\Gamma_-$ are referred to the three
helicity states of $\rho$. This branching ratio is evidently
consistent with the experimental one (\ref{eq:cleo}).

The reported results may be computed for different  $V_{ub}$
values just multiplying by $|V_{ub}/0.0032|^2$ and  choosing a
particular $V_{ub}$. The same holds for $\tau_B$ (sec) multiplying
it by $\tau_B/(1.56\times 10^{-12})$.

A discussion on the theoretical uncertainty of these values is in
order. The large error on  $V^\rho (0)$ reflects in the
determination of $\Gamma_+$ and $\Gamma_-$, but it doesn't affect
$\Gamma_0$ and has only a small effect on the branching ratio.
Theoretical uncertainties for $A_1^\rho (0)$ and $A_2^\rho (0)$
are probably related. The theoretical error on widths comes from
the sum in quadrature of the error on  $V^\rho (0)$  and a $\pm
15\%$ error common to $A_1^\rho (0)$ and $A_2^\rho (0)$.

Let's turn now to the semileptonic channel $B\to a_1\ell\nu $. CQM
predictions for it are the following:
\begin{eqnarray}
{\cal B}(\bar B^0 \to a_1^+ \ell \nu) &=& (8.4 \pm 1.6) \times
10^{-4} \nonumber \\ \Gamma_0 (\bar B^0 \to a_1^+ \ell \nu)&=&
(4.0 \pm 0.7) \times 10^{8} \; s^{-1} \nonumber \\ \Gamma_+ (\bar
B^0 \to a_1^+ \ell \nu)&=& (4.6 \pm 0.9) \times 10^{7} \; s^{-1}
\nonumber \\ \Gamma_- (\bar B^0 \to a_1^+ \ell \nu)&=& (0.98 \pm
0.18) \times 10^{8} \; s^{-1}, \label{ba1}
\end{eqnarray}
where $\Gamma_0$, $\Gamma_+$, $\Gamma_-$  label the three helicity
states of  $a_1$.

In the determination of these decay widths we have included only
the uncertainty arising from $f_a$. Lower values correspond to
$f_a=0.30~{\rm GeV}^2$, while higher values correspond to
$f_a=0.25~{\rm GeV}^2$. The theoretical errors arising from the
form factors at $q^2=0$ are difficult to estimate reliably and are
not included in this analysis. We can guess that the theoretical
uncertainty is larger at least by a factor of two.

CQM predicts a branching ratio for the $\bar{B}^0\to a_1^+\ell\nu$
decay higher than the branching ratio for $\bar{B}^0\to
\rho^+\ell\nu$ and this analysis shows that $\bar{B}^0\to
a_1^+\ell\nu$ could be responsible for almost the $50 \%$ of the
semileptonic channel $B\to X_u \ell\nu$; the $B\to\rho\ell\nu$
takes another $15 \%$.

The form factor $A$ suffers for an analogous cancellation problem
as $V$. The analysis of Aliev and Savci \cite{aliev}, based on the
QCD sum rules method, gives a value for $A$ that is five times
smaller than the CQM predicted one (with the sign reversed, but
this is because of a different overall phase definition of the
hadronic matrix elements).

        \subsection{\label{sec:btopi}$B\to \pi\ell\nu$}

The semileptonic $B\to\pi\ell\nu$ channel is of crucial interest
for the evaluation of the CKM matrix element $V_{ub}$ and it has
been widely investigated in literature, see the reviews \cite{rep}
and \cite{2art4}.

CQM allows for a simple determination of the already known leading
terms in the soft pion emission limit and of a new sub-leading
contribution computed beyond the soft pion limit approximation,
{\it i.e.}, through the technique discussed in {\bf
\ref{sec:thpi}}.

Let's then write the weak current matrix element for the
semileptonic transition amplitude  $B\to \pi$ \cite{rep}:
\begin{eqnarray}
\langle \pi(q_\pi)|V^{\mu}(q)|B(p)\rangle &=& \left[
(p+q_\pi)^{\mu}+\frac{m_{\pi}^2-m_B^2}{q^2}q^\mu \right]\;
F_1(q^2)\nonumber \\ &-& \left[ \frac{m_{\pi}^2-m_B^2}{q^2}q^{\mu}
\right] \; F_0(q^2),
\end{eqnarray}
where $F_1(0)=F_0(0)$.

\subsubsection{\label{sec:nd} The non derivative contribution}

\begin{figure}[t!]
\begin{center}
\epsfig{bbllx=128pt,bblly=16cm,bburx=466pt,bbury=706pt,height=7truecm,
        figure=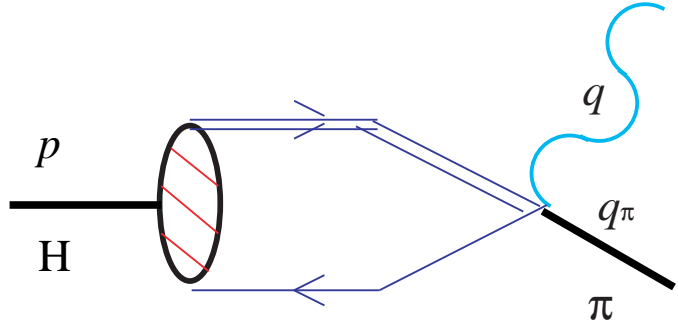}
\caption{\label{fig:ct} \footnotesize
          ``Callan-Treiman"  contribution to the form factors of
          the $B\to\pi$ semileptonic decay.}
\end{center}
\end{figure}

The diagram in fig. \ref{fig:ct} shows a non derivative coupling
of the pion to the $B$ meson and to the weak current. To obtain
the contribution to the form factors deriving from the diagram in
fig. \ref{fig:ct}, one has to expand $\chi=\xi q$, defined in {\bf
\ref{sec:cqmlag}}, up to the first order in $\pi$, neglecting the
term of order zero. In so far we have always considered $\chi=q$,
truncating the expansion at order zero.

Of course the diagram in fig. \ref{fig:ct} produces a result
proportional to the leptonic decay constant of $B$, as one can
easily see comparing fig. \ref{fig:ct} with fig. \ref{fig:lepto}.

On the other hand, the diagram in fig. \ref{fig:ct} makes sense
only for very small $q_\pi$, {\it i.e.}, in the limit in which
almost the entire incoming meson momentum is brought away by the
weak current.

We therefore obtain \cite{art4}:
\begin{eqnarray}
F_{0}^{\rm ND}&=&\frac{f_B}{f_\pi} \\ F_{1}^{\rm
ND}&=&\frac{f_B}{2f_\pi}, \label{eq:catr}
\end{eqnarray}
where $f_B$ was defined in (\ref{eq:effeb}).

\subsubsection{\label{sec:pr} The polar contribution}

Consider now a diagram of the kind of that given in fig.
\ref{fig:polari} where, instead of  $\rho$ or $a_1$, one must
consider the pion attached, through  a derivative coupling, to the
light quarks. The intermediate meson state can be a $1^-$ state,
belonging to the $H$ multiplet, or a $0^+$ state, belonging to the
$S$ multiplet. The latter case is less important since it produces
only a small correction to the chiral limit.

Consider the case in which the intermediate meson state is $1^-$.
Having in mind fig. \ref{fig:polari}, we expect a contribution
proportional to $g\hat{F}$, where $g$, the strong coupling
constant $HH\pi$, has been calculated through CQM in in {\bf
\ref{sec:hhpi}}. The following contribution to $F_1$ has been
obtained \cite{rep,hlwise}:
\begin{equation}
F_{1}^{\rm Pol}(q^2)= \frac{\hat{F}g}{f_{\pi} \sqrt{m_B}}
\frac{1}{1-q^2/m_{B^*}^2}. \label{eq:funopol}
\end{equation}

If instead the polar meson is $0^+$, one has:
\begin{equation}
F_{0}^{\rm Pol}(q^2)=\frac{1}{m_B^2-m_\pi^2}\left( \frac{h m_\pi
\sqrt{m_B} {\hat F}^+}{f_\pi}\right)\frac{1}{1-q^2/m_{B^{**}}^2},
\label{eq:fzeropol}
\end{equation}

where $h$  is the coupling constant $HS\pi$ discussed in {\bf
\ref{sec:thpi}} and by $B^{**}$ we mean the $0^+$ state. The
linear dependence of (\ref{eq:fzeropol}) on $m_\pi$ makes this
term not relevant in the chiral limit.

The polar form factors (\ref{eq:funopol}, \ref{eq:fzeropol}) are
reliable near the poles, {\it i.e.}, for  $q^2\simeq m_B$. We will
discuss later about one way to extrapolate these results  to a
wider $q^2$ range. Numerically, if one uses the CQM results
obtained in so far, one finds:
\begin{eqnarray}
F_{1}^{\rm Pol}(0)&=& 0.52 \pm 0.01\\ F_{0}^{\rm
Pol}(0)&=&0.012\pm 0.001.
\end{eqnarray}

\subsubsection{\label{sec:dr} The direct contribution}

The contributions discussed above are well known in literature.
CQM predicts the existence of a new contribution due to a diagram
of the kind of that depicted in fig. \ref{fig:diretti} where the
pion couples directly to $B$ and to the weak current. This
contribution differs from the non derivative one (ND), since it is
derivative, and from the polar (which instead is derivative),
since it doesn't contain couplings to resonances.

The evaluation of these diagrams proceeds following the rules we
have already discussed many times. The result is:
\begin{eqnarray}
F_{1}^{\rm Dir}(q^2)&=&\frac{2}{f_{\pi}}\sqrt{\frac{Z_H}{m_H}}
\left[q_\pi (C - m A) + m_H \;B\right] \label{eq:primoder}\\
F_{0}^{\rm Dir}(q^2)&=&\frac{2}{f_{\pi}}\sqrt{\frac{Z_H}{m_H}}
\left[\left(1-\frac{q^2}{m_{\pi}^2-m_B^2}\right) q_\pi (C-m A)
\right. \nonumber \\ &+& \left. m_H \;B
\left(1+\frac{q^2}{m_{\pi}^2-m_B^2}\right)\right],
\label{eq:secondoder}
\end{eqnarray}
where:
\begin{eqnarray}
A&=&\frac{1}{2 q_\pi}(I_3(\Delta_H-q_\pi)-I_3(\Delta_H))\\ B&=& m
A - m^2 Z(\Delta_H)\\ C&=&\frac{1}{2 q_\pi} (\Delta_H
I_3(\Delta_H) - (\Delta_H-q_\pi) I_3(\Delta_H-q_\pi)).
\end{eqnarray}
Here we have used the notation  $q_\pi^\mu=(q_\pi,0,0,q_\pi)$,
while $Z$ is given in the Appendix. Observe that the dependence of
(\ref{eq:primoder}) and (\ref{eq:secondoder}) on $q^2$ is due to
the fact that:
\begin{equation}
q_\pi=(m_B^2-q^2)/2m_B.
\end{equation}
Numerically we find:
\begin{equation}
F_{1}^{\rm Dir}(q^2=0)=F_{0}^{\rm Dir}(q^2=0)=0.13 \pm 0.05.
\end{equation}

The error made in the numeric evaluation is due to the variation
of $\Delta_H$ in the range $0.3-0.5$ GeV. The direct diagram
contribution induces a correction to the form factors that ranges
from $10\%$ to $30\%$ depending on $q^2$.

Since the three contributions to $F_0$ and $F_1$ are independent,
one can sum them up with the following result:
\begin{equation}
\hat{F_j}(q^2)=\frac{f_B}{(j+1)f_\pi}+ F_{j}^{\rm Dir}(q^2) +
\frac{F_{j}^{\rm Pol}(0)}{1-q^2/m^2_j}, \label{eq:ff}
\end{equation}
where $m_1=m_{B^*}$, $m_0=m_{B^{**}}$ and we have indicated the
total form factor as $\hat{F}$. $\hat{F}$ is the correct total
form factor only when we are at the maximum value of  $q^2$, {\it
i.e.}, $q^2\simeq m_B^2=q_0^2$. The $\hat{F_j}$ form factors could
be seen as the first terms in an expansion where higher terms are
characterized by an increasing number of derivatives of the pion
field. These terms are subleading only when $q^2$ is high.

Let's try to extrapolate to small  $q^2$ values using the
auxiliary functions $G_j(q^2)$:
\begin{equation}
F_j(q^2)=\hat{F_j}(q^2)G_j(q^2),
\end{equation}
where $j\in(0,1)$. This parameterization  must satisfy the
condition:
\begin{equation} G_j(q_0^2)=1, \label{uno}
\end{equation}
since $q^2 \simeq q_0^2$ is the region where $\hat{F_j}$ are
better approximating the $F_j$.

Since ${F_1}(0)={F_0}(0)$, the following condition must hold:
\begin{equation}
\hat{F_1}(0) G_1(0)=\hat{F_0}(0) G_0(0). \label{due}
\end{equation}

It is reasonable to assume that the corrections to $
G_j(q^2)\equiv 1$ come from terms in which one has one more
derivative of the pion field. Let's put:
\begin{equation}
G_j(q^2)= 1 + \frac{E_\pi}{\alpha_j \Lambda_\chi}=1 +
\frac{(q_\pi\cdot p)}{\alpha_j m_B\Lambda_\chi},\label{eq:gj}
\end{equation}
where $E_\pi$ is the pion energy in the frame where $B$ is at
rest, while $\alpha_j$ are free parameters. Moreover, since
(\ref{eq:gj}) is equivalent to:
\begin{equation}
G_j(q^2)= 1 + \frac{q^2-m_B^2}{2 m_B \Lambda_\chi \alpha_j  },
\end{equation}
the condition (\ref{uno}) is automatically satisfied. Equation
(\ref{due}) implies that $\alpha_0$ and $\alpha_1$ are related
according to:
\begin{equation}
\frac{2 \Lambda_\chi \alpha_1}{m_B}=1- \left( 1
-\frac{m_B}{2\Lambda_\chi \alpha_0} \right)
\frac{\hat{F_0}(0)}{\hat{F_1}(0)}.
\end{equation}

It is not possible to fix the  $\alpha$ constants only from
experimental data, therefore we need to refer to other theoretical
evaluations in literature. Quark models,
\cite{10art4}-\cite{12art4}, predict $F_0(0)=0.20-0.50$, with the
exception of  \cite{13art4}, which gives a value of $F_0(0)=0.09$.
QCD and factorization combined give  $F_0(0)=0.33$ \cite{14art4},
while chiral effective theory and HQET combined give $F_0(0)=0.38$
\cite{rep}. QCD sum rules give $F_0(0)=0.25-0.40$
\cite{15art4}-\cite{17art4} and some lattice QCD computations give
$F_0(0)=0.27-0.35$ \cite{18art4}-\cite{20art4}.

Let's use, as an input parameter, the result obtained with QCD sum
rules in \cite{16art4}, according to which $F_0(0)=0.30\pm 0.04$.
In so doing, one finds $\alpha_0=3.6$. An $\alpha_0>1$ suggests
that the energy scale controlling the  expansion (\ref{eq:gj}) is
not $\Lambda_\chi\simeq 1$ GeV, but larger. At this point one has
to remind that the first term in (\ref{eq:gj}) describes the
situation in which $q^2\simeq q_0^2$ while, in the corrections
depending on the pion momentum, we could have $q_\pi$ values
higher than expected, due to an effective expansion scale
$\alpha\Lambda_\chi>\Lambda_\chi$. In such a way we can extend to
small $q^2$ values the range of validity of the form factor
expressions.

In Table \ref{t:ultima} we show, for some  $q^2$  values, the two
form factors $F_1$ and $F_0$ also containing the CQM correction.
On the obtained results one can compute the error due to a
variation of $\alpha_0$ in the range $(2.9,3.6,4.3)$. What can be
read off from  Table \ref{t:ultima} is that CQM is in good
agreement with the results obtained by other methods. This also
means that the CQM deviations from the dominating contributions
(the Callan-Treiman and the polar one), due to the direct
contributions, are not strong enough to modify qualitatively the
polar behaviour of the form factors predicted by the chiral
effective theory.
\newpage
\begin{table}[h]
\begin{center}
\begin{tabular}{|c|c|c|c|c|}
\hline $q^2$ & 14.9 GeV$^2$ & 17.2 GeV$^2$ & 20 GeV$^2$ & 26.4
GeV$^2$\\ \hline CQM &&&&\\ \hline $F_1^{B\pi}$&
$1.58^{+0.28}_{-0.52}$& $2.06^{+0.27}_{-0.50}$&
$2.96^{+0.26}_{-0.47}$ &$13.78^{+0.13}_{-0.31}$\\ $F_0^{B\pi}$&
$0.59^{+0.10}_{-0.18}$& $0.62^{+0.08}_{-0.14}$&
$0.65^{+0.05}_{-0.10}$ &$0.83 \pm 0.01$\\ \hline IS  \cite{21art4}
&&&&\\ \hline $F_1^{B\pi}$& 0.83& 0.96 & 1.19 & 3.14\\
$F_0^{B\pi}$& 0.48& 0.48 & 0.48 & 0.47\\ \hline GNS \cite{11art4}
&&&&\\ \hline $F_1^{B\pi}$& 0.82& 1.05& 1.45&2.31\\ $F_0^{B\pi}$&
0.38& 0.40& 0.40&0.07\\ \hline LNS \cite{12art4}&&&&\\ \hline
$F_1^{B\pi}$& 0.53& 0.57 & -- & --\\ $F_0^{B\pi}$& 0.69& 0.76 & --
& --\\ \hline Ball  \cite{17art4}&&&&\\ \hline
$F_1^{B\pi}$&$0.85\pm 0.15$ & $1.1 \pm 0.2$& 1.6&--\\
$F_0^{B\pi}$& $0.5\pm0.1$& $0.55 \pm 0.15$& 0.7&--\\ \hline
Lattice (UKQCD) \cite{20art4}&&&&\\ \hline $F_1^{B\pi}$&$0.85 \pm
0.20$ & $1.10 \pm 0.27$&$1.72 \pm 0.50$&--\\ $F_0^{B\pi}$& $0.46
\pm 0.10$& $0.49\pm 0.10$&$0.56\pm 0.12$&--\\ \hline
\end{tabular}
\caption{Form factors $F_1$ and $F_0$ predicted by CQM at high
$q^2$ (near $q^2_{max}\simeq 26.4 {\rm GeV}^2$) and comparison
with other theoretical evaluations. The error quoted for CQM
results is due to a variation of $20 \%$ in the parameter
controlling  the evolution from higher to lower $q^2$ values,
where the results are less reliable. IS, GNS and LNS (from author
names), are quark models, while the Ball's paper makes use of
light cone QCD sum rules.} \label{t:ultima}
\end{center}
\end{table}
\vspace*{0.3truecm}
\newpage

\section{\label{chap:app}Appendix}

This Appendix is completely devoted to the $I_i$ integrals used in
the text and to their linear combinations arising in CQM
applications. These integrals have been computed adopting the
proper time regularization prescription discussed in {\bf
\ref{sec:reg}}. The analytical expressions here listed, have been
numerically treated using Mathematica 3.0. Everywhere $N_c=3$ is
meant.
\begin{eqnarray}
I_0(\Delta)&=& \frac{iN_c}{16\pi^4} \int^{\mathrm {reg}}
\frac{d^4k}{(v\cdot k + \Delta + i\epsilon)} \nonumber \\ &=&{N_c
\over {16\,{{\pi }^{{3/2}}}}} \int_{1/{{{\Lambda}^2}}}^{1/{{{\mu
}^2}}} {ds \over {s^{3/2}}} \; e^{- s( {m^2} - {{\Delta }^2} ) }
\left( {3\over {2\,s}} + {m^2} - {{{\Delta}}^2} \right)
[1+{\mathrm {erf}}(\Delta\sqrt{s})]\nonumber \\ &-& \Delta {{N_c
m^2}\over {16 \pi^2}} \Gamma\left(-1,{{{m^2}} \over
{{{\Lambda}^2}}},{{{m^2}}\over {{{\mu }^2}}}\right)
\\
I_1&=&\frac{iN_c}{16\pi^4} \int^{reg} \frac{d^4k}{(k^2 - m^2)}
={{N_c m^2}\over {16 \pi^2}} \Gamma\left(-1,{{{m^2}} \over
{{{\Lambda}^2}}},{{{m^2}}\over {{{\mu }^2}}}\right)
\\
I_1^{\prime}&=&\frac{iN_c}{16\pi^4} \int^{\mathrm {reg}} d^4
k\frac{k^2}{(k^2 - m^2)} ={{N_c m^4}\over {8 \pi^2}}
\Gamma\left(-2,{{{m^2}} \over {{{\Lambda}^2}}},{{{m^2}}\over
{{{\mu }^2}}}\right)\\ I_2&=&-\frac{iN_c}{16\pi^4} \int^{\mathrm
{reg}} \frac{d^4k}{(k^2 - m^2)^2}= \frac{N_c}{16\pi^2}
\Gamma\left(0,\frac{m^2}{\Lambda^2}, \frac{m^2}{\mu^2}\right)\\
I_3(\Delta) &=& - \frac{iN_c}{16\pi^4} \int^{\mathrm {reg}}
\frac{d^4k}{(k^2-m^2)(v\cdot k + \Delta + i\epsilon)}\nonumber \\
&=&{N_c \over {16\,{{\pi }^{{3/2}}}}}
\int_{1/{{\Lambda}^2}}^{1/{{\mu }^2}} {ds \over {s^{3/2}}} \; e^{-
s( {m^2} - {{\Delta }^2} ) }\; \left( 1 + {\mathrm {erf}}
(\Delta\sqrt{s}) \right)\\
I_4(\Delta)&=&\frac{iN_c}{16\pi^4}\int^{\mathrm {reg}}
\frac{d^4k}{(k^2-m^2)^2 (v\cdot k + \Delta + i\epsilon)}
\nonumber\\ &=&\frac{N_c}{16\pi^{3/2}}
\int_{1/\Lambda^2}^{1/\mu^2} \frac{ds}{s^{1/2}} \;
e^{-s(m^2-\Delta^2)} \; [1+{\mathrm {erf}}(\Delta\sqrt{s})]~.
\end{eqnarray}
In these equations,
\begin{equation}
\Gamma(\alpha,x_0,x_1) = \int_{x_0}^{x_1} dt\;  e^{-t}\;
t^{\alpha-1},
\end{equation}
is the incomplete gamma function, while erf is the error function
defined by:
\begin{equation}
{\rm erf}(z)=\frac{2}{\sqrt{\pi}}\int_{0}^{z}dx e^{-x^2}.
\end{equation}
Let us now introduce:
\begin{equation}
\sigma(x,\Delta_1,\Delta_2,\omega)={{{\Delta_1}\,\left( 1 - x
\right)  + {\Delta_2}\,x}\over {{\sqrt{1 + 2\,\left(\omega -1
\right) \,x + 2\,\left(1-\omega\right) \,{x^2}}}}}
\end{equation}
\begin{eqnarray} I_5(\Delta_1,\Delta_2,\omega) &= &
\frac{iN_c}{16\pi^4} \int^{\mathrm {reg}}
\frac{d^4k}{(k^2-m^2)(v\cdot k + \Delta_1 + i\epsilon ) (v'\cdot k
+ \Delta_2 + i\epsilon )} \nonumber \\
 & = & \int_{0}^{1} dx \frac{1}{1+2x^2 (1-\omega)+2x
(\omega-1)}\times\nonumber\\ &&\Big[
\frac{6}{16\pi^{3/2}}\int_{1/\Lambda^2}^{1/\mu^2} ds~\sigma \;
e^{-s(m^2-\sigma^2)} \; s^{-1/2}\; (1+ {\mathrm {erf}}
(\sigma\sqrt{s})) +\nonumber\\
&&\frac{6}{16\pi^2}\int_{1/\Lambda^2}^{1/\mu^2} ds \;
e^{-s(m^2-2\sigma^2)}\; s^{-1}\Big]
\end{eqnarray}

\begin{eqnarray}
I_6(\Delta_1,\Delta_2,\omega)&=&\frac{iN_c}{16\pi^4} \int^{\mathrm
{reg}} \frac{d^4k}{(v\cdot k + \Delta_1 + i\epsilon) (v'\cdot k +
\Delta_2 + i\epsilon)} \nonumber \\ &=&I_1 \int_{0}^{1} dx
\frac{\sigma}{1+2x^2(1-\omega) + 2x(\omega-1)}\nonumber\\
&-&\frac{N_c}{16 \pi^{3/2}}\int_{0}^{1} dx
\frac{1}{1+2x^2(1-\omega) + 2x(\omega-1)}\times\nonumber\\
&&\int_{1/\Lambda^2}^{1/\mu^2} \frac{ds}{s^{3/2}}
e^{-s(m^2-\sigma^2)}\Big\{\sigma [1+{\mathrm
{erf}}(\sigma\sqrt{s})]\cdot[1+2s(m^2-\sigma^2)]\nonumber\\ & +&2
{\sqrt{\frac{s}{\pi}}} e^{-s\sigma^2}\left[ \frac{3}{2s} +
(m^2-\sigma^2)\right]\Big\}.
\end{eqnarray}
In the  $\tau_{1/2,3/2}$ form factor determination, the following
expressions are needed:
\begin{eqnarray}
S(\Delta_1,\Delta_2,\omega)&=&\Delta_1\,I_3(\Delta_2) + \omega \,
\left( I_1 + \Delta_2 \,I_3(\Delta_2 ) \right)  +
{{\Delta_1}^2}\,I_5(\Delta_1,\Delta_2 ,\omega ) \nonumber \\
T(\Delta_1,\Delta_2,\omega)&=&{m^2}\,I_5(\Delta_1,\Delta_2 ,\omega
) + I_6(\Delta_1,\Delta_2 ,\omega ) \nonumber \\
U(\Delta_1,\Delta_2,\omega)&=&I_1 + \Delta_2 \,I_3(\Delta_2 ) +
\Delta_1\, I_3(\Delta_1) + \Delta_2
\,\Delta_1\,I_5(\Delta_1,\Delta_2 ,\omega ).
\end{eqnarray}

Here are listed the integral $Z$, introduced in {\bf
\ref{sec:hhpi}}, and all the auxiliary combinations used during
the study of $T\to H\pi$, $S\to H\pi$ processes:
\begin{eqnarray}
Z(\Delta) &=&  \frac{iN_c}{16\pi^4} \int^{\mathrm {reg}}
\frac{d^4k}{(k^2-m^2)[(k+q)^2-m^2](v\cdot k + \Delta +
i\epsilon)}\nonumber \\ &=&\frac{N_c}{16\pi^{3/2}}
\int_{1/\Lambda^2}^{1/\mu^2} \frac{ds}{s^{1/2}} e^{-s m^2}
\int_{0}^{1}dx e^{s \Delta^2(x)} [1+{\mathrm {erf}} (\Delta
(x)\sqrt{s})],
\end{eqnarray}
where $q^{\mu}=(q_\pi,0,0,q_\pi)$ is the pion four momentum and
$\Delta (x)=\Delta-xq_\pi$. Let us observe that, in the soft pion
limit $q_\pi \to 0$, one has $Z(\Delta) \to I_4(\Delta)$. The
auxiliary combinations are:
\begin{eqnarray}
\Theta &=& \frac{N_c}{16\pi^{2}}\int_{1/\Lambda^2}^{1/\mu^2}ds
\left( \frac{3-2 q_\pi^2 s}{6 s^2} \right)e^{-s m^2}  \\
R_1(\Delta_T) &=& m^2 Z(\Delta_T)-I_3(\Delta_H)\\ R_2(\Delta_T)
&=& \Delta^2_T
Z(\Delta_T)+\left(\frac{q_\pi}{2}+\Delta_T\right)I_2 \\
R_3(\Delta_T) &=& \frac{q_\pi}{2}(\Delta_T I_3(\Delta_T) -\Delta_H
I_3(\Delta_H))  \\ R_4(\Delta_T) &=&
\frac{\Delta_T}{2}(I_3(\Delta_T)-I_3(\Delta_H)) \\ S_1(\Delta_T)
&=& \Theta - \frac{\Delta_T h}{2} I_2 -\Delta^2_T I_2-\Delta^3_T
Z(\Delta_T)\\ S_2(\Delta_T) &=& I_1 + \Delta_T I_3(\Delta_H) - m^2
I_2 -m^2\Delta_T Z(\Delta_T) \\ S_3(\Delta_T) &=&
q_\pi(I_1+\Delta_H I_3(\Delta_H))+\frac{m^2}{2}
(I_3(\Delta_H)-I_3(\Delta_T))\\ S_4(\Delta_T) &=& \frac{q_\pi}{2}
I_1 + \frac{\Delta^2_T}{2}(I_3(\Delta_H) - I_3(\Delta_T))\\
S_5(\Delta_T) &=& \frac{q_\pi \Delta_T}{2} (\Delta_H
I_3(\Delta_H)- I_3(\Delta_T)),
\end{eqnarray}
where $q_\pi=v\cdot q$.\newline The following polynomial appears
in the $h^\prime$ determination:
\begin{eqnarray}
P(R_i,S_i,q_\pi)&=&-\frac{1}{88 q_\pi^4}\left[8q_\pi^3(11 m R_1 +
4 S_1 - 6 S_2)+ 2q_\pi^2(-176 m R_4 +14 S_1 + S_2 \right.\nonumber
\\ &+& \left.8 S_3 + 48 S_4) +3q_\pi (88 m R_3 + S_3 - 16 S_4 - 32
S_5) + 15 S_5 \right],
\end{eqnarray}
while the following expressions emerge in the determination of $f$
in {\bf \ref{sec:hhpi}}:
\begin{eqnarray}
V&=&\frac{1}{3}[(m^2-\Delta_T^2)I_4(\Delta_T)-I_3(\Delta_T)-\Delta_T
I_2]
\\
Y&=&\frac{3}{32\pi^2}\int_{1/\Lambda^2}^{1/\mu^2}\frac{ds}{s^3}e^{-sm^2}\\
T&=&
\frac{1}{3}[I_1-Y+(\Delta_T^2-m^2)I_2+\Delta_T(1-\Delta_T^2)I_3(\Delta_T)-m^2\Delta_T
I_4(\Delta_T)].
\end{eqnarray}
The processes meson$\to (\rho,a_1)$ requires to compute $Z$ in the
case in which $q^\mu= x v^{\prime\mu}$, $\omega=v\cdot
v^{\prime}$, $\Delta_2=\Delta_1- x ~ \omega$, $x$ being the
$\rho(a_1)$ mass:
\begin{eqnarray}
Z &=&  \frac{iN_c}{16\pi^4} \int^{\mathrm {reg}}
\frac{d^4k}{(k^2-m^2)[(k+q)^2-m^2](v\cdot k + \Delta_1 +
i\epsilon)}\nonumber \\ &=&\frac{I_5(\Delta_1,
x/2,\omega)-I_5(\Delta_2,- x/2,\omega)}{2 x}~.
\end{eqnarray}
The following $\Omega_i$ expressions have been found in the
determination of the strong couplings $HH\rho,...$, see {\bf
\ref{sec:rhoa}}, and have been used in the  determination of the
semileptonic form factors $B\to \rho(a_1)$, see {\bf
\ref{sec:diretti}}:
\begin{eqnarray} K_1&=&m^2 Z
-I_3(\Delta_2) \\ K_2&=&\Delta_1^2 Z -\frac{I_3(x/2)- I_3(-x/2)}{4
x}[\omega ~ x + 2 \Delta_1]
 \\
K_3&=&\frac{x^2}{4} Z +\frac{I_3(\Delta_1)-3
I_3(\Delta_2)}{4}+\frac{\omega}{4}[\Delta_1 I_3(\Delta_1)-
\Delta_2 I_3(\Delta_2)]
 \\
K_4&=&\frac{x \Delta_1}{2} Z +\frac{\Delta_1[I_3(\Delta_1)-
I_3(\Delta_2)]}{2 x}+\frac{I_3(x/2)-I_3(-x/2)}{4}  \\
\Omega_1&=&\frac{ I_3(-x/2)-I_3(x/2)+\omega[I_3(\Delta_1)-
I_3(\Delta_2)]}{2 x (1-\omega^2)} - \frac{[\Delta_1-\omega x/2]Z}
{1-\omega^2}
 \\
\Omega_2&=&\frac{
-I_3(\Delta_1)+I_3(\Delta_2)-\omega[I_3(-x/2)-I_3(x/2)]} {2 x
(1-\omega^2)} - \frac{[x/2- \Delta_1\omega ]Z}{1-\omega^2}
 \\
\Omega_3&=&\frac{K_1}{2}+\frac{2 \omega
K_4-K_2-K_3}{2(1-\omega^2)}
 \\
\Omega_4&=&\frac{-K_1}{2(1-\omega^2) }+\frac{3 K_2-6 \omega
K_4+K_3 (2\omega^2+1)}{2(1-\omega^2)^2}
 \\
\Omega_5&=&\frac{-K_1}{2(1-\omega^2) }+\frac{3 K_3-6 \omega
K_4+K_2 (2\omega^2+1)}{2(1-\omega^2)^2}
 \\
\Omega_6&=&\frac{K_1\omega}{2(1-\omega^2) }+\frac{2
K_4(2\omega^2+1)- 3\omega( K_2+K_3) }{2(1-\omega^2)^2}.
\end{eqnarray}
\section*{Acknowledgements}

I am grateful to A. Deandrea, N. Di Bartolomeo, R. Gatto, and G.
Nardulli, for constant help and discussions, and to P. Colangelo,
F. Feruglio and N. Paver, for comments on the manuscript. I also
acknowledge J. Arponen for his friendly help. This work has been
supported from EU-TMR programme, contract CT98-0169.
                

\begin{thebibliography}{999}


\bibitem{artuso} M. Artuso, hep-ph/9812372.

\bibitem{perret} P. Perret, hep-ex/9811047.


\bibitem{falcorari} A. Falk, hep-ph/9503485.

\bibitem{shif} M. Shifman, hep-ph/9802214.

\bibitem{rothe} H.J. Rothe, {\it Lattice gauge theories}, World Scientific, 1980.

\bibitem{falcorass} A. Falk, hep-ph/9812217.

\bibitem{okun} L.B. Okun, {\it Particle Physics: The Quest for the Substance of Substance},
Harwood, 1984.

\bibitem{martinelli} G. Martinelli, hep-ph/9610455.


\bibitem{rass1} A. Manohar, Lectures at the Schladming Winter School, March 1996,
                hep-ph/9606222.

\bibitem{rass2} D. Kaplan, nucl-th/9506035.

\bibitem{rass3} J. Polchinski, Lectures presented at TASI 92, hep-th/9210046.

\bibitem{rass4} H. Georgi, Annu. Rev. Nuc. Part. Sci. 43, 209 (1994).

\bibitem{rass5} H. Georgi, {\it Weak Interactions and Modern Particle Theory},
                Benjamin/Cummings, 1984.

\bibitem{rass6} S. Weinberg, Physica {\bf 96A}, 327 (1979).

\bibitem{wilson} K.G. Wilson and J.G. Kogut, Phys. Rep. {\bf 12}, 75 (1974).

\bibitem{rass7} M. Neubert, hep-ph/9610385.

\bibitem{rass8} M. Neubert, Phys. Rep. 245, 259 (1994).

\bibitem{stone} S. Stone (Ed.), {\it B-Decays}, World Scientific, 1994.

\bibitem{rass9} B. Grinstein, hep-ph/9508227.

\bibitem{rass10} J.M. Flynn and N. Isgur, J. Phys. {\bf G18}, 1627 (1992).

\bibitem{rass11} A. Falk, hep-ph/9610363.

\bibitem{rass12} M.B. Wise, hep-ph/9805468.

\bibitem{georgi} H. Georgi, Phys. Lett. {\bf B240}, 447 (1990).

\bibitem{isgur} N. Isgur and M.B. Wise, Phys. Lett. {\bf B232}, 113 (1989).

\bibitem{hill} E. Eichten and B. Hill, Phys. Lett. {\bf B234}, 511 (1990).

\bibitem{huang} K. Huang, {\it Quarks, Leptons and Gauge Fields}, 2nd Ed.,
                World Scientific, 1992.

\bibitem{neub}  M. Neubert, Nucl. Phys. {\bf B371}, 149 (1992).

\bibitem{mano}  A. Manohar, hep-ph/9305298.

\bibitem{geomano} A. Manohar and H. Georgi, Nucl. Phys. {\bf B234}, 189 (1984).

\bibitem{galeut} J. Gasser, H. Leutwyler, Annals Phys. {\bf 158}, 142 (1984);
Nucl. Phys. {\bf B250}, 465 (1985).

\bibitem{shuryak} E. Shuryak, {\it The QCD Vacuum, Hadrons and the Superdense Matter},
                  World Scientific, 1988.

\bibitem{coleman} S. Coleman {\it et al.} Phys. Rev. {\bf 177}, 2239 (1969);
                  C.G. Callan jr. {\it et al.} Phys. Rev. {\bf 177}, 2247 (1969); S. Weinberg in
                  Lectures on Elementary Particle Physics and Quantum Field Theory - 1970
                  Brandais University Institute in Theoretical Physics, MIT press, 1970.

\bibitem{weinbook} S. Weinberg, {\it The Quantum Theory of Fields vol. 2}, Cambridge, 1996.


\bibitem{treiman} S. Treiman, R. Jackiw, B. Zumino, E. Witten, {\it Current Algebra
                   and Anomalies}, Princeton, 1985.

\bibitem{rep} R. Casalbuoni {\it et al.} Phys. Rep. {\bf 281}, 145 (1997).

\bibitem{luca} A. Falk and M. Luke, Phys. Lett. {\bf B292}, 219 (1992).

\bibitem{hlwise} M.B. Wise Phys. Rev. {\bf D45}, 2188 (1992); G.~Burdman and J. F.
Donoghue, Phys. Lett. {\bf B280}, 287 (1992); L. Wolfenstein,
Phys. Lett. {\bf B291}, 177 (1992); T. M. Yan, H.Y. Cheng, C. Y.
Cheung, G. L. Lin, Y. C. Lin and H. L. Yu, Phys. Rev. {\bf D46},
1148 (1992).



\bibitem{art1} A. Deandrea, N. Di Bartolomeo, R. Gatto, G. Nardulli and A.D. Polosa
               Phys. Rev. {\bf D58}, 034004 (1998).

\bibitem{art2} A. Deandrea, R. Gatto, G. Nardulli and A.D. Polosa
               Phys. Rev. {\bf D59}, 074012 (1999).

\bibitem{art3} A. Deandrea, R. Gatto, G. Nardulli and A.D. Polosa
               J. High Energy Phys. 02 (1999) 021.

\bibitem{art4} A. Deandrea, R. Gatto, G. Nardulli and A.D. Polosa
               hep-ph/9907225, to appear on Phys. Rev. D.

\bibitem{art5} A.D. Polosa, hep-ph/9909371, to appear in
               Proceedings of the XIth Rencontres de Blois, Blois, France,
               27 June-3 Jul 1999.

\bibitem{njl} Y. Nambu and G. Jona-Lasinio, Phys. Rev. {\bf 122}, 345 (1960).

\bibitem{strocchi} F. Strocchi, {\it Elements of Quantum Mechanics of Infinite Systems},
                   World Scientific, 1985.


\bibitem{ebert} D. Ebert, T. Feldmann, R. Friedrich and H. Reinhardt,
Nucl. Phys. {\bf B434} 619 (1995); D.~Ebert, T.~Feldmann and
H.~Reinhardt, Phys. Lett. {\bf B388} 154 (1996); T.~Feldmann
hep-ph/9606451.

\bibitem{bije} J. Bijnens, Phys. Rep. {\bf 265}, 369 (1996);
T. Hatsuda and T. Kunihiro, {\it ibid.} {\bf 247}, 221 (1994).

\bibitem{miranski} V.A. Miranski {\it Dynamical Symmetry Breaking in Quantum Field
Theories}, World Scientific, 1993.

\bibitem{witten} E. Witten, Commun. Math. Phys. 92, 455 (1984).

\bibitem{pdg} C. Caso {\it et al.} (Particle Data Group), Eur. Phys. Jour. {\bf C3} 1 (1998)
and http://pdg.lbl.gov.

\bibitem{col1} P. Colangelo {\it et al.}, Phys. Rev. {\bf D52}, 5422 (1995).

\bibitem{zoeller} M.M. Zoeller for the CLEO collaboration, talk at American
Physical Society DPF'99,
http://www.physics.ucla.edu/dpf99/trans/3-16.pdf.

\bibitem{fey-phint} R.P. Feynman, {\it Photon-Hadron Interactions},
                    Addison Wesley, 1972.

\bibitem{marshak} R.E. Marshak, Riazuddin, C.P. Ryan, {\it Theory
of Weak Interactions in Particle Physics}, Wiley-Interscience,
1969.

\bibitem{sakurai} J.J. Sakurai, {\it Currents and Mesons}, Chicago Press,
1969.

\bibitem{kroll} N.M. Kroll {\it et al.}, Phys. Rev. {\bf 157}, 1376 (1967).

\bibitem{kugo} M. Bando {\it et al.}, Phys. Rep. {\bf 164}, 217 (1988).

\bibitem{9art2} N. Isgur, C. Morningstar and C. Reader,
Phys. Rev {\bf D39}, 1357 (1989).

\bibitem{10art2} I.A. Shushpanov, hep-ph/9612289.

\bibitem{11art2} M. Wingate, T. DeGrand, S. Collins and U.M.Heller,
Phys. Rev. Lett. {\bf 74}, 4596 (1995).



\bibitem{32art1} A. A. Ovchinnikov, Sov. J. of Nucl. Phys. {\bf 50}, 519 (1989);
P. Colangelo, G. Nardulli, A. Deandrea, N. Di Bartolomeo, R. Gatto
and F. Feruglio, Phys. Lett. {\bf B339}, 151 (1994); V.M. Belyaev,
V.M. Braun, A. Khodzamirian, R. Rueckl, Phys. Rev. {\bf D51}, 6177
(1995); T.M. Aliev, D.A. Demir, E. Iltan and N.K. Pak, Phys. Lett.
{\bf B351}, 339 (1995).

\bibitem{31art1} P. Colangelo, F. De Fazio and G. Nardulli, Phys. Lett.
{\bf B334}, 175 (1994).

\bibitem{33art1} M. Sutherland, B. Holdom, S. Jaimungal, and R. Lewis, Phys. Rev.
{\bf D51}, 5053 (1995).

\bibitem{becirevic} A. Le Yaouanc and Becirevic, J. High Energy
Phys. 021 (1999) 9903.

\bibitem{bernard} C. Bernardini, O. Ragnisco e P.M. Santini, {\it Metodi Matematici della
Fisica}, NIS, 1993

\bibitem{14art3} A.F. Falk and T. Mehen, Phys. Rev. {\bf D53}, 231 (1996).

\bibitem{18art2} T.M. Aliev, D.A. Demir, E. Iltan and N.K. Pak,
Phys. Rev. {\bf D53}, 355 (1996); T.M. Aliev, D.A. Demir, E. Iltan
and N.K. Pak, Z. Phys. {\bf C69}, 481 (1996).

\bibitem{19art2} P. Colangelo, F. De Fazio and G. Nardulli,
Phys. Lett. {\bf B316}, 555 (1993).






\bibitem{1art2} Belle Progress Report, Belle Collaboration,
KEK-PROGRESS-REPORT-97-1.

\bibitem{2art2} Status of the Babar Detector, BaBar Collaboration, SLAC-PUB-7951,
presented at 29th International Conference on High-Energy Physics
(ICHEP 98), Vancouver, Canada, 23-29 July 1998.

\bibitem{4art2} CLEO Collab., J. P. Alexander {\it et al.}, Phys. Rev. Lett.
{\bf 77}, 5000 (1996).

\bibitem{aliev} T.M. Aliev and M. Savci, Phys. Lett. {\bf B456}, 256 (1999).

\bibitem{16art1} M. Neubert, Phys. Rev. {\bf D45}, 2451 (1992).

\bibitem{17art1} P. Colangelo, G. Nardulli and N. Paver, Phys. Lett. {\bf B293},
207 (1992); P. Colangelo, F. De Fazio and N. Paver, Phys. Rev.
{\bf D58}, 116005 (1998).

\bibitem{3art1} J. M. Flynn, C. T. Sachrajda, hep-lat/9710080 and
references therein.

\bibitem{luke} M.E. Luke, Phys. Lett. {\bf B264}, 455 (1991).

\bibitem{agatto} M. Ademollo and R. Gatto, Phys. Rev. Lett. {\bf 13}, 264 (1964).

\bibitem{18art1} B. Blok and M. Shifman, Phys. Rev. {\bf D47}, 2949 (1993); M. Neubert,
Phys. Rep. {\bf 245}, 259 (1994); E. Bagan, P. Ball and P.
Gosdzinsky, Phys. Lett. {\bf B301}, 249 (1993); S. Narison, report
CERN-TH 7103/93, PM-93/45.

\bibitem{19art1} S. Godfrey and N. Isgur, Phys. Rev. {\bf D32}, 189 (1985).

\bibitem{20art1} S. Veseli and I. Dunietz, Phys. Rev. {\bf D54}, 6803 (1996).

\bibitem{21art1} P. Cea, P. Colangelo, L. Cosmai and G. Nardulli, Phys. Lett.
{\bf B 206}, 691 (1988);  P. Colangelo, G. Nardulli and M.
Pietroni, Phys. Rev. {\bf D43}, 3002 (1991).

\bibitem{22art1} N. Isgur, D. Scora, B. Grinstein and M.B. Wise
{\bf D39}, 799 (1989).

\bibitem{23art1} V. Morenas, A. Le Yaouanc, L. Oliver, O. P\`ene and
J.-C. Raynal, Phys. Rev. {\bf D56}, 5668 (1997).


\bibitem{25art1} J. Christensen, T. Draper and C. McNeile,
in {\it Lattice 97} Proceedings of the International Symposyum,
Edinburgh, Scotland, edited by C. Davies {\it et al.} [Nucl. Phys.
(Proc. Suppl.) {\bf B63}, 377 (1998)], hep-lat/9710025.

\bibitem{26art1} N. Isgur and M. B. Wise, Phys. Rev. {\bf D43}, 819 (1991).

\bibitem{27art1} A. K. Leibovich, Z. Ligeti, I. W. Stewart and M. B. Wise,
Phys. Rev. Lett. {\bf 66}, 3995 (1997).

\bibitem{28art1} A. Wambach, Nucl. Phys. {\bf B434}, 647 (1995).

\bibitem{29art1} M. Neubert, hep-ph/9801269, in International Europhysics Conference on
High Energy Physics, Jerusalem, Israel, 19-26 August 1997.

\bibitem{janne} A. Le Yaouanc {\it et al.}, hep-ph/0003087.

\bibitem{faustebe} D. Ebert, R.N. Faustov and V.O. Galkin, Phys.
Rev. {\bf D61}, 014016 (1999).

\bibitem{30art1} D. Buskulic {\it et al.}, ALEPH Collaboration, Z. Phys. {\bf C73}, 60 (1997);
ibidem, Phys. Lett. {\bf B395} 373 (1997); K. Ackerstaff {\it et
al.}, OPAL Collaboration, Z.Phys. {\bf C76}, 425 (1997).

\bibitem{17art2} M. Wirbel, B. Stech, M. Bauer, Z. Phys. {\bf C29}, 637 (1985).

\bibitem{20art2} P. Colangelo, F. De Fazio, M. Ladisa, G. Nardulli, P. Santorelli and
A. Tricarico, hep-ph/9809372.

\bibitem{21art2} P. Ball and V. M. Braun, Phys. Rev. {\bf D58}, 094016 (1998).

\bibitem{22art2} P. Ball, Phys. Rev. {\bf D 48}, 3190 (1993); P. Colangelo, F. De Fazio
and P. Santorelli, Phys. Rev. {\bf D 51}, 2237 (1995); P.
Colangelo, F. De Fazio, P. Santorelli and E. Scrimieri, Phys. Rev.
{\bf D 53}, 3672 (1996); {\bf D 57}, 3186 (1998) (E).

\bibitem{23art2} L. Del Debbio {\it et al.}, (UKQCD Collaboration),
Phys. Lett. {\bf B416}, 392 (1988).


\bibitem{2art4} J.D. Richman and P.R. Burchat, Rev. Mod. Phys. {\bf 67}, 893
(1995).

\bibitem{10art4} M. Wirbel, B. Stech and M. Bauer,
Z. Phys. {\bf C29}, 637 (1985); {\bf C34}, 103 (1987); M. Bauer
and M. Wirbel, Z. Phys. {\bf C42}, 671 (1989); R.N. Faustov, V.O.
Galkin and A.Yu. Mishurov, Phys. Rev. {\bf D53}, 6302 (1996); D.
Melikhov, Phys. Lett. {\bf B394}, 385 (1997); P.J. O'Donnel, Q.P.
Xu and H.K.K. Tung, Phys. Rev. {\bf D52}, 3966 (1995); C.-Y.
Cheung, C.-W. Hwang and W.-M. Zhang, Z.Phys. {\bf C75}, 657
(1997).

\bibitem{11art4} I.L. Grach, I.M. Narodetskii, S. Simula, Phys. Lett. {\bf B385},
317  (1996).

\bibitem{12art4} M. Ladisa, G. Nardulli, P. Santorelli, hep-ph/9903206.

\bibitem{13art4} B. Grinstein, M.B. Wise and N. Isgur,
Phys. Rev. Lett. {\bf 56}, 298 (1986); N. Isgur, D. Scora, B.
Grinstein and M.B. Wise, Phys. Rev. {\bf D39}, 799 (1989).

\bibitem{14art4} D.S. Hwang and B.H. Lee, Eur. Phys. J. {\bf C6}, 663 (1999).

\bibitem{15art4} V.M. Belyaev, A. Khodjamirian and R. R\"uckl,
Z. Phys. {\bf C60}, 349 (1993); A. Khodjamirian and R. R\"uckl, in
{\it Continuous Advances in QCD 1996}, M.I. Polikarpov Ed. (World
Scientific, 1996); S. Narison, Phys. Lett. {\bf B283}, 384 (1992);
S. Narison, Phys. Lett. {\bf B345}, 166 (1995); C.A. Dominguez and
N. Paver, Z. Phys. {\bf C41}, 217 (1988); A. Khodjamirian, R.
R\"uckl, Heavy Flavors, 2nd edition, eds. A.J. Buras and M. Linder
(World Scientific), hep-ph/9801443; P. Ball, Phys. Rev. {\bf D48},
3190 (1993).

\bibitem{17art4} P. Ball, J. High Energy Phys. {\bf 09} (1998) 005.

\bibitem{18art4} APE Collaboration (C.R. Allton {\it et al.}), Phys. Lett. {\bf B345},
513 (1995).

\bibitem{19art4} A. Abada {\it et al.}, Nucl. Phys. {\bf B416}, 675 (1994)

\bibitem{20art4} UKQCD Collaboration (D.R.Burford {\it et al.}), Nucl. Phys {\bf B447} 425 (1995);
L. Lellouch, Nucl. Phys {\bf B479} 353 (1996); L. Lellouch,
private communication; UKQCD Collaboration (L. Del Debbio {\it et
al.}), Phys. Lett. {\bf B416} (1998) 392.

\bibitem{16art4} P. Ball in Proceedings of 33rd Rencontres de Moriond, QCD and High Energy
Hadronic Interactions, Les Arcs, France, 21-28 March 1998,
hep-ph/9803501.

\bibitem{21art4} M.A. Ivanov and P. Santorelli, hep-ph/9903446.




                \end{thebibliography}
\end{document}